\documentclass[11pt]{article}

\usepackage{amsfonts,amssymb,amsthm, amsmath, mathtools,graphicx}
\usepackage{fullpage}
\usepackage{changepage}
\usepackage{enumitem}
\usepackage{scalerel}
\usepackage{accents}
\usepackage[margin=1.in]{geometry}
\usepackage{layout}
\usepackage{printlen}
\usepackage{booktabs}
\usepackage{float}

\usepackage{optidef}

\usepackage{nicefrac,xfrac}

\usepackage[toc,page]{appendix}

\usepackage{makecell}
\usepackage{tablefootnote}
\usepackage{lipsum}
\usepackage{caption}
\usepackage{subcaption}
\usepackage{hyperref}
\usepackage{mfirstuc}
\usepackage{multirow}
\usepackage{array}
\usepackage{complexity}

\usepackage{natbib}
\usepackage{csquotes}

\usepackage{bbm}

\usepackage[noabbrev]{cleveref}
\crefname{claim}{claim}{claims}
\crefname{enumi}{property}{properties}
\crefname{adversary}{adversary}{adversaries}

\Crefname{subfigure}{Figure.}{Figures.}

\usepackage{tikz}
\usepackage{pgfplots}
\pgfplotsset{width=10cm,compat=1.9}
\definecolor{darkpastelgreen}{rgb}{0.01, 0.75, 0.24}
\definecolor{darkorchid}{rgb}{0.6, 0.2, 0.8}
\definecolor{blue(pigment)}{rgb}{0.2, 0.2, 0.6}

\usepackage[ruled,vlined,linesnumbered]{algorithm2e}
\SetKw{Break}{break}

\newcounter{adversarycf}         \newcounter{algocfbackup}        

\makeatletter
\newenvironment{adversary}[1][]{\refstepcounter{adversarycf}\setcounter{algocfbackup}{\value{algocf}}\setcounter{algocf}{\value{adversarycf}}\begin{algorithm}[#1]}{\end{algorithm}
  \setcounter{algocf}{\value{algocfbackup}}}
\makeatother

\SetKwProg{KwInit}{init}{}{end}
\Crefname{algorithm}{Algorithm}{Algorithms}
\allowdisplaybreaks

\usetikzlibrary{arrows, decorations.markings}

\tikzstyle{vecArrow} = [thick, decoration={markings,mark=at position
   1 with {\arrow[semithick]{open triangle 60}}},
   double distance=1.4pt, shorten >= 5.5pt,
   preaction = {decorate},
   postaction = {draw,line width=1.4pt, white,shorten >= 4.5pt}]
\tikzstyle{innerWhite} = [semithick, white,line width=1.4pt, shorten >= 4.5pt]

\allowdisplaybreaks
\theoremstyle{plain}
\newtheorem{theorem}{Theorem}[section]
\newtheorem{lemma}[theorem]{Lemma}

\newtheorem{observation}[theorem]{Observation}
\newtheorem*{theorem*}{Meta-Theorem}

\theoremstyle{plain}
\newtheorem{definition}{Definition}[section] \newtheorem{example}[definition]{Example}
\newtheorem{remark}[definition]{Remark}

\theoremstyle{plain}
\newtheorem{assumption}{Assumption}

\usepackage{xfrac}
\usepackage{thm-restate}

\newcommand{\xhdr}[1]{\vspace{2mm} \noindent{\bf #1}}

\newcommand{\abs}[1]{\left| #1 \right|}
\newcommand{\factorial}[1]{{#1}!}

\newcommand{\action}{a}

\newcommand{\actionAgent}{\action_{\timeslot}}
\newcommand{\actionT}{\action_{\timeslot}}
\newcommand{\actionOptT}{\action^*_{\timeslot}}
\newcommand{\actionTOrder}[1]{\actionT\super{#1}}

\newcommand{\actions}{A}
\newcommand{\actionsT}{\actions_{\timeslot}}
\newcommand{\actionNum}{n}
\newcommand{\actionsKnown}{\actions^{-}}
\newcommand{\actionsKnownT}{\actions^{-}_{\timeslot}}

\newcommand{\actionsUnknownT}{\actions^{+}_t}

\newcommand{\cost}{c}
\newcommand{\costA}[1]{\cost\super{#1}}
\newcommand{\costAT}[1]{\costA{#1}_{\timeslot}}
\newcommand{\costLB}{\underline{C}}

\newcommand{\reward}{r}
\newcommand{\rewardA}[1]{\reward\super{#1}}
\newcommand{\rewardAT}[1]{\rewardA{#1}_{\timeslot}}
\newcommand{\nonDeReward}[1]{\bar{\reward}_{\timeslot}(#1)}

\newcommand{\revenuePrincipal}{V_{\timeslot}}
\newcommand{\revenuePrincipalCon}{{\Pi}}
\newcommand{\revenuePrincipalConPessimistic}{\tilde{\Pi}}
\newcommand{\regretImmediate}{\mathcal{R}_{\timeslot}}
\newcommand{\regret}{\mathcal{R}}

\newcommand{\utiAgent}{U}

\newcommand{\tpre}{\tau}

\newcommand{\volume}[1]{\mathrm{Vol}\left(#1\right)}
\newcommand{\ball}{\mathbb{B}_\dimension}

\newcommand{\vecBall}{\mathbb{B}}

\newcommand{\contractSimple}{x}
\newcommand{\contract}[1]{\contractSimple_{#1}}
\newcommand{\contracts}{\mathcal{X}}
\newcommand{\nonDeContract}[1]{\bar{\contractSimple}_{\timeslot}(#1)}
\newcommand{\contractOptT}{\contract{\timeslot}^*}
\newcommand{\optContract}[1]{\contractSimple\paren{#1}}
\newcommand{\contractTOrder}[1]{\contractSimple_{\timeslot}\super{#1}}

\newcommand{\hvec}{\theta} \newcommand{\hvecTrue}{\hat{\hvec}}
\newcommand{\hvecOptT}{\hvec_{\timeslot}^*}
\newcommand{\hvecAOptT}[1]{\hvec_{\timeslot}\super{#1}}
\newcommand{\hvecs}{\Theta}
\newcommand{\hvecsPess}{\hat{\hvecs}}

\newcommand{\context}{\mu}
\newcommand{\contextTA}[2]{\context_{#1}^{(#2)}}
\newcommand{\contextAT}[1]{\context_{\timeslot}^{(#1)}}
\newcommand{\contextA}[1]{\context^{(#1)}}
\newcommand{\timeslot}{t}
\newcommand{\timeHorizon}{T}

\newcommand{\instance}[2]{\instanceSimple_{\timeslot}\left(#1,#2\right)}
\newcommand{\instanceSimple}{\mathcal{I}}
\newcommand{\instanceKnown}{\instanceSimple(\actionsKnownT)}
\newcommand{\optProfit}[1]{\texttt{OPT}\left(#1\right)}
\newcommand{\optProfitFromA}[2]{\texttt{OPT}\left(#1,#2\right)}

\newcommand{\leftCostGap}[2]{\Delta\left(#1;#2\right)}

\newcommand{\leftGapThreshold}{\Delta}

\newcommand{\dimension}{d}

\newcommand{\informationWidth}[3]{\mathtt{InfoWidth}\left(#1,#2;#3\right)}
\newcommand{\intrinsicVolume}[2]{V_{#1}\left(#2\right)}
\newcommand{\potential}[2]{\phi_{#1}\left(#2\right)}
\newcommand{\potentialSum}[1]{\Phi\left(#1\right)}

\newcommand{\jumpVec}[2]{\context_{\timeslot}\super{#1,#2}}
\newcommand{\jumpSet}{\mathcal{J}_{\timeslot}}

\newcommand{\gap}{\Delta}

\newcommand{\mincost}{\cost_{\min}}

\newcommand{\nullAction}
{\blacklozenge}
\newcommand{\knownOptAction}{\clubsuit}
\newcommand{\restrictingAction}{\spadesuit}
\newcommand{\adversarialAction}{\heartsuit}
\newcommand{\radius}{R}
\newcommand{\rewardLargestPossible}{\beta}
\newcommand{\equalRevenueReward}{\gamma}

\newcommand{\contractThreshold}{X}
\newcommand{\determineThreshold}{\mathbf{K}}

\newcommand{\sphericalCode}{\Omega_{\dimension}}
\newcommand{\sphericalCodedim}[1]{ \Omega_{#1}}
\newcommand{\codeAngle}{\alpha}
\newcommand{\codeWord}[1]{\omega\super{#1}}

\newcommand{\agentRevenueFunc}[1]{h(#1)}
    \newcommand{\unknownRewardFunc}[1]{f(#1)}
    \newcommand{\unknownRewardFuncDerivative}[1]{f'(#1)}

\newcommand{\convexBody}{\hvecs}
\newcommand{\convexBodyNext}{\hvecs^+}
\newcommand{\cone}{C}
\newcommand{\coneBase}{B}
\newcommand{\coneApex}{\hvec_p}

\newcommand{\regretPessimistic}{\tilde{\regret}}

\newcommand{\actionNull}{0}
\newcommand{\contractGeneral}{\pi}
\newcommand{\contractGeneralA}[1]{\pi\super{#1}}
\newcommand{\actionBest}[1]{\action_{#1}}

\newcommand{\ALG}{\texttt{ALG}}
\newcommand{\PA}{Principal-agent game\xspace}

\newcommand{\val}{v}
\newcommand{\price}{p}

\newcommand{\klx}{-1.7} \newcommand{\klxNPoffset}{.8} \newcommand{\NPy}{3.1} \newcommand{\NPyOffset}{1.1}

\newcommand{\klxPoffset}{1.8} \newcommand{\pathA}{[scale=1.05]
  (0,2) .. controls (0,3.8) and (-1.2, 4) .. (-2,4)
          .. controls (-3.8,4) and (-6.5,2.8) .. (-6.5,2)
          .. controls (-6.5,1.2) and (-3.8,0) .. (-2,0)
          .. controls (-1.2,0) and (0,.2) .. (0,2)
}
\definecolor{mypink}{RGB}{255,204,204}  \definecolor{mybrown}{RGB}{210,180,140}  \definecolor{mygreen}{RGB}{153,179,153}  \definecolor{myblue}{RGB}{160,200,235}  \definecolor{mylineblue}{RGB}{65,105,255}  \definecolor{mylinegreen}{RGB}{34,139,34}

\newcommand{\leftmostX}{-6.85}
\newcommand{\safeConx}{-4.3}
\newcommand{\infoYUpper}{4.3}
\newcommand{\infoYLower}{0}
\newcommand{\klxdash}{\klx-3.8}
\newcommand{\myfigboxOne}{4.5cm}
\newcommand{\myfigboxTwo}{3.8cm}
\newcommand{\klYUpper}{4.7}
\newcommand{\klYLower}{-0.5}
\newcommand{\myscale}{.8}

\newcommand{\LLS}{\texttt{LLS}}
\newcommand{\timeset}{\mathcal{T}} 
\newcommand{\set}[1]{\left\{#1\right\}}

\newcommand{\lrangle}[1]{\langle #1 \rangle}
\newcommand{\paren}[1]{\left(#1\right)}
\newcommand{\parenfix}[1]{(#1)}

\newcommand{\super}[1]{^{(#1)}}

\DeclareMathOperator*{\argmax}{arg\,max}
\DeclareMathOperator*{\argmin}{arg\,min}

\newcommand{\reals}{\mathbb{R}}
\newcommand{\nnreals}{\reals_{+}}
\newcommand{\naturals}{\mathbb{N}}

\newcommand{\deq}{\triangleq}

\newcommand{\innerproduct}[1]
{\langle#1\rangle}

\newcommand{\condition}{\,\mid\,}

\newcommand{\prob}[2][]{\text{Pr}\ifthenelse{\not\equal{}{#1}}{_{#1}}{}\!\left[{\def\givenn{\middle|}#2}\right]}
\newcommand{\expect}[2][]{\mathbb{E}\ifthenelse{\not\equal{}{#1}}{_{#1}}{}\!\left[{\def\givenn{\middle|}#2}\right]}
\newcommand{\norm}[1]{\left\|#1\right\|}

\newcommand{\tparen}{\big}
\newcommand{\tprob}[2][]{\text{Pr}\ifthenelse{\not\equal{}{#1}}{_{#1}}{}\tparen[{\def\given{\tparen|}#2}\tparen]}
\newcommand{\texpect}[2][]{\mathbb{E}\ifthenelse{\not\equal{}{#1}}{_{#1}}{}\tparen[{\def\given{\tparen|}#2}\tparen]}

\newcommand{\sprob}[2][]{\text{Pr}\ifthenelse{\not\equal{}{#1}}{_{#1}}{}[#2]}
\newcommand{\sexpect}[2][]{\mathbb{E}\ifthenelse{\not\equal{}{#1}}{_{#1}}{}[#2]}

\newcommand{\realNumbers}{\mathbb{R}}

\newcommand{\eulerNumber}{e}

\newcommand{\clipX}{-1} \newcommand{\clipY}{-2.5} \newcommand{\clipRectX}{15} \newcommand{\clipRectY}{5} \newcommand{\xxScale}{8}
\newcommand{\xScale}{.9} \newcommand{\yScale}{.7} \newcommand{\cOne}{0.19} \newcommand{\cTwo}{1.1} \newcommand{\cThree}{4.4} \newcommand{\cFour}{2.1} \newcommand{\cFive}{\cFour} \newcommand{\ROne}{1.25} \newcommand{\RTwo}{3.8} \newcommand{\RThree}{8.8} \newcommand{\RFour}{5.9} \pgfmathsetmacro{\RFive}{((\cTwo-\cThree)/(\RTwo-\RThree)*\RTwo-\cTwo + \cFive)/( (\cTwo-\cThree)/(\RTwo-\RThree) )} \pgfmathsetmacro{\alphaOne}{\cOne/\ROne}
\pgfmathsetmacro{\alphaTwo}{(\cTwo-\cOne)/(\RTwo-\ROne)}
\pgfmathsetmacro{\alphaThree}{(\cThree-\cTwo)/(\RThree-\RTwo)}
\pgfmathsetmacro{\intersectionOneTwo}{(\ROne*\cTwo-\RTwo*\cOne)/(\RTwo-\ROne)}
\pgfmathsetmacro{\intersectionTwoThree}{(\RTwo*\cThree-\RThree*\cTwo)/(\RThree-\RTwo)}  
\title{Contextual Search in Principal-Agent Games: \\ The Curse of Degeneracy}

\author{Yiding Feng\thanks{Hong Kong University of Science and Technology, China. Email: \url{ydfeng@ust.hk}}
\and
Mengfan Ma\thanks{Central China Normal University, China. Email: \url{mengfanma1@gmail.com}}
\and
Bo Peng\thanks{Shanghai University of Finance and Economics, China. Email: \url{ahqspb@163.sufe.edu.cn}}
\and 
Zongqi Wan\thanks{Great Bay University, China. Email: \url{zongqiwan98@gmail.com}}
}
\date{}

\begin{document}

\maketitle

\begin{abstract}

In this work, we introduce and study contextual search in general principal-agent games, where a principal repeatedly interacts with agents by offering contracts based on contextual information and historical feedback, without knowing the agents' true costs or rewards. Our model generalizes classical contextual pricing by accommodating richer agent action spaces. Over $T$ rounds with $\dimension$-dimensional contexts, we establish an asymptotically tight exponential $\timeHorizon^{1 - \Theta(1/\dimension)}$ bound in terms of the \emph{pessimistic Stackelberg regret}, benchmarked against the best utility for the principal that is consistent with the observed feedback.
    
We also establish a lower bound of $\Omega(T^{\frac{1}{2}-\frac{1}{2\dimension}})$ on the \emph{classic Stackelberg regret} for principal-agent games, demonstrating a surprising double-exponential hardness separation from the contextual pricing problem (a.k.a, the principal-agent game with two actions), which is known to admit a near-optimal $O(\dimension\log\log \timeHorizon)$ regret bound~\citep{KL-03,LS-18, LLS-21}. In particular, this double-exponential hardness separation occurs even in the special case
with three actions and two-dimensional context.
We identify that this significant increase in learning difficulty arises from a structural phenomenon that we call \emph{contextual action degeneracy}, where adversarially chosen contexts can make some actions strictly dominated (and hence unincentivizable), blocking the principal’s ability to explore or learn about them, and fundamentally limiting learning progress.

 \end{abstract}

\thispagestyle{empty}
\newpage
\setcounter{page}{1}

\section{Introduction}
\label{sec:intro}
The \emph{contextual pricing problem} 
\citep{LLV-18,LS-18,CLP-20,LLS-21}
provides a quintessential example of learning under partial feedback in a market setting. In this model, a seller (learner) repeatedly offers prices for products described by observed context vectors, and only \emph{binary feedback} is received in each round – the buyer either purchases the product at the offered price or not. The buyer’s valuation is assumed to be an unknown linear function of the context, so the seller faces the challenge of \emph{learning the buyer’s valuation} while setting prices. Regret is measured as the revenue loss relative to a clairvoyant seller who knows the buyer’s valuation in advance. A sequence of works culminating in \citet{LLS-21} established that this pricing problem admits an exceedingly small optimal regret of order $O(\dimension\log \log \timeHorizon+\dimension\log\dimension)$ over $\timeHorizon$ rounds, where $\dimension$ is the dimension of the context vector. In other words, with an optimal pricing strategy, the seller’s regret grows only double-logarithmically in $\timeHorizon$, a nearly negligible rate.

In this paper, we introduce a new model of \emph{contextual search in principal-agent games} that extends the classical framework of contextual pricing to accommodate richer action spaces and information structure. In this setting, a principal interacts with a sequence of agents over $\timeHorizon$ rounds. In each round, the principal observes contextual information associated with each available action of the arriving agent and offers a \emph{linear contract}, specifying a fraction of the reward to be transferred from the principal to the agent.\footnote{While the contract design literature often assumes hidden-action models, we consider a deterministic setting—within the broader principal-agent framework—where outcomes reveal the agent's action. Some other works also consider settings where the agent’s action is either directly observed or deterministically inferred by the principal, including \citet{LR-25, LMTJR-25, STBCMJEO-24, DSA-23, HAX-24, BMMT-24}. We thus retain the term ``contract'' for consistency with the literature. See more discussion in \Cref{footnote:prelim:linear contract} and \Cref{remark:linear contract interpretation}.} Based on this contract, the agent selects one of the available actions to maximize their own utility, which depends on the payment (i.e., the fraction of reward transferred) from the principal and the action's cost. The agent receives the payment minus the incurred cost, while the principal gains the reward associated with the chosen action, net of the payment made to the agent.
In our contextual search model, the principal does not observe the actions' costs or rewards directly. Instead, each action is associated with a context vector, and the principal observes these contexts. We consider two canonical settings: the \emph{cost-context setting} and the \emph{reward-context setting}. In the cost-context setting, the cost of each action is given by the inner product between its context and an unknown hidden vector, while the reward is known to the principal. Conversely, in the reward-context setting, the reward is the inner product of the context and the hidden vector, and the cost is known.

It is not hard to observe that the pricing problem is a special case of our principal-agent model with two actions (``purchase'' vs.\ ``no purchase''), and that the contextual pricing problem corresponds to a special case of our proposed contextual search problem under the principal-agent model with two actions and cost context.\footnote{Besides dynamic pricing, our contextual model of principal-agent game admits many realistic applications such as crowdsourcing markets, product development, and R\&D. See \Cref{apx:practical application} for more discussion.} Notably, although the contextual pricing problem admits a relatively simple mathematical formulation, a variety of novel algorithmic ideas (e.g., asymmetric exploration due to asymmetric loss functions, \citealp{KL-03}) and analytical techniques (e.g., cylindrification operation in \citealp{LLV-18}, potential-based argument leveraging intrinsic volumes in \citealp{LS-18}) have been developed for it, and, as mentioned above, it remarkably admits double-logarithmic regret guarantees.

From a theoretical perspective, our model arguably offers one of the most natural generalizations of contextual pricing--extending from two actions to multiple actions. In addition, from a practical perspective, it captures a broad range of real-world applications. Motivated by these considerations, we aim to investigate the following two questions:

\vspace{-2mm}
\begin{displayquote}
\emph{Does the (double-)logarithmic regret guarantee established for contextual pricing with binary actions extend to contextual search in general principal-agent games with multiple actions? Can the algorithmic ideas and analytical techniques developed for contextual pricing be generalized to this broader contextual search setting?}
\end{displayquote}
\vspace{-5mm}

\begin{table}[t]
    \centering
\resizebox{\textwidth}{!}{\begin{tabular}{lccc}
        \toprule
        \textbf{Setting} 
        & 
        \makecell{\textbf{Pessimistic Stackelberg} \\ \textbf{regret (upper bound)}} 
        & 
        \makecell{\textbf{Pessimistic Stackelberg} \\ \textbf{regret (lower bound)}}  
        & 
        \makecell{\textbf{Stackelberg} \\ \textbf{regret (lower bound)}}    
        \\
        \midrule
        Contextual pricing
            & \makecell{$O(\dimension\log\log \timeHorizon+\dimension\log\dimension)$ \\ \citep{LLS-21}}
            & \makecell{$\Omega(\dimension\log\log \timeHorizon)$ \\ \citep{LS-18}}
            & \makecell{$\Omega(\dimension\log\log \timeHorizon)$ \\ \citep{LS-18}} 
            \\
            \midrule
        \PA without context
            & \makecell{$O(\log \log \timeHorizon)$\\ \citep{DGSW-23}\footnotemark[2]}
            & N/A
            & \makecell{$\Omega(\log \log \timeHorizon)$ \\ \citep{KL-03}}
            \\
            \midrule
        \PA with cost context 
            & \makecell{$O\parenfix{\dimension^{\frac{1}{4}}\cdot\timeHorizon^{1 - \frac{1}{2\dimension+1}}}$ \\ {[}Thm.~\ref{thm:cost context upper bound}{]}}
            & \makecell{$\Omega\parenfix{\timeHorizon^{1 - \frac{1}{\dimension}}}$ \\ {[}Thm.~\ref{thm:lower-bound-cost-opti}{]}}
            & \makecell{$\Omega(T^{\frac{1}{2}-\frac{1}{2d}})$ \\ {[}Thm.~\ref{thm:lower-bound-true}{]}}
            \\
            \midrule
        \PA with reward context
            & \makecell{$O\parenfix{\dimension^{\frac{1}{4}}\cdot\timeHorizon^{1 - \frac{1}{2\dimension}}+\dimension}$ \\ {[}Thm.~\ref{thm:reward context upper bound}{]}}
            & \makecell{$\Omega\parenfix{\timeHorizon^{1- \frac{1}{\dimension}}}$ \\ {[}Thm.~\ref{thm:lower-bound-reward-opti}{]}}
            & \makecell{$\Omega(T^{\frac{1}{2}-\frac{1}{2d}})$ \\ {[}Thm.~\ref{thm:lower-bound-reward-true}{]}}
            \\
        \bottomrule
\end{tabular} }
    \caption{Comparison of regret bounds across contextual pricing and contextual search in principal-agent games. The upper bounds for the principal-agent settings correspond to pessimistic Stackelberg regret. Since classic Stackelberg regret is always no larger than pessimistic regret, our upper bounds also apply to the classic regret. Notably, contextual pricing corresponds to a special case of our setting with only two actions. Our lower bound for $\dimension = 2$ holds even for three-action instances, while for $\dimension > 2$, it is established using four-action instances.
\label{tab:regret-comparison}
    }
\end{table}

\footnotetext[2]{The model in \citet{DGSW-23} specializes the principal-agent game to a setting without context, i.e., the costs and rewards are fixed in all rounds. As a result, the pessimistic Stackelberg regret is not defined, and their results apply only to the classic Stackelberg regret.}
\stepcounter{footnote}

\subsection{Our Contributions and Techniques}\label{sec:contributions}

In this paper, we provide compelling answers to both of these questions. In a nutshell, we show that the (double-)logarithmic regret guarantee breaks down as soon as we move beyond the two-action setting (i.e., contextual pricing). However, several algorithmic ideas and technical tools developed in the literature can still be adapted to our more general setting. Below, we summarize our main results and techniques. For a summary of the regret guarantees established in this work and a comparison with prior literature, see \Cref{tab:regret-comparison}.

\xhdr{Model setups.}
In the first part of the paper, we focus on the \emph{cost-context setting}. In each round, when a new agent arrives, the principal (online algorithm) observes the action set and the reward associated with each action. Both the action set and the reward profile can vary across rounds. However, the cost associated with each action is not directly observed. Instead, each action is linked to a $\dimension$-dimensional cost context, which is observable by the principal. The cost of each available action for a given agent follows a linear structure, i.e., it is the inner product of the corresponding context and an unknown cost hidden vector $\hvecTrue$. We assume that all actions from all agents follow the same linear structure, meaning that $\hvecTrue$ is both agent-independent and action-independent.
Although the cost hidden vector is not directly known to the principal (as implied by its name), it can be learned through interaction with the agents over multiple rounds. The principal's goal is to design an online algorithm that maximizes her long-term utility. Following the contextual search literature, we impose no stochastic assumptions on the arrival pattern. Specifically, both the agents' action set, reward profiles, and cost context profiles are designed by an adaptive adversary.

In the second part of the paper, we study the \emph{reward-context setting}, which is symmetric to the cost-context setting. In this case, the cost profile of each agent is directly observed, while the reward for a given agent and action follows a linear structure: it is the inner product of an observed reward context and an unknown reward hidden vector. A key difference in the reward-context setting is that, once an action is selected by the agent, the principal not only observes the action but also the induced reward, i.e., the value of the inner product between the reward context and the unknown reward. In contrast, in the cost-context setting, the cost of the selected action is not observable.\footnote{Although the reward-context setting may initially appear to be an easier learning problem, we obtain the same regret upper and lower bounds in both settings. }

\xhdr{Double-exponential separation from contextual pricing.}
Our problem is closely related to several classic problems in the literature. As mentioned earlier, contextual dynamic pricing \citep{KL-03,LLS-21} is a special case of our model with two actions (see \Cref{sec:contextual pricing connection} for a formal reduction). When the contexts remain identical (i.e., the online algorithm faces the same principal-agent game) across all rounds (or equivalently, the non-contextual version of our model), our problem reduces to the task of optimizing a one-sided Lipschitz function with a domain between $[0, 1]$, which was recently studied in \citet{DGSW-23}. Notably, in both contextual dynamic pricing (a special case of our model with two actions) and optimizing a one-sided Lipschitz function (which generalizes to the non-contextual version of our model), it has been shown that the optimal Stackelberg regret exhibits a $\log\log T$ dependence on the number of rounds, $T$ (see \Cref{tab:regret-comparison}).

As the first (conceptual) contribution of this work, we show that the (double-)logarithmic regret observed in the prior models is fragile and quickly breaks down in our more general model. Even in the setting with three actions and two-dimensional contexts, we show that the Stackelberg regret of every online algorithm is at least $\Omega(T^{\frac{1}{4}})$ (\Cref{thm:warm-up:three action regret lower bound}), which is \emph{double-exponentially} worse than the regret guarantees in the prior models. For more general setting with $\dimension$-dimensional contexts, we prove a Stackelberg regret lower bound of $\Omega(T^{\frac{1}{2}-\frac{1}{2d}})$ with only four actions (\Cref{thm:lower-bound-true}).

\xhdr{Adversary construction with contextual action degeneracy.}
The primary driving force behind this double-exponential hardness separation arises from a structural phenomenon in our model, which we refer to as \emph{contextual action degeneracy}. In the principal-agent game, an action is considered degenerated if there is no contract that can incentivize the agent to choose this action, meaning its payoff for the agent is dominated by other actions in all linear contracts. The contextual action degeneracy occurs when a key unknown action remains degenerated for most possible values of the hidden vector $\hvecTrue$. This structure creates a \emph{``needle in a haystack'' challenge}: an exploration is only informative if it targets one of the very few non-degenerated hidden vectors. Consequently, the principal is forced to explore exhaustively, with each exploration action only capable of eliminating a single, small ``spherical cap'' of the uncertainty space (aka., hypothesis set), all while incurring regret. Loosely speaking, this provides the adversary with sufficient leverage to prevent online algorithms from achieving (double-)logarithmic regret.

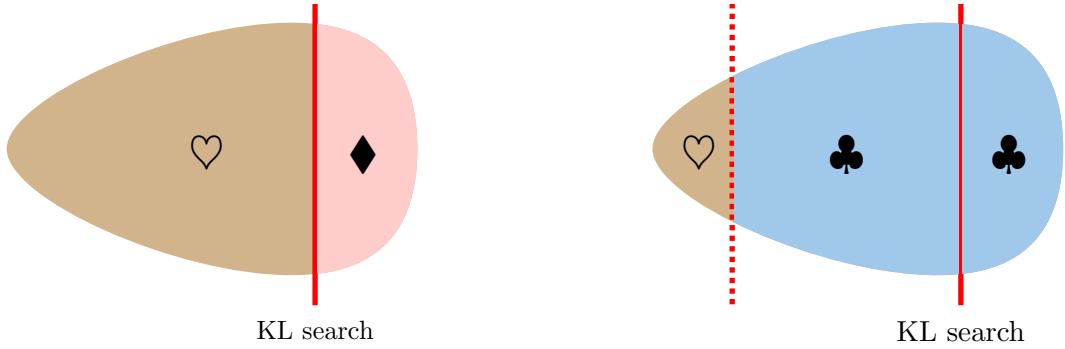
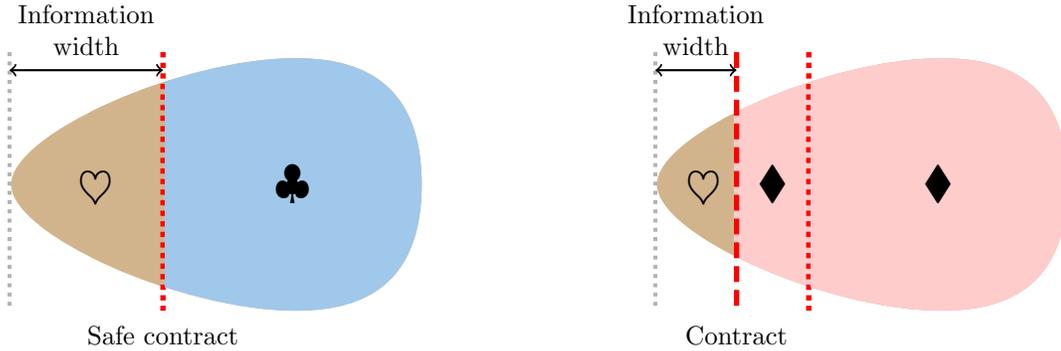
\begin{figure}
  \centering
  \begin{subfigure}[t]{0.48\textwidth}
    \centering
    \begin{tikzpicture}[scale=\myscale]

\clip (-7.5,-1.1) rectangle (2.5,4.5);
  
\fill[mybrown] \pathA;

\draw[line width=2pt, color=red, font=\small] (\klx,\klYUpper)--(\klx,\klYLower);
    \node[below=2pt,  font=\small] at (\klx,\klYLower) {KL search};

\begin{scope}
        \clip (\klx+0.035,-2) rectangle (2,6); \fill[mypink] \pathA;   \end{scope}
  
\node[font=\LARGE] at (\klx+\klxNPoffset,\NPy-\NPyOffset) {$\nullAction$};

\node[font=\LARGE] at (\klx-\klxPoffset,\NPy-\NPyOffset) {$\adversarialAction$};

\end{tikzpicture}     \parbox[t][\myfigboxOne][t]{\textwidth}{\caption{KL search in the contextual pricing problem: when ``purchase'' action $\adversarialAction$ (``no purchase'' action~$\nullAction$) is incentivized (i.e., preferred by the agent) and observed, the hypothesis set shrinks to the brown (pink) subregion, which has high (low) intrinsic volume and low (high) immediate regret approximately equal to the contextual width (a constant).}
    \label{fig:KL search in dynamic pricing}
    }
    
  \end{subfigure}
  \hfill
  \begin{subfigure}[t]{0.48\textwidth}
    \centering
    \begin{tikzpicture}[scale=\myscale]
\clip (-7.5,-1.1) rectangle (2.5,4.5);

\fill[mybrown] \pathA;

\draw[line width=2pt, color=red, font=\small] (\klx,\klYUpper)--(\klx,\klYLower);
    \node[below=2pt] at (\klx,\klYLower) {KL search};

\draw[line width=2pt, color=red, dotted] (\klxdash,\klYUpper)--(\klxdash,\klYLower);
    \node[below=2pt,  font=\small] at (\klxdash,\klYLower) {};

\begin{scope}
        \clip (\klx+0.035,-2) rectangle (2,6); \fill[myblue] \pathA;   \end{scope}
  
\node[font=\LARGE] at (\klx+\klxNPoffset,\NPy-\NPyOffset) {$\knownOptAction$};

\begin{scope}
        \clip (\klxdash+0.035,-2) rectangle (\klx-0.035,6); \fill[myblue] \pathA;   \end{scope}

	\node[font=\LARGE] at ({(\klx+\klxdash)/2},\NPy-\NPyOffset) {$\knownOptAction$};

    \node[font=\LARGE] at (\klxdash-.55,\NPy-\NPyOffset){$\adversarialAction$};

\end{tikzpicture}     \parbox[t][\myfigboxOne][t]{\textwidth}{\caption{Failure of KL search in our adversary: under carefully constructed contexts, the hidden vector lies to the left of the solid red line (KL search) iff ``purchase'' action $\adversarialAction$ is preferred to ``no purchase'' action $\nullAction$. However, dominating action~$\knownOptAction$ is preferred to ``purchase'' action $\adversarialAction$ iff the hidden vector lies to the right of the dotted red line. Thus, when dominating action $\knownOptAction$ is incentivized and observed, the hypothesis set shrinks to the blue subregion, which has large intrinsic volume and incurs large regret.}
    \label{fig:failure of KL search}
    }
  \end{subfigure}

  \vspace{1em}

    \begin{subfigure}[t]{0.48\textwidth}
    \centering
    \begin{tikzpicture}[scale=\myscale]
\clip (-7.5,-0.6) rectangle (0.1,5.1);
    
\fill[mybrown] \pathA;
    
\begin{scope}
        \clip (\safeConx+0.035,-2) rectangle (\klxdash-0.035,6); \fill[mybrown] \pathA;   \end{scope}
    
\draw[dotted, line width=2pt, color=red] (\safeConx,\infoYUpper)--(\safeConx,\infoYLower);
    \node[below=2pt, font=\small] at (\safeConx,\infoYLower) {Safe contract};

\draw[line width=1.5pt, color=white!70!black, dotted] (\leftmostX,\infoYUpper)--(\leftmostX,\infoYLower);

\begin{scope}
        \clip (\safeConx+0.035,-2) rectangle (2,6); \fill[myblue] \pathA;   \end{scope}
  
    \node[font=\LARGE] at ({\safeConx/2},\NPy-1) {$\knownOptAction$};

    \node[font=\LARGE] at ({(\leftmostX+\safeConx)/2+0.15},\NPy-\NPyOffset){$\adversarialAction$}; 

\draw[<->, thick] (\leftmostX,\infoYUpper-0.3) -- (\safeConx,\infoYUpper-0.3);
    \node[align=center, above=2pt, font=\small] at ({(\leftmostX+\safeConx)/2},\infoYUpper-0.3) {Information\\width};

\end{tikzpicture}     \parbox[t][\myfigboxTwo][t]{\textwidth}{\caption{Tradeoff in our adversary construction: by deploying the ``safe'' contract (red dotted line), either dominating action $\knownOptAction$ or ``purchase'' action $\adversarialAction$—whichever yields higher utility—will be incentivized. The immediate regret approximately equal to the information width.}
    \label{fig:adversary tradeoff}
    }
  \end{subfigure}
    \hfill 
    \begin{subfigure}[t]{0.48\textwidth}
    \centering
    \begin{tikzpicture}[scale=\myscale]
\clip (-7.5,-0.6) rectangle (0.1,5.1);
    
\fill[mybrown] \pathA;
    
\begin{scope}
        \clip (\safeConx+0.035,-2) rectangle (\klxdash-0.035,6); \fill[mypink] \pathA;   \end{scope}
    
\draw[dotted, line width=2pt, color=red] (\safeConx,\infoYUpper)--(\safeConx,\infoYLower);
    \node[below=2pt, font=\small] at (\safeConx,\infoYLower) {};

\draw[dash pattern=on 6pt off 3pt, line width=2pt, color=red] (\klxdash,\infoYUpper)--(\klxdash,\infoYLower);
    \node[below=2pt,  font=\small] at (\klxdash,\infoYLower) {Contract};

\draw[line width=1.5pt, color=white!70!black, dotted] (\leftmostX,\infoYUpper)--(\leftmostX,\infoYLower);

\begin{scope}
        \clip (\safeConx+0.035,-2) rectangle (2,6); \fill[mypink] \pathA;   \end{scope}
  
    \node[font=\LARGE] at ({\safeConx/2},\NPy-1) {$\nullAction$};

\node[font=\LARGE] at ({(\safeConx+\klxdash)/2},\NPy-1) {$\nullAction$};

    \node[font=\LARGE] at (\klxdash-.55,\NPy-\NPyOffset){$\adversarialAction$}; 

\draw[<->, thick] (\leftmostX,\infoYUpper-0.3) -- (\klxdash,\infoYUpper-0.3);
    \node[align=center, above=2pt, font=\small] at ({(\leftmostX+\klxdash)/2},\infoYUpper-0.3) {Information\\width};

\end{tikzpicture}     \parbox[t][\myfigboxTwo][t]{\textwidth}{\caption{Suppose a contract (red dashed line) that is more risky than the safe contract (red dotted line) is deployed.
    Then dominating action~$\knownOptAction$ has non-positive utility and will never be incentivized. When ``purchase'' action $\adversarialAction$ (``no purchase'' action~$\nullAction$) is incentivized and observed, the hypothesis set shrinks to the brown (pink) subregion which has low (high) immediate regret approximately equal to the information width (constant). 
}
    \label{fig:adversary general contract}
    }
  \end{subfigure}

  \caption{Illustration of the KL search in contextual pricing problem and the three-action adversary (\Cref{adv:warm up:three-action setting}) in our contextual principal–agent problem. The actions $\nullAction$, $\adversarialAction$, and $\knownOptAction$ correspond to the ``no purchase'', ``purchase'', and ``dominating'' actions, respectively. The convex region represents the current hypothesis set for the hidden vector. Colored subregions indicate the updated hypothesis sets conditioned on the different actions incentivized and observed by the principal.}
  \label{fig:kl-pair}
\end{figure}

For readers familiar with the contextual pricing literature, recall that a Stackelberg regret of $O(d\log T)$ can be easily achieved using the Ellipsoid method. By incorporating the asymmetric structure of the pricing loss function, the optimal regret of $O(d\log\log T)$ can be achieved. In both of these results, a crucial algorithmic idea is ensuring the desired shrinkage of the hypothesis set for the hidden vector. For instance, to achieve $O(d\log T)$, it suffices to shrink the hypothesis set by half with each round. To achieve $O(d\log\log T)$, a more aggressive shrinkage is required when the agent does not purchase (leading to constant immediate regret), and a less aggressive shrinkage when the agent purchases (resulting in immediate regret proportional to the ``width'' of the hypothesis set). We refer to this adaptive search strategy as \emph{KL search} \citep{KL-03}; see \Cref{fig:KL search in dynamic pricing} for an illustration.

However, in our more general model, the adversary can limit the online algorithm's ability to shrink the hypothesis set by making some actions degenerated for most points in the hypothesis set. This creates an \emph{insufficient learning challenge} for the algorithm. Concretely, in \Cref{sec:contextual pricing connection}, we construct an adversary (\Cref{adv:warm up:three-action setting}) with three actions: $\{\nullAction, \knownOptAction, \adversarialAction\}$. Here, $\nullAction$ and $\adversarialAction$ correspond respectively to the known ``no purchase'' and unknown ``purchase'' actions from contextual pricing, while $\knownOptAction$---which we call the \emph{dominating action}---is a known, high-performing action introduced by the adversary to induce {contextual action degeneracy}. By carefully choosing the context, the adversary ensures that, for a large fraction of the hypothesis set, the dominating action $\knownOptAction$ strictly outperforms the unknown ``purchase'' action $\adversarialAction$.
Consequently, if the online algorithm employs a linear contract corresponded to KL search, it fails to achieve the same shrinkage guarantees as in contextual pricing. In standard pricing, one can always observe whether the ``purchase'' action $\adversarialAction$ is preferable to the no purchase action $\nullAction$, enabling informative feedback for hypothesis elimination. In our setting, however, the presence of the dominating action $\knownOptAction$ masks this comparison: for many parameter values, the algorithm observes $\knownOptAction$ being chosen instead, yielding no information about the relative ordering of $\adversarialAction$ and $\nullAction$. As a result, the hypothesis set cannot be effectively refined. See \Cref{fig:failure of KL search} for an illustration of this phenomenon.

Nonetheless, the constructed adversary cannot make the shrinkage of the hypothesis set arbitrarily small. In fact, our adversarial construction inherently embodies a tradeoff. At a high level, while the dominating action $\knownOptAction$ obscures information about the hidden parameter (by masking comparisons between ``purchase'' action $\adversarialAction$ and ``no purchase'' action $\nullAction$), it simultaneously provides the algorithm with a ``safe'' contract: since $\knownOptAction$ is known and consistently high-performing, the algorithm can always incentivize this action without incurring large regret. More precisely, in the presence of $\knownOptAction$, the algorithm may deploy a contract that incentivizes either the dominating action $\knownOptAction$ or the unknown ``purchase'' action $\adversarialAction$—whichever yields higher utility.
Quantitatively, there is an inverse relationship between the fraction of the hypothesis set masked by $\knownOptAction$ and the regret achievable by such a safe contract: the larger the masked region, the lower the regret the algorithm can guarantee by defaulting to $\knownOptAction$. See \Cref{fig:adversary tradeoff} for an illustration. Our final adversarial construction (\Cref{adv:warm up:three-action setting} in \Cref{sec:contextual pricing connection}) carefully balances this tradeoff by selecting an optimal sequence of masking fractions over time, and thereby establishes a regret lower bound of $\Omega(T^{1/4})$ for this three-action setting with two-dimensional contexts.

Finally, to achieve our targeted dependence on the context dimension $\dimension$ in the regret lower bounds across all settings (see \Cref{tab:regret-comparison}), our adversarial constructions for these more general results introduce a fourth action and incorporate a key technical ingredient from coding theory: \emph{spherical codes} (\Cref{def:spherical codes}). The core idea is to leverage the geometric properties of these codewords to force any algorithm into a prolonged and costly exploration phase. Specifically, the exponentially large number of codewords defines a correspondingly large number of uncertainty directions in the hypothesis set. To learn the true hidden vector, the principal must explore along each of these directions, which ensures that the overall hypothesis set shrinks at an exceedingly slow rate. This slow shrinkage results in a significant regret dependence on the context dimension $\dimension$.

\xhdr{Benchmark-optimal contextual search algorithm.}
Given the challenges posed by contextual action degeneracy, it seems unlikely to naively apply the algorithmic ideas and the regret analysis framework from the literature to our settings. Specifically, we can not apply the KL search approach used by previous works \citep{KL-03,LS-18,LLS-21} to our problem. Instead, we can only expect to cut a small region from the current hypothesis set by carefully selecting the contract.
Beyond the failure of KL-search, other natural approaches also face significant obstacles. For instance, one might view our problem as a contextual continuous bandit problem, where contracts constitute a continuous action space. However, a straightforward discretization of this space followed by the application of classical online learning algorithms—such as the multiplicative weights update (MWU)—fails to yield meaningful guarantees. The primary difficulty lies in controlling the discretization error. Crucially, neither the optimal payoff nor the optimal contract exhibits a clear Lipschitz dependence on the hidden parameter vector, which undermines standard discretization arguments.\footnote{In the online contract design literature, some works establish ``weaker-than-Lipschitz'' regularity properties; however, these are defined over the contract space rather than the hidden vector space that governs the contextual structure of our setting \citep{ZBYWJJ-23,DGSW-23}. It remains unclear whether such techniques can be extended or adapted to our framework.}

We design our online algorithms following the \emph{principle of optimism}. In the cost-context setting, at each round, our proposed algorithm (Algorithm~\ref{alg: unknown cost}) first identifies the \emph{optimistic hidden vector}—that is, the value of the hidden vector (consistent with the observed history) that maximizes the principal's payoff given the current context. It then computes the optimal contract under this optimistic hidden vector and perturbs it slightly by sharing a bit more revenue with the agent. Our proposed algorithm (Algorithm~\ref{alg: single stage}) for the reward-context setting follows a similar but more complicated construction. To understand the rationale behind this procedure, consider the three-action adversary described earlier. If the adversary selects a context such that dominating action~$\knownOptAction$ masks the majority of the hypothesis set, the resulting contract closely approximates the ``safe'' contract, thereby guaranteeing low regret. Conversely, when dominating action~$\knownOptAction$ does not mask most of the hypothesis set, the slight revenue-sharing perturbation ensures a carefully controlled shrinkage of the hypothesis set—enabling effective learning over time.

One key technical ingredient of our algorithms is a new notion we introduce: the \emph{information width} (\Cref{def:information width}), defined as the width of the portion of the hypothesis set excluded by the algorithm in the contextual direction. Informally, the information width governs the immediate regret in a round; in the worst case (e.g., against our adversary), it reflects the degree of contextual action degeneracy—namely, the fraction of the hypothesis set not masked by dominating action~$\knownOptAction$. (Also see \Cref{fig:adversary tradeoff} for an illustration.)
As in prior work, we use the \emph{intrinsic volume} (\Cref{def:intrinsic volume}) to define a potential function over the hidden vector space, enabling a regret analysis via the potential method. However, unlike previous approaches that rely on multiplicative volume reduction, we establish a direct relationship between the decay of intrinsic volume and the information width of the contract selected by the algorithm. Through this relationship, we build our regret upper bounds for both cost-context and reward-context settings using a potential argument. 

To evaluate the performance of our proposed online algorithms, we consider the classic \emph{Stackelberg regret} (\Cref{def:stackelberg regret}), which is defined against the \emph{optimum-in-hindsight benchmark}, where the hidden vector is known and the optimal contract is designed for each agent individually. Additionally, we introduce a stronger regret measure, termed the \emph{pessimistic Stackelberg regret} (\Cref{def:pessimistic stackelberg regret}), which provides a more stringent comparison. This stronger notion evaluates the long-term utility of the principal under the algorithm, relative to an ``optimistic benchmark''. Specifically, for each round, the optimistic benchmark is defined as the maximum achievable principal utility over all possible contracts, considering all potential values of the hidden vector that are consistent with the algorithm’s observed history (i.e., agents' information, proposed contracts, and the agents' best responses in all previous rounds). By construction, since the true hidden vector is always consistent with the observed history, the optimistic benchmark serves as an upper bound to the optimum-in-hindsight benchmark. Consequently, an upper bound on the pessimistic Stackelberg regret implies the same upper bound on the classic Stackelberg regret. Notably, the analysis framework used in many prior works \citep[e.g.,][]{KL-03,LLV-18,LS-18,LLS-21,KLPS-21,FTX-22} in contextual pricing and related problems implicitly compares the algorithm’s utility against the optimistic benchmark's gain in each round. Therefore, although not explicitly stated, the regret guarantees in these prior works are directly applicable to the pessimistic Stackelberg regret definition.

As shown in \Cref{tab:regret-comparison}, for both the cost-context and reward-context settings, our proposed algorithms (Algorithms~\ref{alg: unknown cost} and ~\ref{alg: single stage}) achieve pessimistic Stackelberg regrets with a regret dependence on $T$ of $T^{1-\Theta(1/\dimension)}$. We complement these results with pessimistic regret lower bounds of $\Omega(T^{1-1/\dimension})$, certifying that our algorithms are ``benchmark-optimal'' under the pessimistic Stackelberg regret.\footnote{Proving the optimality of our proposed algorithms under the classic Stackelberg regret or designing improved algorithms is an interesting future direction. Since most prior work essentially computes the pessimistic Stackelberg regret, achieving an improved upper bound for the classic Stackelberg regret in our model would likely require additional algorithmic and analytical approaches.}

In summary, our work reveals that contextual search in principal-agent games is fundamentally more complex than the well-studied contextual pricing problem. The paper establishes both the possibility of dramatically larger regret in the general model (even under mild extensions of action space and context dimension) and provides matching or complementary bounds that characterize the growth of regret in both cost-context and reward-context settings. These findings yield new insights for contextual learning in principal-agent environments, highlighting how additional actions and hidden parameters can sharply increase the learning difficulty.

\subsection{Related Work}\label{sec:related}
Our work connects to several lines of research, including contextual search and pricing, learning in principal-agent settings, and incentive exploration. Each of these areas has developed distinct approaches to learning and decision-making under uncertainty and strategic interaction. We review the most relevant developments below.

\xhdr{Contextual search and pricing.}
\citet{KL-03} initiated the study of dynamic pricing in non-contextual settings, showing that doubly logarithmic regret can be attained and establishing a foundational $\Omega(\log\log \timeHorizon)$ lower bound for regret in adversarial posted-price auctions. Extending this result to contextual scenarios yields a $\Omega(\dimension\log\log \timeHorizon)$ lower bound \citep{LS-18}. 
A sequence of works \citep{CLP-20,CLP-17,LLV-18,LS-18} studied the contextual pricing problem and progressively improved this to $O_{\dimension}(\log \timeHorizon)$ and eventually to $O_{\dimension}(\log\log \timeHorizon)$ using tools from integral geometry, particularly intrinsic volumes. These advances, culminating in the near-optimal regret bound of $O(\dimension \log\log \timeHorizon + \dimension\log \dimension)$ by \citet{LLS-21}, confirm that the $\Omega(\dimension\log\log \timeHorizon)$ lower bound is tight up to additive and polynomial factors in $\dimension$. Other variants of this problem have been explored—for example, contextual search with irrational agents \citep{KLPS-21}, contextual pricing with noise \citep{LSS-22,TGMP-24}, stochastic context \citep{ARS-14, JH-19, LSS-22}, and non-myopic agents \citep{GJM-19}.

\xhdr{Learning in principal-agent game.}
Learning in contract design has drawn growing interest, particularly under hidden-action models.
\citet{HSV-14} initiated the study of bandit algorithms for repeated contract design with bounded, monotone contracts, later extended to risk-averse agents by \citet{CDK-22}. \citet{ZBYWJJ-23} showed that removing the monotonicity assumption can lead to exponential sample complexity, motivating efficient algorithms under structural assumptions by \citet{CCDH-24} and under small action spaces by \citet{BCMG-24}. \citet{DGSW-23} reduced optimal linear contract learning to one-sided Lipschitz optimization, achieving optimal $\Theta(\log\log T)$ regret under deterministic feedback. Several studies consider learning under more complex agent behaviors where actions remain unobservable: \citet{STBCMJEO-24,WCWWX-24,GKSTVWW-24,BBCMG-24} investigate contract learning with reinforcement-learning agents, \citet{CGR-24} accounts for repeated interactions with non-myopic agents, and \citet{Zuo-24} addresses multitask settings with unobserved individual efforts. \citet{DSA-23} considers agents with perfect knowledge of rewards under hidden principal information. In contrast, other works consider settings where the agent’s action is either directly observed or deterministically inferred by the principal, as in \citet{LR-25,LMTJR-25,STBCMJEO-24,DSA-23, HAX-24, BMMT-24}. 

Finally, our work is conceptually related to the incentivized exploration literature \citep{FKKK-14, KMP-14, MSS-15, SS-21, MSSW-22}. In both models, the online algorithm must incentivize strategic participants to take desired actions. However, while in the incentivized exploration setting the algorithm is focused on learning the true state of nature, in our model the algorithm is learning the participants' types (i.e., their preferences).

\section{Preliminaries}
\label{sec:prelim}

In this section, we introduce the classic model of the principal-agent game (\Cref{subsec:prelim:p-a game}), the contextual learning variant studied in this work (\Cref{subsec:prelim:contextual p-a learning}), and two geometric concepts that are crucial to our regret analysis (\Cref{subsec:prelim:technical concept}).

\subsection{The Principal-Agent Game}
\label{subsec:prelim:p-a game}

The principal-agent model describes the interaction between a \emph{principal}, 
who delegates a task, and an \emph{agent}, who performs the task on the principal's behalf.
The agent's set of possible actions is denoted by 
$\actions \deq \set{0, 1,\ldots,\actionNum - 1}$,
where $\actionNum$ represents the total number of available actions.
Each action $\action \in \actions$ has an associated cost $\costA{\action} \in \nnreals$ 
and \emph{deterministically} leads to an outcome yielding a reward $\rewardA{\action} \in \nnreals$ for the principal.
We use the terms ``action'' and ``outcome'' interchangeably.
The set of actions includes a trivial action $\actionNull$, for which both the cost and reward are zero.

\xhdr{(Linear) Contracts.} To incentivize the agent to undertake costly actions, the principal offers a contract $\contractGeneral = (\contractGeneralA{1}, \contractGeneralA{2}, \dots, \contractGeneralA{\actionNum})$, where $\contractGeneralA{\action}$ specifies the payment to the agent if action $\action$ is chosen, resulting in reward $\rewardA{\action}$.
A contract is said to be \emph{linear} if there exists a parameter $\contractSimple \in [0,1]$ such that
$\contractGeneralA{\action} = \contractSimple \cdot \rewardA{\action}$
for all actions $\action \in \actions$. Namely, the principal transfers a fixed percentage of the reward to the agent. We denote a linear contract simply by its parameter $\contractSimple$.\footnote{\label{footnote:prelim:linear contract}Linear contracts are widely studied due to their simplicity and tractability, even in settings where they are not without loss of generality~\citep{ZBYWJJ-23,DDPP-24,CCL-25}. 
If the principal is allowed to specify action-dependent payments, the model becomes (almost) equivalent to contextual multi-item dynamic pricing with unit-demand buyers. In that case, one can achieve regret with a double-logarithmic dependence on $T$; see \Cref{apx:act-dep contract and pricing}.}

\begin{remark}
\label{remark:linear contract interpretation}
Our model can be viewed as a simplification of the classic binary-outcome hidden-action principal–agent setting, where each action induces a successful outcome with some probability, and the action’s reward equals this success probability. In this formulation, linear contracts are without loss of generality; thus, our negative results—specifically, the double-exponential separation in regret—carry over directly. Meanwhile, our positive results also extend to scenarios where the principal observes repeated interactions with the same agent and estimates expected rewards from empirical averages.
\end{remark}

Both the principal and agent have quasi-linear utilities. Specifically, given a contract $\contractSimple \in [0, 1]$ and an action $\action$, the utility of the principal is $\revenuePrincipalCon(\contractSimple,\action) = (1 - \contractSimple)\cdot \rewardA{\action}$, and the utility of the agent is $\utiAgent(\contractSimple, \action) = \contractSimple \cdot \rewardA{\action} - \costA{\action}$.
Let $\actionBest{\contractSimple}$ represent the agent's best response under the contract $\contractSimple$, i.e., $\actionBest{\contractSimple} \in \argmax_{\action\in\actions} \utiAgent(\contractSimple,\action)$.\footnote{Following the literature, we assume that the agent breaks ties in favor of the principal.} With a slight abuse of notation, we also denote by $\revenuePrincipalCon(\contractSimple)$ and $\utiAgent(\contractSimple)$ the principal and agent's \emph{indirect utilities} when the agent best respond to a given contract $\contractSimple$, i.e., $\revenuePrincipalCon(\contractSimple)\triangleq \revenuePrincipalCon(\contractSimple,\action_\contractSimple)$ and $\utiAgent(\contractSimple)\triangleq \utiAgent(\contractSimple,\action_\contractSimple)$.
We highlight the following structural observation regarding the two players' indirect utilities as functions of the contract~$\contractSimple$.
\begin{observation}[\citealp{DEFK-21}]
\label{obv:indirect utility structure}
The agent's indirect utility $\utiAgent(\contractSimple)$ is continuous, non-decreasing, convex, and piecewise linear in $\contractSimple \in [0, 1]$.
Each linear segment of $\utiAgent$ has slope $\rewardA{\action_\contractSimple}$ and $y$-intercept $-\costA{\action_\contractSimple}$.

The principal's indirect utility $\revenuePrincipalCon(\contractSimple)$ is right-continuous and piecewise linear in $\contractSimple \in [0, 1]$.
Each linear segment of $\revenuePrincipalCon$ has slope $-\rewardA{\action_\contractSimple}$, $y$-intercept $\rewardA{\action_\contractSimple}$, and $x$-intercept $1$.
\end{observation}
We next define the degeneracy of actions.
\begin{definition}[Degeneracy]
\label{def:action degeneracy}
    An action $\action \in \actions$ is said to be \emph{implementable} if there exists a contract $\contractSimple \in [0,1]$ such that $\action = \actionBest{\contractSimple}$, i.e., it is the agent's best response under that contract.
    Otherwise, we say the action is \emph{degenerated}---it cannot be implemented by any contract.
\end{definition}

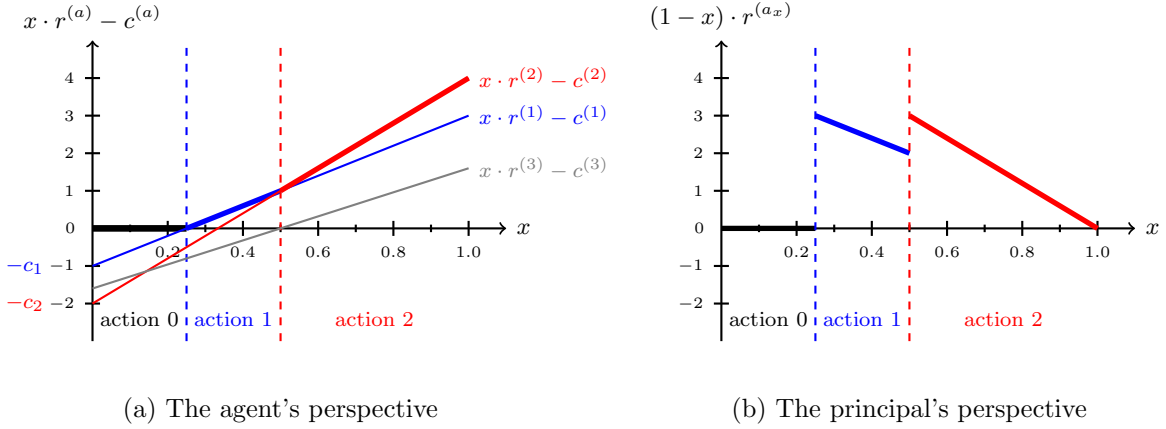
\begin{figure}
    \centering
    \begin{subfigure}[t]{0.4\textwidth}
        \centering
        \begin{tikzpicture}
    \useasboundingbox (0,-2) rectangle (5,3);
\draw[thick,->] (0,-1.5) -- (0,2.5) node[above] {\footnotesize $\contractSimple\cdot \rewardA{\action} - \costA{\action}$};
\draw[thick,->] (0,0) node[left] {\tiny $0~$} -- (5.5,0) node[right] {\footnotesize $\contractSimple$};
\draw[thick,-] (0.5,0.05) -- (0.5,-0.05); 
    \draw[thick,-] (1,0.1) -- (1,-0.1) node[below] {\tiny $0.2$}; 
    \draw[thick,-] (1.5,0.05) -- (1.5,-0.05);
    \draw[thick,-] (2,0.1) -- (2,-0.1) node[below] {\tiny $0.4$};
    \draw[thick,-] (2.5,0.05) -- (2.5,-0.05);
    \draw[thick,-] (3,0.1) -- (3,-0.1) node[below] {\tiny $0.6$};
    \draw[thick,-] (3.5,0.05) -- (3.5,-0.05);
    \draw[thick,-] (4,0.1) -- (4,-0.1) node[below] {\tiny $0.8$};
    \draw[thick,-] (4.5,0.05) -- (4.5,-0.05);
    \draw[thick,-] (5,0.1) -- (5,-0.1) node[below] {\tiny $1.0$};
\draw[thick,-] (0.1,-1.0) -- (-0.1,-1.0) node[left] {\tiny $-2$};
    \draw[thick,-] (0.1,-0.5) -- (-0.1,-0.5) node[left] {\tiny $-1$};
    \draw[thick,-] (0.1,0.5) -- (-0.1,0.5) node[left] {\tiny $1$};
    \draw[thick,-] (0.1,1.0) -- (-0.1,1.0) node[left] {\tiny $2$};
    \draw[thick,-] (0.1,1.5) -- (-0.1,1.5) node[left] {\tiny $3$};
    \draw[thick,-] (0.1,2.0) -- (-0.1,2.0) node[left] {\tiny $4$};

\draw[line width=2.5pt,black] (0,0) -- (1.25,0); \node at (0.625,-1.2) {\scriptsize \textcolor{black}{action $\actionNull$}};

\draw[thick,blue,dashed] (1.25,-1.5) -- (1.25,2.5); 
    \node at (-0.9,-0.5) {\textcolor{blue}{\scriptsize{$-\cost_1$}}};
    \draw[thick,blue] (0,-0.5) -- (5,1.5) node[right] {\scriptsize{$\contractSimple\cdot \rewardA{1} - \costA{1}$}}; \draw[line width=2pt,blue] (1.25,0) -- (2.5,0.5); \node at (1.875,-1.2) {\scriptsize \textcolor{blue}{action $1$}};

\draw[thick,red,dashed] (2.5,-1.5) -- (2.5,2.5);
    \draw[thick,red] (0,-1) -- (5,2) node[right] {\scriptsize{$\contractSimple\cdot \rewardA{2} - \costA{2}$}}; \node at (-0.9,-1) {\textcolor{red}{\scriptsize{$-\cost_2$}}};
    \draw[line width=2pt,red] (2.5,0.5) -- (5,2); \node at (3.75,-1.2) {\scriptsize \textcolor{red}{action $2$}};
    
\draw[thick,gray] (0,-0.8) -- (5,.8) node[right] {\scriptsize{$\contractSimple\cdot \rewardA{3} - \costA{3}$}}; \end{tikzpicture}         \caption{The agent's perspective}
        \label{fig:upper-envelope-agent}
    \end{subfigure}
    \hspace{1.5cm}
    \begin{subfigure}[t]{0.4\textwidth}
        \centering
        \begin{tikzpicture}
    \useasboundingbox (0,-2) rectangle (5,3);
\draw[thick,->] (0,-1.5) -- (0,2.5) node[above] {\footnotesize $(1-\contractSimple)\cdot\rewardA{\actionBest{\contractSimple}}$};
\draw[thick,->] (0,0) node[left] {\tiny $0~$} -- (5.5,0) node[right] {\footnotesize $\contractSimple$};
\draw[thick,-] (0.5,0.05) -- (0.5,-0.05); 
    \draw[thick,-] (1,0.1) -- (1,-0.1) node[below] {\tiny $0.2$}; 
    \draw[thick,-] (1.5,0.05) -- (1.5,-0.05);
    \draw[thick,-] (2,0.1) -- (2,-0.1) node[below] {\tiny $0.4$};
    \draw[thick,-] (2.5,0.05) -- (2.5,-0.05);
    \draw[thick,-] (3,0.1) -- (3,-0.1) node[below] {\tiny $0.6$};
    \draw[thick,-] (3.5,0.05) -- (3.5,-0.05);
    \draw[thick,-] (4,0.1) -- (4,-0.1) node[below] {\tiny $0.8$};
    \draw[thick,-] (4.5,0.05) -- (4.5,-0.05);
    \draw[thick,-] (5,0.1) -- (5,-0.1) node[below] {\tiny $1.0$};
\draw[thick,-] (0.1,-1.0) -- (-0.1,-1.0) node[left] {\tiny $-2$};
    \draw[thick,-] (0.1,-0.5) -- (-0.1,-0.5) node[left] {\tiny $-1$};
    \draw[thick,-] (0.1,0.5) -- (-0.1,0.5) node[left] {\tiny $1$};
    \draw[thick,-] (0.1,1.0) -- (-0.1,1.0) node[left] {\tiny $2$};
    \draw[thick,-] (0.1,1.5) -- (-0.1,1.5) node[left] {\tiny $3$};
    \draw[thick,-] (0.1,2.0) -- (-0.1,2.0) node[left] {\tiny $4$};
\draw[black,line width=2pt] (0,0) -- (1.25,0);
    \node at (0.625,-1.2) {\scriptsize \textcolor{black}{action $0$}};
\draw[thick,blue,dashed] (1.25,-1.5) -- (1.25,2.5);
    \node at (1.875,-1.2) {\scriptsize \textcolor{blue}{action $1$}};
    \draw[blue,line width=2pt] (1.25,1.5) -- (2.5,1);
\draw[thick,red,dashed] (2.5,-1.5) -- (2.5,2.5);
    \draw[red,line width = 2pt] (2.5,1.5) -- (5,0);
    \node at (3.75,-1.2) {\scriptsize \textcolor{red}{action $2$}};
\end{tikzpicture}         \caption{The principal's perspective}
        \label{fig:upper-envelope-principal}
    \end{subfigure}
    \caption{
        The agent's indirect utility as a function of the contract $\contractSimple$ \textbf{(left)}, 
        and the principal's indirect utility as a function of $\contractSimple$ \textbf{(right)}, 
        for the principal-agent setting in \Cref{ex:example-one-new}.
    }
    \label{fig:upper-envelope}
\end{figure}

To conclude this subsection, we illustrate the structure of indirect utilities and action degeneracy using \Cref{ex:example-one-new} and \Cref{fig:upper-envelope}.

\begin{example}
    \label{ex:example-one-new}
    Consider a principal-agent setting with four actions $\actions = \{ 0, 1, 2, 3\}$, 
    characterized by the following costs and rewards:
    \begin{align*}
        \costA{0} = \rewardA{0} = 0,\quad
        \costA{1} = 1,\ \rewardA{1} = 4,\quad
        \costA{2} = 2,\ \rewardA{2} = 6,\quad
        \costA{3} = 1.6,\ \rewardA{3} = 3.2.
    \end{align*}
    The agent’s utility for each action, as a function of the contract $\contractSimple$, is depicted by the four straight lines (black, blue, red, and gray) in \Cref{fig:upper-envelope-agent}. The black line, corresponding to the trivial action $\actionNull$, is horizontal at $0$. The agent's best response is given by the upper envelope of these lines, highlighted by the thick segments in \Cref{fig:upper-envelope-agent}. Note that action~3 is degenerated, as its utility line does not appear on the upper envelope; consequently, it cannot be implemented by any contract.
\end{example}

\subsection{Contextual Search in Principal-Agent Games}
\label{subsec:prelim:contextual p-a learning}
In this section, we introduce the \emph{contextual search} setting for the principal-agent game. Unlike the classical setting, where the principal has full knowledge of the agent's cost profile over all actions,\footnote{In \Cref{sec:unknown reward}, we introduce and study a variant where the principal is instead uncertain about the reward profile over actions.} the principal in this variant has no prior information about costs and must learn through repeated interactions with a sequence of possibly heterogeneous agents.

There are $T \in \naturals$ sequential and discrete rounds. In each round $t \in [T]$, a new agent indexed by $t$ arrives. For this agent, both the action set $\actionsT$ and the reward profile $\{\rewardAT{\action}\}_{\action \in \actionsT}$ are observed by the principal. However, the cost profile $\{\costAT{\action}\}_{\action \in \actionsT}$ is not directly observable. Instead, the principal observes a \emph{cost context profile} $\{\contextAT{\action}\}_{\action \in \actionsT}$, where $\contextAT{\action} \in \vecBall_\dimension$ is a $\dimension$-dimensional vector lying in the unit ball. For each action $\action \in \actionsT$, the corresponding cost satisfies the following linear structure:
\begin{align}
\label{eq:linear cost structure}
\tag{linear cost structure}
\costAT{\action} = \innerproduct{\contextAT{\action}, \hvecTrue},
\end{align}
where $\hvecTrue \in \vecBall_{\dimension}$ is the \emph{cost hidden vector}, an unknown vector in the $\dimension$-dimensional unit ball.

We assume that the action sets $\{\actionsT\}_{t \in [T]}$, reward profiles $\{\rewardAT{\action}\}_{\action \in \actionsT, t \in [T]}$, cost context profiles $\{\contextAT{\action}\}_{\action \in \actionsT, t \in [T]}$, and the hidden cost vector $\hvecTrue$ are all chosen by an \emph{adaptive} adversary.
Moreover, we assume that the linear cost structure ensures $\costAT{\action} = \innerproduct{\contextAT{\action}, \hvecTrue} \geq 0$ for all actions $\action \in \actionsT$, with $\rewardAT{0}=\costAT{0} = 0$ for the trivial action $0$.

In each round $t\in[T]$, the principal proposes a contract $\contract{\timeslot} \in [0,1]$, and the agent selects a best-response action
\begin{align*}
    \actionT\in \argmax\nolimits_{\action\in \actionsT}~ \contract{\timeslot}\cdot \rewardAT{\action}-\lrangle{\contextAT{\action},\hvecTrue}.
\end{align*}
The principal receives utility $\revenuePrincipalCon_\timeslot(\contract{\timeslot})=(1-\contract{\timeslot})\cdot \rewardAT{\actionT}$, and the agent receives utility $\utiAgent_t(\contract{\timeslot})=\contract{\timeslot}\cdot\rewardAT{\actionT}-\innerproduct{\contextAT{\actionT},\hvecTrue}$. The principal then observes the chosen best-responding action $\actionT$. The goal of the principal is to design an online algorithm that proposes contracts $\{\contract{\timeslot}\}_{\timeslot\in[T]}$ to maximize her long-term utility $\sum_{\timeslot\in[T]}\revenuePrincipalCon_t(\contract{\timeslot})$.

\xhdr{Benchmark and regret definitions.} We evaluate the performance of an online policy by its Stackelberg regret \citep{CLP-20b} against the optimum-in-hindsight benchmark. The \emph{optimum-in-hindsight benchmark} knows the cost hidden vector $\hvecTrue$ and proposes the optimal contracts 
\begin{align*}
    \contract{\timeslot}^* \in \argmax\nolimits_{\contractSimple\in[0,1]}~
    \revenuePrincipalCon_\timeslot(\contractSimple)
\end{align*}
for each arriving agent $\timeslot\in[T]$ separately. 
\begin{definition}[Stackelberg regret]\label{def:stackelberg regret}
    Given cost hidden vector $\hvecTrue$ and a sequence of $T$ agents with action sets $\{\actionsT\}_{t \in [T]}$, reward profiles $\{\rewardAT{\action}\}_{\action \in \actionsT, t \in [T]}$, and cost context profiles $\{\contextAT{\action}\}_{\action \in \actionsT, t \in [T]}$, the \emph{Stackelberg regret} of an online algorithm $\ALG$ is 
    \begin{align*}
        \regret(\ALG) \triangleq \sum\nolimits_{\timeslot\in[T]}
        \revenuePrincipalCon_\timeslot(\contract{\timeslot}^*) - 
        \revenuePrincipalCon_\timeslot(\contract{\timeslot}),
    \end{align*}
    where $\contract{\timeslot}^*$ and $\contract{\timeslot}$ are the contracts proposed in the optimum-in-hindsight benchmark and algorithm $\ALG$ for each agent $\timeslot\in[T]$.
\end{definition}

In addition to the standard optimum-in-hindsight benchmark and its corresponding Stackelberg regret, we are also interested in the following \emph{optimistic benchmark} and the corresponding \emph{pessimistic Stackelberg regret}.

Fix an online algorithm $\ALG$. For each round $\timeslot \in [T]$, define the \emph{hypothesis set} $\hvecsPess_t \subseteq \vecBall_\dimension$ as the set of vectors consistent with the contract history $\set{\contract{\tpre}}_{\tpre \in [\timeslot - 1]}$ and observed best-response actions history $\{\action_{\tpre}\}_{\tpre \in [\timeslot - 1]}$.\footnote{That is, a vector $\hvec \in \vecBall_{\dimension}$ belongs to the hypothesis set $\hvecsPess_\timeslot$ if and only if the first $\timeslot - 1$ agents' best-response actions are $\{\action_{\tpre}\}_{\tpre \in [\timeslot - 1]}$ when the hidden cost vector is $\hvecTrue = \hvec$.}
The optimistic benchmark is defined with respect to the execution of the algorithm, and in particular, the hypothesis sets $\{\hvecsPess_\timeslot\}_{\timeslot \in [T]}$ it induces. In each round $\timeslot\in[T]$, the benchmark receives a payoff $\revenuePrincipalConPessimistic_\timeslot(\hvecsPess_\timeslot)$ equal to the maximum achievable principal utility over all vectors in the hypothesis set $\hvecsPess_\timeslot$, i.e.,\footnote{Throughout the paper, we use tilde notation (e.g., $\tilde{\cdot}$) to denote concepts related to the optimistic benchmark and pessimistic Stackelberg regret.}
\[
\revenuePrincipalConPessimistic_\timeslot(\hvecsPess_\timeslot)
    \triangleq \max\limits_{\hvec\in\hvecsPess_\timeslot}
    \max\limits_{\contractSimple\in[0, 1]}~\revenuePrincipalCon_\timeslot(\contract{} | \hvec),
\]
where $\revenuePrincipalCon_\timeslot(\contract{} |\hvec)$ is principal's utility when proposing a contract $x$ under the hidden vector $\theta$, i.e.,
\begin{align*}
    \revenuePrincipalCon_\timeslot(\contract{} | \hvec)
    \triangleq 
    (1 - \contractSimple) \cdot \rewardAT{\action_{\timeslot,\hvec}}
    \;\;
    \mbox{s.t.}
    \;\;
    \action_{\timeslot,\hvec} \in \argmax\limits_{\action\in\actionsT}~ 
    \contractSimple\cdot \rewardAT{\action} - 
    \innerproduct{\contextAT{\action},\hvec}.
\end{align*}
\begin{definition}[Pessimistic Stackelberg regret]\label{def:pessimistic stackelberg regret}
    Given cost hidden vector $\hvecTrue$ and a sequence of $T$ agents with action sets $\{\actionsT\}_{t \in [T]}$, reward profiles $\{\rewardAT{\action}\}_{\action \in \actionsT, t \in [\timeHorizon]}$, and cost context profiles $\{\contextAT{\action}\}_{\action \in \actionsT, t \in [\timeHorizon]}$, the \emph{pessimistic Stackelberg regret} of an online algorithm $\ALG$ is 
    \begin{align*}
        \regretPessimistic(\ALG) \triangleq \sum\nolimits_{\timeslot\in[T]} \regretPessimistic_\timeslot(\ALG)\triangleq\sum\nolimits_{\timeslot\in[T]}
        \revenuePrincipalConPessimistic_\timeslot(\hvecsPess_\timeslot) - 
        \revenuePrincipalCon_\timeslot(\contract{\timeslot}),
    \end{align*}
    where $\revenuePrincipalConPessimistic_\timeslot(\hvecsPess_\timeslot)$ is the payoff received in the optimistic benchmark, and $\contract{\timeslot}$ is the contracts proposed in algorithm $\ALG$ for each agent $\timeslot\in[T]$.
\end{definition} 
By construction, the optimistic benchmark upper bounds the optimum-in-hindsight benchmark, and thus the pessimistic Stackelberg regret upper bounds the classic Stackelberg regret. Since it depends only on revealed history, the pessimistic Stackelberg regret also serves as a more robust measure of algorithmic performance.

\begin{remark}
    Although not explicitly stated, the regret guarantees in most prior work \citep[e.g.,][]{KL-03,LLV-18,LS-18,LLS-21,KLPS-21,FTX-22,LL-25} on (contextual) pricing and related problems are applicable to the pessimistic Stackelberg regret definition.
\end{remark}

\subsection{Geometric Concepts for Regret Analysis}
\label{subsec:prelim:technical concept}

In this section, we introduce two geometric concepts—\emph{intrinsic volume} and \emph{spherical code}—that will be used in our subsequent regret analysis.

\xhdr{Intrinsic volumes.}
For any convex body $\convexBody \subseteq \realNumbers^{\dimension}$, the \emph{intrinsic volumes} $(\intrinsicVolume{0}{\convexBody}, \intrinsicVolume{1}{\convexBody}, \dots, \intrinsicVolume{\dimension}{\convexBody})$ 
are a sequence of geometric quantities that generalize volume, surface area, and other notions of size. For example,
$\intrinsicVolume{\dimension}{\convexBody}$ equals the standard $\dimension$-dimensional volume of $\convexBody$, while $\intrinsicVolume{\dimension - 1}{\convexBody}$ corresponds (up to scaling) to its surface area. The formal definition is as follows.
\begin{definition}[Intrinsic volumes]
\label{def:intrinsic volume}
    Let $\convexBody \subseteq \realNumbers^\dimension$ be a convex body, and let $\ball$ denote the unit ball in $\realNumbers^\dimension$. 
    Given any convex body $\convexBody \subseteq \realNumbers^\dimension$, its \emph{intrinsic volumes} $\{\intrinsicVolume{j}{\convexBody}\}_{j\in[0:\dimension]}$ are defined as the coefficients in the \emph{Steiner formula}:\footnote{It has been shown that $\volume{\convexBody + \epsilon \ball}$ is a polynomial in $\epsilon$ \citep{Sch-86}.}
    \begin{align*}
        \volume{\convexBody + \epsilon \ball} = \sum\nolimits_{j\in[0:\dimension]}
        \kappa_{\dimension - j} \, \intrinsicVolume{j}{\convexBody} \, \epsilon^{\dimension - j},
    \end{align*}
    where $\volume{\!\cdot\!}$ is the $\dimension$-dimensional volume and $\kappa_{\dimension-j}$ is the volume of the $(\dimension\!-\!j)$-dimensional unit ball.
\end{definition}

Our regret analysis utilizes the following properties of the intrinsic volumes.

\begin{lemma}[\citealp{LS-18,SW-08}]
\label{lem:intrinsic volume properties}
The intrinsic volumes satisfy the following properties:
    \begin{itemize}
        \item \textbf{(Monotonicity):} For any two convex bodies $S,T\subseteq \reals^\dimension$ and index $j\in[0:\dimension]$, if $S \subseteq T$, then 
        $
            \intrinsicVolume{j}{S} \leq \intrinsicVolume{j}{T}.
        $

        \item \textbf{(Additivity):} For any two convex bodies $S,T\subseteq \reals^\dimension$ and index $j\in[0:\dimension]$,
        $
            \intrinsicVolume{j}{S} + \intrinsicVolume{j}{T}
            = \intrinsicVolume{j}{S \cup T} + \intrinsicVolume{j}{S \cap T}.
        $

        \item \textbf{(Cone inequality):} For any convex cone $\cone\in\reals^\dimension$ and index~$j\in[0:\dimension]$,
        $
            \intrinsicVolume{j}{\cone} \geq \frac{h}{j} \cdot \intrinsicVolume{j-1}{\coneBase},
        $
        where $\coneBase$ and $h$ are the base and height of convex cone $\cone$, respectively.

        \item \textbf{(Isoperimetric inequality):} For any convex body $\convexBody\in\reals^\dimension$ and index $j\in[0:\dimension]$,
        \begin{align*}
            \left( \factorial{(j+1)} \cdot \intrinsicVolume{j+1}{\convexBody} \right)^{1/(j+1)}
            \leq \left( \factorial{j} \cdot \intrinsicVolume{j}{\convexBody} \right)^{1/j}.
        \end{align*}
        
        \item \textbf{(Intrinsic volume of unit ball):} For the $\dimension$-dimensional unit ball $\ball$, 
$\intrinsicVolume{1}{\ball} \leq \sqrt{2\pi \dimension}$.
\end{itemize}
\end{lemma}

\xhdr{Spherical code.} The second geometric concept used in our regret lower bound analysis is the notion of \emph{spherical code}. We present its formal definition below, along with the specific property that will be used in our analysis.

\begin{definition}[Spherical code, \citealp{conway2013sphere}]\label{def:spherical codes}
    Fix any $\dimension\in\naturals$ and $\codeAngle \in\reals_+$.
    Let $\sphericalCode$ be a finite subset of $(\dimension-1)$-dimensional sphere $\mathbb{S}_{\dimension-1}$. We say $\sphericalCode$ is a $\dimension$-dimensional \emph{spherical code} with minimal angle $\codeAngle$ if $\codeAngle$ is the largest number such that for any $x \neq y\in \sphericalCode,$
    $\innerproduct{x, y} \leq \cos (\codeAngle).$
    The elements of a spherical code are referred to as \emph{codewords}. 
\end{definition}

\begin{lemma}[Lower bound about the maximal size of a spherical code, \citealp{JJP-18}]\label{lem:code lower bound}
    Fix any $\dimension\in\naturals$ and $\codeAngle \in\reals_+$.
    Let $N(\dimension,\codeAngle)$ be the maximal size of $\dimension$-dimensional spherical code with minimal angle $\codeAngle$. Then
    \begin{equation*}
        N(\dimension,\codeAngle)\geq C \frac{ \sqrt{d} \cos{\codeAngle}}{\sin^{d-1} \codeAngle}   
    \end{equation*}
    where $C$ is an absolute constant.
\end{lemma}
 
\section{Double Exponential Separation from Contextual Pricing}
\label{sec:contextual pricing connection}
In this warm-up section, we establish the connection between our model and the classical contextual pricing model \citep{LLV-18, LS-18}. We begin by showing that a special case of our model with only two actions is equivalent to contextual pricing. As a result, the optimal regret bound of $O(\dimension \log \log T + \dimension \log \dimension)$ in this setting can be directly implied by existing result \citep{LLS-21} from the contextual pricing literature.
We then highlight a fundamental difference between the two models. Specifically, we demonstrate that even with just three actions in a two-dimensional setting, the optimal regret in our model becomes $\Omega(T^{1/4})$, representing a double-exponential gap compared to the regret achieved in classical contextual pricing.

\subsection{Contextual Search in Principal-Agent Games with Two Actions}

The pricing problem can be viewed as a special case of the principal-agent game with two actions. The principal is the seller, and the agent is the buyer. Without loss of generality, we assume that the valuation of the buyer is in $[0,1]$. There is only one non-trivial action in pricing, which represents ``purchase''. The reward of ``purchase'' is $\reward=1$ and the cost $\cost$ of the ``purchase'' action is $\cost=1-\val$ where $\val$ is the value of the buyer. The trivial action represents ``no purchase'' with reward and cost both being $0$. Posting a price $\price$ is equivalent to proposing a linear contract $\contract{}=1-\price$.

To understand why the above correspondence is reasonable, let us further investigate the utility functions of the principal (seller) and the agent (buyer). In the pricing view, the buyer purchases the item only when the price $\price \leq \val$, in which case their utility is $\val - \price$. When $\price > \val$, the buyer does not purchase the item, and their utility is zero.
This corresponds to the principal-agent view, where the agent will choose the action ``purchase'' only when $\contract{} \cdot \reward - \cost \geq 0$, which is equivalent to the condition $\price \leq \val$ according to our definition of actions.
As for the principal's utility, if the agent chooses ``purchase'', the principal receives a utility of $(1 - \contract{}) \cdot \reward = \price$, which aligns with the pricing view.

By the above discussion, it is not surprising to see that the contextual principal-agent game with two actions (i.e., a single non-trivial action) has an $O(\dimension \log \log T + \dimension \log \dimension)$ regret since contextual pricing has such a regret bound by \citet{LLS-21}. 
In fact, we prove the following theorem by giving a reduction to the contextual pricing problem. 
Its formal analysis 
is deferred to \Cref{apx:pricing connection proof}.

\begin{restatable}{theorem}{thmTwoActionRegret}
\label{thm:warm-up:pricing connection}
    In the setting where agents have at most two actions, there exists an online algorithm $\ALG$ with a pessimistic Stackelberg regret of 
    \begin{align*}
        \regretPessimistic(\ALG) = \Theta(\dimension \log \log T + \dimension \log \dimension).
    \end{align*}
    Moreover, it matches the lower bound of the (classic) Stackelberg regret, and thus is regret-optimal.
\end{restatable}

To better illustrate how this reduction works, we next outline the key idea behind the proof. Let action~$\adversarialAction$ denote the ``purchase'' action in the contextual pricing problem, and let $\contextAT{\adversarialAction}$ be the only context. Then, we note that the value $\val_t = \innerproduct{\contextAT{\adversarialAction}, \hvecTrue}$ follows a linear structure.
By invoking the correspondence mentioned earlier, we obtain a variant of the contextual search in the principal-agent game, where the context is defined based on the welfare $\rewardAT{\adversarialAction} - \costAT{\adversarialAction}$, rather than the cost $\costAT{\adversarialAction}$. Nevertheless, it is straightforward to construct a reduction to contextual pricing even for the cost context, which can be achieved by increasing the dimension of the context by one to encode $\rewardAT{\adversarialAction}$.

\subsection{Regret Lower Bound for Contextual Search with Three Actions}

In this section, we illustrate the difference between our model and the contextual pricing model. As we will demonstrate, the optimal regret in our model (even with three actions) grows double exponentially worse compared to the optimal regret in the contextual pricing model.

One important feature of the contextual pricing model is the implementability of the ``purchase'' action. Namely, the ``purchase'' action is never degenerated (\Cref{def:action degeneracy}). 
Although not explicitly stated, this simple feature allows the contextual pricing algorithms \citep{LS-18, LLS-21} to identify an ideal price for each buyer. As a result, whenever a high immediate regret occurs, the hypothesis set for the hidden vector $\hvecTrue$ shrinks significantly.

Unfortunately, in our principal-agent model with three or more actions, the adversary can exploit certain actions to intentionally degenerate a targeted action under the true cost hidden vector $\hvecTrue$. Since the algorithm does not know the hidden vector $\hvecTrue$ but only has a hypothesis set for it, it cannot determine whether such an action is degenerated. This gives the adversary sufficient leverage, leading to the following lower bound on the regret.

\begin{restatable}{theorem}{thmThreeActionRegretLB}
\label{thm:warm-up:three action regret lower bound}
    In the setting where agents have at most three actions and $2$-dimensional cost contexts, no online algorithm $\ALG$ can achieve a Stackelberg regret of 
    \begin{align*}
        \regret(\ALG) \leq \frac{1}{32}\cdot \timeHorizon^{\frac{1}{4}}.
    \end{align*}
\end{restatable}
In the remainder of this section, we first sketch the high-level intuition by describing a concrete scenario in which no online algorithm can achieve the ideal balance between immediate regret and the shrinkage of the hypothesis set, as is possible with contextual pricing algorithms with double logarithmic regret. We then extend the insights from this scenario to formally construct an adversary and prove \Cref{thm:warm-up:three action regret lower bound}.

\xhdr{Insufficient learning due to contextual action degeneracy.}
Consider action set $\actions = \set{\nullAction, \knownOptAction, \adversarialAction}$, where $\nullAction$ is the trivial action with zero cost and zero reward, $\knownOptAction$ and $\adversarialAction$ is associated with a $2$-dimensional context $\contextA{\knownOptAction}$ and $\contextA{\adversarialAction}$. The reward of actions $\knownOptAction$ and $\adversarialAction$ are $\rewardA{\knownOptAction}=\frac{1}{2}$ and $\rewardA{\adversarialAction} = \frac{1}{2} + \gap$, respectively. Here $\gap$ is a small parameter to be determined by the adversary.
The cost hidden vector $\hvecTrue$ is promised to take the form of $\hvecTrue = (\frac{1}{2},\hvecTrue_2)$, where $\hvecTrue_2$ is in $[ \frac{\sqrt{3}}{4},\frac{\sqrt{3}}{2}]$.

Suppose the cost contexts for actions $\knownOptAction$ and $\adversarialAction$ are given by $\contextA{\knownOptAction} = \left( \frac{1}{2}, 0 \right)$ and $\contextA{\adversarialAction} = \left( 0, \frac{\sqrt{3}}{3} \right)$, respectively. By construction, the cost of action $\knownOptAction$ is known to be $\costA{\knownOptAction} = \frac{1}{4}$, while the cost $\costA{\adversarialAction}$ of action $\adversarialAction$ ranges over $[\frac{1}{4}, \frac{1}{2}]$.
It can be verified that action~$\adversarialAction$ is fully dominated by action $\knownOptAction$ and becomes \emph{degenerated} if $\costA{\adversarialAction} > \frac{1}{4} + \gap$, which is very close to $\frac{1}{4}$. As a result, the principal cannot determine whether the cost of action $\adversarialAction$ exceeds any given threshold greater than $\frac{1}{4} + \gap$.
Even when proposing the contract $\contract{} = 1$, the algorithm can only decide whether $\costA{\adversarialAction} \leq \frac{1}{4} + \gap$ or not. Under this contract, the algorithm incurs $\Omega(1)$ immediate regret, while only shrinking the hypothesis set by a small $O(\gap)$ region.

This illustrates a key difference compared with contextual pricing, where the contextual pricing algorithm can determine whether the valuation exceeds any threshold in the range $[0, 1]$. In this case, the algorithm can shrink a much larger region while suffering $O(1)$ immediate regret, leading to a low cumulative regret.

\xhdr{Adversary for the three-action setting.} 
We now extend the insights from the discussion above to prove \Cref{thm:warm-up:three action regret lower bound}. 
(The pseudocode for our constructed adversary can be found in \Cref{adv:warm up:three-action setting}.)

\begin{adversary}[t!]
\caption{\textsc{Adversary for Three Actions}}
\label[adversary]{adv:warm up:three-action setting}
\SetAlgoLined
\KwData{$\ALG, \timeHorizon, \gap$}

 $\mincost \gets  \frac{1}{4}+\gap-\frac{\gap}{4\gap +2} - \frac{\gap^2}{4+8\gap}$

Initialize $\contextAT{\knownOptAction} = (\frac{1}{2},0), \rewardAT{\knownOptAction}=\frac{1}{2}, \rewardAT{\adversarialAction} = \frac{1}{2}+\gap, \forall \timeslot\geq 1$

Initialize $\radius\super{1} = \frac{\sqrt{3}}{2}$, $\hvecsPess_{1}=\set{\frac{1}{2}}\times   [\radius\super{1},\frac{\sqrt{3}}{2}]$

$\contractThreshold\gets \frac{4\mincost-1}{4\gap}$

$\timeslot\gets 1$

\While{$\timeslot\leq \timeHorizon$}{
    Set     $\contextAT{\adversarialAction}  =  \left(0, \frac{\mincost}{\radius\super{\timeslot}}\right)$
    
    Show contexts to $\ALG$ and receive contract $\contract{\timeslot}$.

    \If{$\contractSimple_{\timeslot}<\contractThreshold$}{
     $\radius\super{\timeslot+1} \gets \radius\super{\timeslot}$

     \If{
$\contract{\tpre} < \contractThreshold$ for all $\tpre \in [\timeslot - \sqrt{\timeHorizon} + 1 : \timeslot]$}
        {
        Determine $\hvecTrue = (\frac{1}{2},\radius\super{\timeslot})\in \hvecsPess_{\timeslot}$ and \textbf{halt}
}
    }
    \If{$\contractSimple_{\timeslot}\geq\contractThreshold$}{
    $\radius\super{\timeslot+1}\gets \min\set{\frac{1+4\gap\contract{\timeslot}}{4\mincost}\radius\super{\timeslot},\frac{\sqrt{3}}{2}}$    
    }
    $\hvecsPess_{\timeslot+1} \gets \set{\frac{1}{2}}\times [\radius\super{\timeslot+1},\frac{\sqrt{3}}{2}]$
    
    $\timeslot \gets \timeslot+1$
}
Determine $\hvecTrue = (\frac{1}{2},\frac{\sqrt{3}}{2})$
\end{adversary}

The setup follows the same construction as the scenario illustrated above: 
all agents share the same action set $\actionsT \equiv \actions = \set{\nullAction, \knownOptAction, \adversarialAction}$. Action $\nullAction$ is the trivial action with zero reward and zero cost. Action $\knownOptAction$ has a fixed reward of $\rewardAT{\knownOptAction} \equiv \rewardA{\knownOptAction} = \frac{1}{2}$ and a fixed cost context of $\contextAT{\knownOptAction} \equiv \contextA{\knownOptAction} = (\frac{1}{2}, 0)$. Action $\adversarialAction$ has a fixed reward of $\rewardAT{\adversarialAction} \equiv \rewardA{\adversarialAction} = \frac{1}{2} + \gap$, and its cost context changes over rounds, always taking the form $\contextAT{\adversarialAction} = (0, \eta_t)$, where $\eta_t \in [0, 1]$.
Finally, the cost hidden vector $\hvecTrue$ is guaranteed to take the form $\hvecTrue = (\frac{1}{2}, \hvecTrue_2)$, where $\hvecTrue_2 \in [ \frac{\sqrt{3}}{4}, \frac{\sqrt{3}}{2}]$. By construction, the cost of action $\knownOptAction$ remains fixed over rounds, with $\costAT{\knownOptAction} \equiv \costA{\knownOptAction} = \frac{1}{4}$.

To complete the construction of the adversary, it remains to specify the following components:
(i) the cost context of action $\adversarialAction$ for each agent,
(ii) each agent’s best response to any contract $\contract{\timeslot}$ proposed by the online algorithm, and
(iii) the selection of the cost hidden vector~$\hvecTrue$.
We describe each of these components in turn below.

\smallskip
\noindent
\emph{\underline{[Part (i)] Cost context construction.}} Our construction ensures that the online algorithm faces almost the same insufficient learning scenario in each round. Let $\hvecsPess_{\timeslot} = \set{\frac{1}{2}} \times [\radius\super{\timeslot}, \frac{\sqrt{3}}{2}]$ be the hypothesis set in round $\timeslot$ under the given online algorithm. 
The adversary selects $\contextAT{\adversarialAction} = (0, \frac{\mincost}{\radius\super{t}})$. By construction, $\mincost$ is the smallest cost of action $\adversarialAction$ consistent with the current hypothesis set. Here, $\mincost$ is a parameter chosen by the adversary, which will be determined later in the regret analysis.

\smallskip
\noindent
\emph{\underline{[Part (ii)] Agent's best response construction.}} We now specify the agent's best response before the adversary finalizes the cost hidden vector $\hvecTrue$ (discussed below). (After the cost hidden vector is finalized, the agent's best response is uniquely determined.) The idea here is also simple: the adversary will always let the agent take the known action $\knownOptAction$. More specifically, define the threshold $\contractThreshold \triangleq \frac{4\mincost - 1}{4\gap}$, which is round-independent. Next, we explain the construction depending on whether the proposed contract $\contract{\timeslot}$ is below or above this threshold.

When the proposed contract $\contract{\timeslot}$ satisfies $\contract{\timeslot} < \contractThreshold$, action $\adversarialAction$ is degenerated (due to action~$\knownOptAction$) for sure. Namely, even if the cost of action~$\adversarialAction$ takes its best (i.e., smallest) value, $\costAT{\adversarialAction} \equiv \mincost$, the agent in round $\timeslot$ will not take this action, since the profit sharing (i.e., $\contract{\timeslot}$) is not enough. Therefore, the hypothesis set $\hvecsPess_{\timeslot+1}$ will not be updated, i.e., 
\begin{align}
\label{eq:warm up:hypothesis set update no learning}
    \hvecsPess_{\timeslot+1} = \hvecsPess_{\timeslot}.
\end{align}

In contrast, when the proposed contract $\contract{\timeslot}$ satisfies $\contract{\timeslot} \geq \contractThreshold$, the agent will take action~$\adversarialAction$ if the true cost $\costAT{\adversarialAction}$ is close to $\mincost$. However, in this case, the adversary will let the agent take action~$\knownOptAction$ instead, and consequently allow the algorithm to update its hypothesis set by increasing the lower bound $\radius\super{\timeslot+1}$ of the hypothesis set $\hvecsPess_{\timeslot+1} = \set{\frac{1}{2}} \times [\radius\super{\timeslot+1}, \frac{\sqrt{3}}{2}]$ as  
\begin{align}
\label{eq:warm up:hypothesis set update under learning}
    \radius\super{\timeslot+1} = \min\left\{ \frac{1 + 4\gap \contract{\timeslot}}{4\mincost} \radius\super{\timeslot}, \frac{\sqrt{3}}{2} \right\}.
\end{align}

We formalize the agent's best response and the evolution of the hypothesis set in the following lemma.

\begin{lemma}\label{lem:warm up:update rule}
    Suppose \Cref{adv:warm up:three-action setting} is employed. 
    Fix any online algorithm $\ALG$. For each round $t\in[T]$:
    \begin{itemize}
        \item[(i)] If $\ALG$ proposes contract $\contract{\timeslot}\geq \contractThreshold$, then the best-responding action of agent is $\nullAction$ or $\knownOptAction$ supposing the current hypothesis set $\hvecsPess_{\timeslot} = \set{\frac{1}{2}} \times [\radius\super{\timeslot}, \frac{\sqrt{3}}{2}]$ satisfies $\radius\super{\timeslot} < \frac{\sqrt{3}}{2}\cdot \frac{4\mincost}{1+4\gap\contract{\timeslot}}$.
        \item[(ii)] If $\ALG$ proposes contract $\contract{\timeslot}<\contractThreshold$, then the best-responding action of agent is $\nullAction$ or $\knownOptAction$ without extra condition.
    \end{itemize}
    Moreover, the update of the hypothesis set defined in \cref{eq:warm up:hypothesis set update no learning,eq:warm up:hypothesis set update under learning} is consistent with the agent's best-responding action.
\end{lemma}
\begin{proof}
    Fix an arbitrary round $t\in[T]$. We consider two cases depending on the proposed contract $\contract{\timeslot}$ in the algorithm $\ALG$ separately. 

    \xhdr{Case (i): $\contract{\timeslot} \geq \contractThreshold$.} In this case, agent $t$'s utility from taking action~$\adversarialAction$ is
    \begin{align*}
        \contract{\timeslot}\cdot \rewardAT{\adversarialAction} - 
        \costAT{\adversarialAction} = 
        \contract{\timeslot}\cdot \rewardAT{\adversarialAction}
        -
        \innerproduct{\contextAT{\adversarialAction}, \hvecTrue}
        =
        \contract{\timeslot}\cdot \left(\frac{1}{2} + \gap\right)
        -
        \frac{\mincost}{\radius\super{\timeslot}}\cdot \hvecTrue_2.
    \end{align*}
    In contrast, agent $t$'s utility from taking action~$\knownOptAction$ is
    \begin{align*}
        \contract{\timeslot}\cdot \rewardAT{\knownOptAction} - \costAT{\knownOptAction} = \contract{\timeslot}\cdot \frac{1}{2} - \frac{1}{4}.
    \end{align*}
    Combining the two bounds above, agent $t$ takes action $\knownOptAction$ if and only if 
    \begin{align*}
        \hvecTrue_2 \geq \frac{1+4\gap\contract{\timeslot}}{4\mincost}\cdot \radius\super{\timeslot}.
    \end{align*}
    Due to the statement assumption that $\radius\super{\timeslot} < \frac{\sqrt{3}}{2}\cdot \frac{4\mincost}{1+4\gap\contract{\timeslot}}$, it is valid for the adversary to update $\radius\super{\timeslot+1}>\frac{1+4\gap\contract{\timeslot}}{4\mincost}\cdot \radius\super{\timeslot}$ to make agent $t$ take action~$\knownOptAction$, and thus the lemma statement holds. 

    \xhdr{Case (ii): $\contract{\timeslot} < \contractThreshold$.} In this case, applying the same calculation as Case (i), it is guaranteed that agent $t$'s utility from taking action~$\adversarialAction$ is strictly smaller than his utility from taking action~$\knownOptAction$ regardless of the cost hidden vector $\hvecTrue\in \hvecsPess_{\timeslot}$. Thus, the lemma statement holds.

    Combining the analysis for the two cases above, we complete the proof of \Cref{lem:warm up:update rule}.\qedhere

\end{proof}

\smallskip
\noindent
\emph{\underline{[Part (iii)] Choice of cost hidden vector.}} Finally, we explain how the adversary determines the cost hidden vector $\hvecTrue$. Here, we consider two cases.

The first case corresponds to the scenario where the online algorithm does not conduct learning for consecutive $\sqrt{T}$ rounds. Specifically, by \Cref{lem:warm up:update rule}, there exists $\timeslot \in [T]$ such that for every $\tpre \in [\timeslot - \sqrt{T} +1: \timeslot]$, the proposed contract satisfies $\contract{\tpre} < \contractThreshold$. In this case, the adversary finalizes the cost hidden vector as $\hvecTrue = \left(\frac{1}{2}, \radius^{(\timeslot)}\right)$. This choice of $\hvecTrue$ ensures that in the previous $\sqrt{T}$ rounds, the principal's optimal payoff is achieved by incentivizing action $\adversarialAction$ with the contract $\contract{\timeslot}^* = \contractThreshold$. By calculation, the cumulative regret from those $\sqrt{T}$ rounds would be $\Omega(\sqrt{T} \cdot \gap)$. See \Cref{lem:warm up:no enough learning} for the formal regret statement and analysis.

The second case is complementary to the first. In this scenario, there are at least $\sqrt{T}$ rounds where the algorithm conducts learning, i.e., the proposed contract satisfies $\contract{\timeslot} \geq \contractThreshold$. In this case, the adversary finalizes the cost hidden vector as $\hvecTrue = \left(\frac{1}{2}, \frac{\sqrt{3}}{2}\right)$. This choice of $\hvecTrue$ ensures that the principal's optimal payoff is achieved by incentivizing action $\knownOptAction$ with the optimal contract $\contract{\timeslot}^* = \frac{1}{2}$, which is smaller than the actual contract $\contract{\timeslot} \geq \contractThreshold = \frac{1}{2} + \Omega(\gap)$ implemented by the algorithm in those rounds.
Notably, in this case, the regret is worse than the contextual pricing setting, since the online algorithm cannot shrink the hypothesis set aggressively due to the contextual action degeneracy.
See \Cref{lem:warm up:too much learning} for the formal regret statement and analysis.

\begin{lemma}
\label{lem:warm up:no enough learning}
    Suppose \Cref{adv:warm up:three-action setting} 
    is employed with $\mincost \triangleq \frac{1}{4}+\gap-\frac{\gap}{4\gap +2} - \frac{\gap^2}{4+8\gap}$.
    For any online algorithm $\ALG$, if it proposes contract $\contract{\tpre}<\contractThreshold$ for consecutive $\sqrt{T}$ rounds, then its Stackelberg regret is at least 
    \begin{align*}\regret(\ALG)\geq \frac{\gap}{8}\cdot \sqrt{T}.\end{align*}
\end{lemma}
\begin{proof}
    Suppose algorithm $\ALG$ proposes contracts $\contract{\tpre} < \contractThreshold, \forall \tpre\in [\timeslot - \sqrt{T} +1:\timeslot]$ for some $\timeslot\in[T]$. 
    Invoking \Cref{lem:warm up:update rule}, the hypothesis set remains unchanged in those rounds, i.e., $\hvecsPess_\tpre \equiv \hvecsPess_\timeslot = [\radius\super{\timeslot},\frac{\sqrt{3}}{2}]$ for every round $\tpre \in [\timeslot - \sqrt{T}+1:\timeslot]$.

    By construction, \Cref{adv:warm up:three-action setting} finalizes cost hidden vector $\hvecTrue = (\frac{1}{2},\radius\super{\timeslot})$. Consequently, the optimal utility of the principal in each round $\tpre \in [\timeslot - \sqrt{T}+1:\timeslot]$ is at least 
    \begin{align*}
        \max\nolimits_{\contract{}\in[0, 1]}~
        \revenuePrincipalCon_\tpre(\contract{}) \geq 
        \revenuePrincipalCon_\tpre(\contractThreshold) =
        (1 - \contractThreshold) \cdot \rewardAT{\adversarialAction}
        =
        \left(1 - 
         \frac{4\mincost - 1}{4\gap}
         \right)
         \cdot 
         \left(\frac{1}{2} + \gap\right) = \frac{1}{4} + \frac{\gap}{8},
    \end{align*}
    where $\contractThreshold = \frac{4\mincost - 1}{4\gap}$ is the smallest and most profitable contract to incentivize action $\adversarialAction$, and the last equality holds due to the assignment of $\mincost$. 

    In contrast, since the proposed contract $\contract{\tpre} < \contractThreshold$ in algorithm $\ALG$, action~$\nullAction$ or action~$\knownOptAction$ is taken by agent $\tpre$, which induces the principal's utility of at most $0$ and $\frac{1}{4}$, respectively. 

    Combining the above two pieces, we lower bound the Stackelberg regret of algorithm $\ALG$ as 
    \begin{align*}
        \regret(\ALG) = \sum\nolimits_{\tpre\in[T]}
        \left(\left(\max\nolimits_{\contract{}\in [0,1]}\revenuePrincipalCon_\tpre(\contract{})\right) - 
        \revenuePrincipalCon_\tpre(\contract{\tpre})\right)
        \geq 
        \sum\nolimits_{\tpre \in[\timeslot - \sqrt{T} + 1:\timeslot]}
        \left(\frac{1}{4} + \frac{\gap}{8} - \frac{1}{4}\right)
        = \frac{\gap}{8}\cdot \sqrt{T}
    \end{align*}
    which completes the proof of \Cref{lem:warm up:no enough learning}.
\end{proof}

\begin{lemma}\label{lem:warm up:too much learning}
    Suppose \Cref{adv:warm up:three-action setting} 
    is employed with $\gap < \frac{1}{4}$ and $\mincost \triangleq \frac{1}{4}+\gap-\frac{\gap}{4\gap +2} - \frac{\gap^2}{4+8\gap}$.
    For any online algorithm $\ALG$, if it never proposes contract $\contract{\tpre}<\contractThreshold$ for consecutive $\sqrt{T}$ rounds, then its Stackelberg regret is at least 
    \begin{align*}\regret(\ALG)\geq \min \set{\frac{\gap}{4}\cdot \sqrt{T},~ \frac{1}{8\gap}\cdot \ln \frac{23}{16}}.\end{align*}
\end{lemma}
\begin{proof}
    By the statement assumption, algorithm $\ALG$ proposes $\contract{\timeslot} \geq \contractThreshold$ for at least $\sqrt{T}$ rounds. Our analysis consider two cases depending on the hypothesis set $\hvecsPess_{T+1} = [\radius\super{T + 1},\frac{\sqrt{3}}{2}]$ after observing the best-responding action of agent $T$ in the last round.

    \xhdr{Case (i): $\radius\super{T + 1} < \frac{\sqrt{3}}{2}$.} In this case, the hypothesis set at the end does not collapse to a singleton. By \Cref{lem:warm up:update rule}, the agent takes action $\knownOptAction$ for all rounds.
    By construction, \Cref{adv:warm up:three-action setting} finalizes cost hidden vector $\hvecTrue = (\frac{1}{2},\frac{\sqrt{3}}{2})$. Consequently, the optimal utility of the principal in each round $\timeslot$ is at least 
    \begin{align*}
        \max\nolimits_{\contract{}\in[0, 1]}~
        \revenuePrincipalCon_\timeslot(\contract{}) \geq 
        \revenuePrincipalCon_\timeslot\left(\frac{1}{2}\right) =
        \frac{1}{2} \cdot \rewardAT{\knownOptAction}
        =
        \frac{1}{4}
    \end{align*}
    where $\contract{} = \frac{1}{2}$ is the smallest and most profitable contract to incentivize action $\knownOptAction$. 

    In contrast, since the proposed contract $\contract{\timeslot} \geq \contractThreshold$ in algorithm $\ALG$ for at least $\sqrt{T}$ rounds and $\radius\super{T + 1} < \frac{\sqrt{3}}{2}$, action~$\knownOptAction$ is taken by agent $\timeslot$ but the contract shares too much profit to the agent. Namely,
    \begin{align*}
        \revenuePrincipalCon_\timeslot(\contract{\timeslot}) = 
        (1 - \contract{\timeslot}) \cdot \rewardAT{\knownOptAction}
        \leq 
        (1 - \contractThreshold) \cdot \rewardAT{\knownOptAction}
        =
        \frac{2+\gap}{8+16\gap}
    \end{align*}
    Combining the above two pieces, we lower bound the Stackelberg regret of algorithm $\ALG$ as 
    \begin{align*}
        \regret(\ALG) = \sum\nolimits_{\timeslot\in[T]}
        \left(\left(\max\nolimits_{\contract{}\in[0,1]}\revenuePrincipalCon_\timeslot(\contract{})\right) - 
        \revenuePrincipalCon_\timeslot(\contract{\timeslot})\right)
        \geq 
        \sqrt{T}
        \cdot \left(
        \frac{1}{4} - \frac{2+\gap}{8+16\gap}
        \right)
        \geq \frac{\gap}{4}
    \end{align*}
    where the last inequality holds due to statement assumption that $\gap \leq \frac{1}{4}$.
    
    \xhdr{Case (ii): $\radius\super{T + 1} = \frac{\sqrt{3}}{2}$.} In this case, the hypothesis set at the end collapses to a singleton $\hvecsPess_{T + 1} = {\frac{\sqrt{3}}{2}}$. By construction, \Cref{adv:warm up:three-action setting} finalizes cost hidden vector $\hvecTrue = (\frac{1}{2},\frac{\sqrt{3}}{2})$. Suppose there are $L$ rounds in which the hypothesis set strictly shrinks, and denote these rounds as $t_1, t_2, \dots, t_L$. Thus, $\contract{\timeslot_\ell} >\contractThreshold$ for any $\ell\in [1:L]$.
For each of these rounds $t_\ell$ with index $\ell\in[L - 1]$, the agent takes action $\knownOptAction$ by \Cref{lem:warm up:update rule}, and the immediate regret is 
    \begin{align*}
        \left(\max\nolimits_{\contract{\in[0,1]}}\revenuePrincipalCon_{\timeslot_\ell}(\contract{})\right) - 
        \revenuePrincipalCon_{\timeslot_\ell}(\contract{{\timeslot_\ell}})
        &\overset{(a)}{\geq}
        \frac{1}{4} - (1- \contract{{\timeslot_\ell}})\cdot \frac{1}{2} 
        \overset{(b)}{\geq} 
        (\contract{{\timeslot_\ell}} - \contractThreshold)\cdot \frac{1}{2} \overset{(c)}{=}  \frac{4\mincost}{8\gap}\cdot\left(\frac{1+4\gap \contract{\timeslot_\ell}}{4\mincost}-1\right)
        \\
        &\overset{(d)}{\geq}
        \frac{1}{8\gap}\cdot 
        \left(\frac{1+4\gap\contract{\timeslot_\ell}}{4\mincost} - 1\right)
\overset{(e)}{=}
        \frac{1}{8\gap}
        \cdot \left(
        \frac{\radius\super{\timeslot_\ell + 1}}{\radius\super{\timeslot_\ell}} - 1
        \right)
        \overset{(f)}{\geq}
        \frac{1}{8\gap}\cdot \ln\left(\frac{\radius\super{\timeslot_\ell + 1}}{\radius\super{\timeslot_\ell}}\right)
    \end{align*}
    where inequality~(a) holds due to the analysis in Case (i), inequality~(b) holds since $\contractThreshold \geq \frac{1}{2}$ due to the assignment of $\gap$ and $\mincost$, equality~(c) is by the definition of $\contractThreshold$, inequality~(d) holds due to $\mincost\geq \frac{1}{4}$ and $\contract{\timeslot_\ell}> \contractThreshold$, equality~(e) holds due to the construction of \Cref{adv:warm up:three-action setting} and its induced hypothesis set evolution (\Cref{lem:warm up:update rule}), and inequality~(f) holds by algebra.

    Aggregating over the immediate regret for all rounds, we lower bound the Stackelberg regret of algorithm $\ALG$ as 
    \begin{align*}
        \regret(\ALG) 
        &= \sum\nolimits_{\timeslot\in[T]}
        \left(\left(\max\nolimits_{\contract{\in[0,1]}}\revenuePrincipalCon_\timeslot(\contract{})\right) - 
        \revenuePrincipalCon_\timeslot(\contract{\timeslot})\right)
        \geq 
        \sum\nolimits_{\ell\in[L-1]}
        \left(\max\nolimits_{\contract{\in[0,1]}}\revenuePrincipalCon_{\timeslot_\ell}(\contract{})\right) - 
        \revenuePrincipalCon_{\timeslot_\ell}(\contract{\timeslot_\ell})
        \\
        &\geq 
        \frac{1}{8\gap}\cdot \sum\nolimits_{\ell\in[L-1]} \ln\left(\frac{\radius\super{\timeslot_\ell + 1}}{\radius\super{\timeslot_\ell}}\right)
        \overset{(a)}{=}
        \frac{1}{8\gap}\cdot 
        \ln\left(\frac{\radius\super{\timeslot_L}}{\radius\super{1}}\right)
        \overset{(b)}{\geq} \frac{1}{8\gap}\cdot \ln \frac{23}{16},
    \end{align*}
    where equality~(a) holds since the hypothesis set is not updated from round $\timeslot_{L-1} + 1$ to round $\timeslot_L - 1$, and inequality~(b) holds since $\radius\super{\timeslot_L} \geq \radius\super{\timeHorizon+1}\cdot \frac{4\mincost}{1+4\gap\contract{\timeslot_L}} \geq \frac{\sqrt{3}}{2}\cdot \frac{4\mincost}{1+4\gap}$ (due to our updating rule of the hypothesis set and the fact that $L$ is last round to make strict hypothesis shrinkage), $\gap \leq \frac{1}{4}$ (due to statement assumption), and $\radius\super{1} = \frac{\sqrt{3}}{4}$ (due to the construction of \Cref{adv:warm up:three-action setting}).

    Combining the analysis for the two cases above, we complete the proof of \Cref{lem:warm up:too much learning}.\qedhere

\end{proof}

\begin{proof}[\bf{Proof of \Cref{thm:warm-up:three action regret lower bound}}]
    Consider \Cref{adv:warm up:three-action setting} with parameters $\gap \triangleq  \frac{1}{4}\timeHorizon^{-\frac{1}{4}}$ and $\mincost \triangleq \frac{1}{4}+\gap-\frac{\gap}{4\gap +2} - \frac{\gap^2}{4+8\gap}$.
    Invoking \Cref{lem:warm up:no enough learning,lem:warm up:too much learning} completes the proof.
\end{proof}

In \Cref{sec:unknown cost,sec:unknown reward} and \Cref{apx:classic lower bound}, we extend \Cref{adv:warm up:three-action setting} from the special setting with three actions and an (essentially) one-dimensional hidden vector to more general settings with an arbitrary number of actions and a hidden vector of arbitrary dimension. First, our constructed adversaries set the arriving context so that the online algorithm faces almost the same insufficient learning scenario in each round as in \Cref{adv:warm up:three-action setting} (Part (i)).
To obtain regret lower bounds with our targeted dependence on the dimension of the hidden vector (e.g., an exponential dependence in the pessimistic Stackelberg regret),
we utilize the spherical code (\Cref{def:spherical codes}), which forces the online algorithm to learn for a sufficient number of rounds in an exponential number of directions in high-dimensional space.
Finally, in addition to extending the regret analysis in \Cref{lem:warm up:no enough learning,lem:warm up:too much learning} (which is essentially a potential function argument based on $\radius^{(\timeslot)}$), we introduce one additional action in the adversaries setup to gain better control over the evolution of the high-dimensional hypothesis set. 
\section{Principal-Agent Games with Cost Context}
\label{sec:unknown cost}
In this section, we present our results on the contextual principal-agent game with cost context. We present our upper bound results in \Cref{sec:upper bound cost} and lower bound results in \Cref{sec:lower-bound-cost-opt}.

\subsection{Upper Bound for Cost Context}\label{sec:upper bound cost}
For the contextual principal-agent game with cost contexts, we first establish the following upper-bound result.

\begin{theorem}\label{thm:cost context upper bound}
    For the contextual principal-agent game with cost contexts, let $\dimension$ be the dimension of the context. Let $\ALG$ be Algorithm~\ref{alg: single stage} 
    with parameter $\gap = \min\set{(2\pi)^{\frac{\dimension}{4\dimension+2}}\dimension^{\frac{\dimension+4}{4\dimension+2}}T^{-\frac{1}{2\dimension+1}},\frac{1}{2}}$, then we have the following pessimistic Stackelberg regret bound of $\ALG$,
\[\regretPessimistic(\ALG)\leq 5(2\pi)^{\frac{\dimension}{4\dimension+2}} \dimension^{\frac{\dimension+4}{4\dimension+2}} T^{1-\frac{1}{2\dimension+1}}.\]
    Since the classic Stackelberg regret is always no larger than the pessimistic Stackelberg regret, our regret upper bound also holds for classic Stackelberg regret.
\end{theorem}

We first present our algorithm, introduce the information width and potential function in \Cref{sec:information width and potential} 
and complete 
its regret analysis in \Cref{sect: cost upper bound}. 
Before formally illustrating our algorithm, we need to bring in the concept of \emph{indifference contexts} which is used in our algorithm design.
For $\timeslot$-th round, given an agent with cost context $\context_{\timeslot}\deq\set{\contextAT{\action}}_{\action\in \actionsT}$ and the corresponding reward $\set{\rewardAT{\action}}_{\action\in \actionsT}$.
For $\action,\action^{\dagger}\in\actionsT$ with $\rewardAT{\action}\neq \rewardAT{\action^{\dagger}}$, we define the their indifference context $\jumpVec{\action}{\action^{\dagger}} \deq \frac{\contextAT{\action}-\contextAT{\action^{\dagger}}}{\rewardAT{\action}-\rewardAT{\action^{\dagger}}}\in \realNumbers^{\dimension}$. 
It is readily verifiable that $\jumpVec{\action}{\action^{\dagger}}$ is the vector such that the agent gets the same utility from action $\action$ and $\action^{\dagger}$ (ignoring other actions) when accepting the linear contract $\contractSimple=\lrangle{\jumpVec{\action}{\action^{\dagger}},\hvecTrue}$. 
Moreover, we define the indifference set $\jumpSet$ to be the set containing all pairs $(\action,\action^{\dagger})$ for any $\action, \action^{\dagger}\in \actionsT$ with $\action\neq \action^{\dagger}$. 

\begin{remark}
    Obviously, $\jumpVec{\action}{\action^{\dagger}} = \jumpVec{\action^{\dagger}}{\action}$. We do not distinguish between these two indifference contexts and use the notation where the latter action has a larger reward compared with the former action for simplicity. That is, $\rewardAT{\action^{\dagger}}> \rewardAT{\action}$ in notation $\jumpVec{\action}{\action^{\dagger}}$.
\end{remark}

\xhdr{Algorithm.} Our algorithm employs an optimistic approach. In each round $\timeslot$, it first identifies the most optimistic hypothesis $\hvecOptT$ within the current hypothesis set $\hvecs_{\timeslot}$—that is, the hypothesis that would yield the maximum possible principal's utility:
\[\hvecOptT  \in \argmax\limits_{\hvec\in\hvecs_\timeslot}\max\limits_{\contractSimple\in[0, 1]}~\revenuePrincipalCon_\timeslot(\contract{}| \hvec)\]
Given this optimistic hypothesis $\hvecOptT$, we can determine the corresponding optimal contract $\contractOptT$ and the action $\actionOptT$ that the agent would select under this contract. 
Note that $\hvecOptT, \contractOptT, \actionOptT$ can be calculated efficiently, see \Cref{remark:efficient calculation} for details.
This pair $(\contractOptT, \actionOptT)$ represents the best-case scenario for the principal in round $\timeslot$. However, proposing $\contractOptT$ directly is risky. If the true hypothesis $\hvecTrue$ is not $\hvecOptT$, the agent might choose a different action. In a pessimistic scenario, this could lead to significant immediate regret while providing minimal information—for instance, only learning that $\hvecTrue \neq \hvecOptT$, which corresponds to removing a zero-measure set from $\hvecs_{\timeslot}$.
To mitigate this risk, our algorithm proposes a more conservative contract, $\contract{\timeslot} = \min\{\contractOptT + \gap, 1\}$, where $\gap > 0$ is a tunable parameter. This ``padded" contract is designed to balance the trade-off between exploiting the optimistic outcome and ensuring substantial learning in the pessimistic one. The analysis then proceeds based on the agent's response, leading to two distinct cases:
\begin{itemize}
    \item High-Reward Outcome ($\rewardAT{\actionAgent} \geq \rewardAT{\actionOptT}$): If the agent selects an action whose reward is at least as high as the optimistic target's reward, the outcome is favorable. In this case, we will prove that the principal's immediate regret is small, bounded by $O(\gap)$. As the observation is consistent with our optimistic outlook, we keep the hypothesis set unchanged for the next round: $\hvecs_{\timeslot+1} = \hvecs_{\timeslot}$.
    \item Low-Reward Outcome ($\rewardAT{\actionAgent} < \rewardAT{\actionOptT}$): If the agent selects an action with a lower reward, it signals that our optimistic hypothesis was incorrect. While the immediate regret in this case can be a large constant, the agent's choice provides significant information. We leverage this information to shrink the hypothesis set. As shown in Lemma~\ref{lem:non-degenerate} and Lemma~\ref{lem:hypothesis valid}, the set can be updated as follows: $\hvecs_{\timeslot+1} \gets \hvecs_{\timeslot} \cap \set{\hvec \mid \lrangle{\jumpVec{\actionT}{\actionOptT}, \hvec} \geq \contract{\timeslot}}$.
\end{itemize}

We include a formal pseudo-code in Algorithm~\ref{alg: unknown cost}.
To formally measure the knowledge acquired in the second case, we use the concept of \emph{information width} introduced in \Cref{def:information width} of \Cref{sec:information width and potential}. We will show in Lemma~\ref{lem:information width cost} that each time the low-reward outcome occurs, the information width gained is at least $\Omega(\gap)$. Subsequently, in Section~\ref{sec:information width and potential}, we introduce a potential function based on the intrinsic volume of the hypothesis set. Lemma~\ref{lem:potential sum information width} establishes that the decay in this potential is lower-bounded by the information width. Since the initial potential is bounded, the number of times the high-regret, high-information case can occur is upper-bounded. By carefully tuning the parameter $\gap$, we balance these two sources of regret to establish the final regret upper bound for our algorithm.

\begin{algorithm}[t]
\caption{\textsc{Learning Indifferent Point}}
\label{alg: unknown cost}
\SetAlgoLined
\KwData{$\gap$}
Initialize $\hvecs_1 = \mathbb{B}_{\dimension}$

\For{$\timeslot=1\rightarrow T$}{
    Receive context $\context_{\timeslot}\deq\set{\contextAT{\action}}_{\action\in \actionsT}$

    Calculate the optimism hidden vector
    $\hvecOptT  \in \argmax_{\hvec\in\hvecs_\timeslot}\max\limits_{\contractSimple\in[0, 1]}~\revenuePrincipalCon_\timeslot(\contract{}|\hvec)$

    Let $\contractOptT$ be the optimal linear contract in $\instance{\actionsT}{\hvecOptT}$ and $\actionOptT$ be the action taken by the agent when the contract is $\contractOptT$

    Propose $\contract{\timeslot}=\min\set{\contractOptT + \gap, 1}$

    Observe the action $\actionT$ taken by the agent
    
    \If{$\rewardAT{\actionT}\geq \rewardAT{\actionOptT}$}{
        $\hvecs_{\timeslot+1}\gets \hvecs_{\timeslot}$
    } 
    \If{$\rewardAT{\actionT}< \rewardAT{\actionOptT}$}{
        $\hvecs_{\timeslot+1}\gets \hvecs_{\timeslot}\cap \set{\hvec \condition \lrangle{\jumpVec{\actionT}{\actionOptT},\hvec}\geq \contract{\timeslot}}$
    }
}
\end{algorithm}

\begin{remark}\label{remark:efficient calculation}
    The optimism hidden vector $\hvecOptT$ as well as $\actionOptT, \contractOptT$ can be efficiently calculated by solving $n$ convex programs, where $n$ is the number of actions. Concretely, for each action $\action$, we solve the following programming problem:
    \begin{mini!}
    {\hvec\in \hvecs_{\timeslot}, \contractSimple\geq 0}
    {\contractSimple}
    {}
    {} 
    \addConstraint{\contractSimple \cdot \rewardAT{\action}-\lrangle{\contextAT{\action},\hvec}}{\geq \contractSimple \cdot \rewardAT{\action^{\dagger}}-\lrangle{\contextAT{\action^{\dagger}},\hvec}, \quad \quad \forall \action^{\dagger}\in \actionsT}
\end{mini!}
    In our algorithm, $\hvecs_{\timeslot}$ is always the intersection of a unit ball and a polytope. Thus, $\hvecs_{\timeslot}$ is convex for any $\timeslot$. Therefore, the above program is a convex program and can be solved efficiently.
    The convex program calculates the ``min-pay'' contract to implement $\action$ under any possible $\hvec$. We then select the action $\actionOptT$ with best principal's utility under the corresponding ``min-pay'' contract. Then we can obtain $\contractOptT$ and $\hvecOptT$ from its ``min-pay'' program.
\end{remark}

Our analysis of Algorithm~\ref{alg: unknown cost} use a potential argument. In \Cref{sec:information width and potential}, we introduce the potential function and a concept of information width to quantify the knowledge obtained by cutting the hypothesis set using a hyperplane.

\subsubsection{Information Width and Potential Function}\label{sec:information width and potential}

\begin{definition}[Information width]\label{def:information width}
    Given a convex body $\hvecs \subseteq \realNumbers^{\dimension}$, a context vector $\context\in \realNumbers^{\dimension}$ and a real number $\reward\in [\min_{\hvec\in\hvecs}\lrangle{\context,\hvec}, \max_{\hvec\in\hvecs}\lrangle{\context,\hvec}]$, we define the \emph{information width} as 
    \begin{equation}
        \informationWidth{\context}{\reward}{\hvecs} \deq \max_{\hvec\in\hvecs}\lrangle{\context,\hvec}-\reward.
    \end{equation}
\end{definition}

\begin{definition}[Potential function]\label{def:potential function}
    Let $\potential{j}{\convexBody} \deq j!\intrinsicVolume{j}{\convexBody}$ where $\intrinsicVolume{j}{\convexBody}$ is $j$-th intrinsic volume of $\convexBody$. The potential function of $\convexBody$ is defined as $\potentialSum{\convexBody} = \sum_{j=1}^{\dimension} \potential{j}{\convexBody}$.
\end{definition}

\begin{lemma}[Initial potential]\label{lem:initial potential}
    $\potentialSum{\mathbb{B}_{\dimension}}\leq \dimension (2\pi\dimension)^{\dimension/2}$.
\end{lemma}
\begin{proof}
    \Cref{lem:intrinsic volume properties}.(4) and \Cref{lem:intrinsic volume properties}.(5) directly give $\potential{j}{\mathbb{B}_{\dimension}}\leq \left(\potential{1}{\mathbb{B}_{\dimension}}\right)^j\leq (2\pi\dimension)^{j/2}, \forall j\geq 1$.
    Lemma follows from $(2\pi \dimension)^{j/2} \leq (2\pi \dimension)^{\dimension}$ and $\potentialSum{\mathbb{B}_{\dimension}} = \sum_{j=1}^d \potential{j}{\mathbb{B}_{\dimension}}$.
\end{proof}

The following two lemmas build the connection between information width and the decay of the potential function.

\begin{lemma}\label{lem:potential information width}
     Given a vector $\context\in \realNumbers^{\dimension}$ and a real number $\reward\in\realNumbers$, define 
     \[\convexBodyNext \deq \convexBody \cap \set{\hvec\condition \lrangle{\hvec, \context}\leq \reward}.\]
     Then there exist $1\leq j\leq \dimension$ such that,
     \[\potential{j}{\convexBody}-\potential{j}{\convexBodyNext}\geq \frac{1}{d}\left(\frac{1}{\|\context\|}\informationWidth{\context}{\reward}{\convexBody}\right)^j.\]
\end{lemma}
\begin{proof}
    \newcommand{\hight}{H}
    Firstly, we define a convex cone. Let its apex be $\coneApex\in \argmax_{\hvec\in\convexBody} \lrangle{\hvec,\context}$. Let its base be 
    \[\coneBase \deq \convexBody \cap \set{\hvec \condition \lrangle{\hvec,\context} = \reward}.\]
    Then the cone is defined as
    \[\cone \deq \bigcup_{\lambda \in [0,1]} \left(\lambda \coneApex + (1-\lambda)\coneBase\right).\]
    Since $\coneApex\in \convexBody$, $\coneBase\subseteq \convexBody$ and $\convexBody$ is convex, we have $\cone\subseteq \convexBody$. By the modularity of intrinsic volume, we have 
    \[\intrinsicVolume{j}{\cone}+\intrinsicVolume{j}{\convexBodyNext} = \intrinsicVolume{j}{\cone\cup \convexBodyNext}+\intrinsicVolume{j}{\cone\cap \convexBodyNext}.\]
    Note that $\cone \cup \convexBodyNext \subseteq \convexBody$ and $\cone \cap \convexBodyNext = \coneBase$, we have 
    \[ \intrinsicVolume{j}{\cone}+\intrinsicVolume{j}{\convexBodyNext} \leq \intrinsicVolume{j}{\convexBody}+\intrinsicVolume{j}{\coneBase}.\]
    The inequality is due to the monotonicity of the intrinsic volume. 
    
    Note that the height of $\cone$ is $ \hight\deq\frac{1}{\|\context\|}\informationWidth{\context}{\reward}{\convexBody}\leq 1$.
    By Lemma~\ref{lem:intrinsic volume properties}.(3), $\intrinsicVolume{j}{\cone}\geq \frac{1}{j} \hight\cdot\intrinsicVolume{j-1}{\coneBase}$ and we have 
    \[ \frac{1}{j} \hight\cdot\intrinsicVolume{j-1}{\coneBase}+\intrinsicVolume{j}{\convexBodyNext} \leq \intrinsicVolume{j}{\convexBody}+\intrinsicVolume{j}{\coneBase}.\]
    By rearranging, 
    \begin{equation}\label{eq: intrinsic volume decay}
        \intrinsicVolume{j}{\convexBody}-\intrinsicVolume{j}{\convexBodyNext}\geq \frac{1}{j} \hight \cdot \intrinsicVolume{j-1}{\coneBase}-\intrinsicVolume{j}{\coneBase}. 
    \end{equation}
    Note that inequality \eqref{eq: intrinsic volume decay} holds for all $j\geq 1$. 

    Firstly, we show that, there exist a $1\leq j\leq d$ such that the following condition holds: 
    \[\intrinsicVolume{j-1}{\coneBase}\geq \frac{1}{\factorial{(j-1)}}\left(1-\frac{j-1}{\dimension}\right)\hight^{j-1} \text{ and } \intrinsicVolume{j}{\coneBase}\leq \frac{1}{\factorial{j}}\left(1-\frac{j}{\dimension}\right)\hight^j.\]
    We will show this by contradiction. Suppose for all $1\leq j\leq d$, the above condition does not hold, then $\forall 1\leq j \leq d$, 
    \[\intrinsicVolume{j-1}{\coneBase} < \frac{1}{\factorial{(j-1)}}\left(1-\frac{j-1}{\dimension}\right)\hight^{j-1} \text{ or } \intrinsicVolume{j}{\coneBase}> \frac{1}{\factorial{j}}\left(1-\frac{j}{\dimension}\right)\hight^j.\]
    
    Note that $\intrinsicVolume{0}{\coneBase}=1\geq \hight$, so it must holds that $ \intrinsicVolume{1}{\coneBase}> \left(1-\frac{1}{\dimension}\right)\hight$ since the the condition does not hold for $j=1$. Using the reduction argument, suppose we have 
    \[\intrinsicVolume{j-1}{\coneBase}> \frac{1}{\factorial{j-1}}\left(1-\frac{j-1}{\dimension}\right)\hight^{j-1},\]
    then it must hold that $\intrinsicVolume{j}{\coneBase}> \frac{1}{\factorial{j}}\left(1-\frac{j}{\dimension}\right)\hight^j$ since the condition does not hold for $j$. By reduction, \[\intrinsicVolume{d}{\coneBase}> \frac{1}{\factorial{d}}\left(1-\frac{d}{\dimension}\right)\hight^d = 0.\]
    However, $\coneBase$ can be embodied in $\realNumbers^{\dimension-1}$ so $\intrinsicVolume{d}{\coneBase}=0$ which leads to a contradiction.

    Let $1\leq j\leq \dimension$ be the integer satisfying
    \begin{equation}\label{eq:noname1}
        \intrinsicVolume{j-1}{\coneBase}\geq \frac{1}{\factorial{(j-1)}}\left(1-\frac{j-1}{\dimension}\right)\hight^{j-1} \text{ and } \intrinsicVolume{j}{\coneBase}\leq \frac{1}{\factorial{j}}\left(1-\frac{j}{\dimension}\right)\hight^j.
    \end{equation}
    Then we have
    \begin{align*}
        \potential{j}{\convexBody}-\potential{j}{\convexBodyNext} &\overset{(a)}{=} \factorial{j} \left(\intrinsicVolume{j}{\convexBody}-\intrinsicVolume{j}{\convexBodyNext}\right)
        \\&\overset{(b)}{\geq}\factorial{j} \left(\frac{1}{j} \hight\cdot \intrinsicVolume{j-1}{\coneBase}-\intrinsicVolume{j}{\coneBase}\right)
        \\&\overset{(c)}{\geq}\factorial{j}\left(\frac{1}{\factorial{j}}\left(1-\frac{j-1}{\dimension}\right)\hight^j - \frac{1}{\factorial{j}}\left(1-\frac{j}{\dimension}\right)\hight^j\right)
        \\&\overset{(d)}{=}\frac{1}{\dimension}\left(\frac{1}{\|\context\|}\informationWidth{\context}{\reward}{\convexBody}\right)^j.
    \end{align*}
    (a) is by the definition of $\potential{j}{\convexBody}$ and $\potential{j}{\convexBodyNext}$; (b) is due to Lemma~\ref{lem:intrinsic volume properties}.(3); (c) is due to \eqref{eq:noname1}; (d) is by identical transformation and $\hight = \frac{1}{\|\context\|}\informationWidth{\context}{\reward}{\convexBody}$.
    
\end{proof}

\begin{lemma}\label{lem:potential sum information width}
    Given a context vector $\context\in \realNumbers^{\dimension}$ and a real number $\reward\in\realNumbers$, define $\convexBodyNext$ as in \Cref{lem:potential information width}, then there exist an integer $1\leq j\leq d$ such that
    \[\potentialSum{\convexBody}-\potentialSum{\convexBodyNext}\geq \frac{1}{\dimension}\left(\frac{1}{\|\context\|}\informationWidth{\context}{\reward}{\convexBody}\right)^{j}.\]
\end{lemma}
\begin{proof}
    By \Cref{lem:potential information width}, there exist $j\leq d$ such that $\potential{j}{\convexBody}-\potential{j}{\convexBodyNext}\geq \frac{1}{d}\left(\informationWidth{\context}{\reward}{\convexBody}\right)^j$. Then,
    \begin{align*}
        \potentialSum{\convexBody}-\potentialSum{\convexBodyNext}&\overset{(a)}{=} \sum_{i=1}^{\dimension} \left(\potential{i}{\convexBody}-\potential{i}{\convexBodyNext}\right)
        \\&\overset{(b)}{\geq} \potential{j}{\convexBody}-\potential{j}{\convexBodyNext}
        \\&\overset{(c)}{\geq} \frac{1}{\dimension}\left(\frac{1}{\|\context\|}\informationWidth{\context}{\reward}{\convexBody}\right)^j,
    \end{align*}
    where (a) is by definition; (b) is because that $\potential{i}{\convexBody}\geq \potential{i}{\convexBodyNext}, \forall i$; (c) is by our choice of $j$.
\end{proof}

\subsubsection{Regret Analysis of Algorithm~\ref{alg: unknown cost}}
\label{sect: cost upper bound}

Before proving \Cref{thm:cost context upper bound}, we first present some supporting lemmas. \Cref{lem:non-degenerate} specifies the update of the hypothesis set we can make by observing the feedback. It helps us to prove \Cref{lem:hypothesis valid}, which shows that our hypothesis update in Algorithm~\ref{alg: unknown cost} is valid. In \Cref{lem:information width cost}, we lower bound the decay of the potential function when round $\timeslot$ falls into the case of the low-reward outcome.

\begin{lemma}\label{lem:non-degenerate}
    In instance $\instance{\actionsT}{\hvec}$, if the principal proposes contract $\contractSimple$ and an action $\action\in \actionsT$ is taken by the agent, then 
    \begin{itemize}
        \item For $\action^{\dagger} \in \actionsT$ with $\rewardAT{\action^{\dagger}} > \rewardAT{\action}$, we have $\lrangle{\jumpVec{\action}{\action^{\dagger}},\hvec}\geq \contractSimple$.
        \item For $\action^{\dagger} \in \actionsT$ with $\rewardAT{\action^{\dagger}}<\rewardAT{\action}$, we have $\lrangle{\jumpVec{\action^{\dagger}}{\action},\hvec}\leq \contractSimple$.
    \end{itemize}
\end{lemma}
\begin{proof}
    Since $\action$ is taken by the agent,
    \[\contractSimple\cdot \rewardAT{\action} - \lrangle{\contextAT{\action},\hvec} \geq \contractSimple \cdot \rewardAT{\action^{\dagger}}-\lrangle{\contextAT{\action^{\dagger}},\hvec}, \forall \action^{\dagger} \in \actionsT.\]
    By rearranging,
    \begin{equation}\label{eq:unknown cost 1}
        \lrangle{\contextAT{\action^{\dagger}}-\contextAT{\action},\hvec}\geq \contractSimple\cdot \left(\rewardAT{\action^{\dagger}}-\rewardAT{\action}\right).
    \end{equation}
    If $\rewardAT{\action^{\dagger}}>\rewardAT{\action}$, then \eqref{eq:unknown cost 1} becomes $\lrangle{\jumpVec{\action}{\action^{\dagger}},\hvec}\geq \contractSimple$. If $\rewardAT{\action^{\dagger}}<\rewardAT{\action}$, then \eqref{eq:unknown cost 1} becomes $\lrangle{\jumpVec{\action}{\action^{\dagger}},\hvec}\leq \contractSimple$.
\end{proof}

\begin{lemma}\label{lem:hypothesis valid}
    During the running of Algorithm~\ref{alg: unknown cost}, $\hvecTrue\in \hvecsPess_\timeslot\subseteq \hvecs_{\timeslot}$ holds for all $\timeslot\in [\timeHorizon]$.
\end{lemma}
\begin{proof}
    Obviously $\hvecTrue\in \hvecsPess_\timeslot$, so we only need to show $\hvecsPess_\timeslot\subseteq \hvecs_\timeslot$ for all $\timeslot\in [\timeHorizon]$.
    
    Suppose that we already have $\hvecsPess_\timeslot \in \hvecs_{\timeslot}$. Since $\hvecsPess$ can only get shrink as $\timeslot$ grows and $\hvecs_{\timeslot}$ remains unchanged in round $\timeslot$ where $\rewardAT{\actionT}\geq \rewardAT{\actionOptT}$, it is enough for us to consider the round $\timeslot$ with $\rewardAT{\actionT}<\rewardAT{\actionOptT}$. 
    In this case, by Lemma~\ref{lem:non-degenerate}, $\lrangle{\jumpVec{\actionT}{\actionOptT},\hvecTrue}\geq \contract{\timeslot}$. 
    Thus, 
    \[\hvecsPess_{\timeslot+1} \subseteq \hvecsPess_{\timeslot} \cap \set{\hvec \condition \lrangle{\jumpVec{\actionT}{\actionOptT},\hvec}\geq \contract{\timeslot}} \overset{(a)}{\subseteq} \hvecs_{\timeslot}\cap \set{\hvec \condition \lrangle{\jumpVec{\actionT}{\actionOptT},\hvec}\geq \contract{\timeslot}}=\hvecs_{\timeslot+1},\]
    Step (a) is due to our induction assumption. Therefore, the proof concludes by an induction argument since $\hvecsPess_1 = \hvecs_1 = \vecBall_\dimension$.
\end{proof}

\begin{lemma}\label{lem:information width cost}
    In round $t$ of Algorithm~\ref{alg: unknown cost}, if $\rewardAT{\actionT}<\rewardAT{\actionOptT}$ and $\contract{\timeslot}<1$, then there exists integer $j\leq \dimension$ such that
    \[\potentialSum{\hvecs_{\timeslot}} -\potentialSum{\hvecs_{\timeslot+1}} \geq \frac{1}{\dimension}\left(\frac{{\rewardAT{\actionOptT}-\rewardAT{\actionAgent}}}{2}\gap\right)^j.\]
\end{lemma}
\begin{proof}
    Since the agent takes $\actionOptT$ in $\instance{\actionsT}{\hvecOptT}$ when the principal proposes $\contractOptT$, we have
    \begin{equation}\label{eq:noname2}
        \lrangle{\jumpVec{\actionT}{\actionOptT},\hvecOptT} \leq \contractOptT.
    \end{equation}
    On the other hand, by Lemma~\ref{lem:non-degenerate} we have
    \begin{equation}\label{eq:noname3}
        \contract{\timeslot}\leq \lrangle{\jumpVec{\actionAgent}{\actionOptT},\hvecTrue}.
    \end{equation}
    Since $\contract{\timeslot}<1$, we have $\contract{\timeslot}=\contractOptT+\gap$ by its definition. Then
    \[\lrangle{\jumpVec{\actionT}{\actionOptT},\hvecOptT}\overset{(a)}{\leq} \contractOptT< \contract{\timeslot}\overset{(b)}{\leq} \lrangle{\jumpVec{\actionAgent}{\actionOptT},\hvecTrue},\]
    where (a) is by \eqref{eq:noname2}, (b) is by \eqref{eq:noname3}.
    
    Because that $\hvecOptT,\hvecTrue\in \hvecs_{\timeslot}$ by Lemma~\ref{lem:hypothesis valid}, we have $\contract{\timeslot}\in [\min_{\hvec \in \hvecs_{\timeslot}}\lrangle{\jumpVec{\actionAgent}{\actionOptT},\hvec}, \max_{\hvec\in \hvecs_{\timeslot}}\lrangle{\jumpVec{\actionAgent}{\actionOptT}, \hvec}]$. Note that 
    \[\hvecs_{\timeslot+1} = \hvecs_{\timeslot}\cap \set{\hvec \condition \lrangle{\jumpVec{\actionAgent}{\actionOptT}, \hvec}\geq \contract{\timeslot}} = \hvecs_{\timeslot}\cap \set{\hvec \condition \lrangle{-\jumpVec{\actionAgent}{\actionOptT}, \hvec}\leq -\contract{\timeslot}}. \]
    Therefore, let $\convexBody= \hvecs_{\timeslot}$ and $\convexBodyNext=\hvecs_{\timeslot+1}$ in Lemma~\ref{lem:potential information width}, there exist integer $j\leq \dimension$ such that 
    \begin{align*}
        \potentialSum{\hvecs_{\timeslot}}-\potentialSum{\hvecs_{\timeslot+1}}&\overset{(a)}{\geq} \frac{1}{\dimension}\left(\frac{1}{\|\jumpVec{\actionAgent}{\actionOptT}\|}\informationWidth{-\jumpVec{\actionT}{\actionOptT}}{-\contract{\timeslot}}{\hvecs_{\timeslot}}\right)^j
        \\& \overset{(b)}{=} \frac{1}{\dimension}\left(\frac{1}{\|\jumpVec{\actionAgent}{\actionOptT}\|}\left(\max_{\hvec\in \hvecs_{\timeslot}}\lrangle{-\jumpVec{\actionT}{\actionOptT},\hvec} + \contractOptT + \gap\right)\right)^j
        \\&\overset{(c)}{\geq} \frac{1}{\dimension} \left(\frac{1}{\|\jumpVec{\actionAgent}{\actionOptT}\|}\gap\right)^j
        \\&\overset{(d)}{\geq} \frac{1}{\dimension} \left(\frac{\abs{\rewardAT{\actionOptT}-\rewardAT{\actionAgent}}}{2} \gap\right)^j,
    \end{align*}
    where (a) is because of Lemma~\ref{lem:potential information width}; (b) is by the definition of information width; (c) is by \eqref{eq:noname2}; (d) is due to $\jumpVec{\actionAgent}{\actionOptT} = \frac{\contextAT{\actionOptT}-\contextAT{\actionAgent}}{\rewardAT{\actionOptT}-\rewardAT{\actionAgent}}$ and $\|\contextAT{\actionOptT}-\contextAT{\actionAgent}\|\leq 2$. 
\end{proof}

Now we are ready to prove \Cref{thm:cost context upper bound}.
\begin{proof}[\bf{Proof of \Cref{thm:cost context upper bound}}]
    First, we show the immediate regret bound for round $\timeslot$ with $\contract{\timeslot}=1$. In this case, we have $\contractOptT+\gap \geq 1$ by definition. Therefore 
    \begin{align*}
        \regretPessimistic_{\timeslot}(\ALG) &= \revenuePrincipalConPessimistic_\timeslot(\hvecsPess_\timeslot) - 
        \revenuePrincipalCon_\timeslot(\contract{\timeslot})\overset{(a)}{\leq} \revenuePrincipalConPessimistic_\timeslot(\hvecs_\timeslot)=
        \left(1-\contractOptT\right) \rewardAT{\actionOptT} \leq \gap \rewardAT{\actionOptT}\leq \gap,
    \end{align*}
    where (a) holds because $\revenuePrincipalCon_\timeslot(\contract{\timeslot})\geq 0$ and $\hvecsPess_\timeslot \subseteq\hvecs_\timeslot$ by \Cref{lem:hypothesis valid}.
    
    Next, we only consider the rounds with $\contract{\timeslot} <1$. We further discuss two cases: $\rewardAT{\actionOptT}-\rewardAT{\action}\geq 2\gap$ and $\rewardAT{\actionOptT}-\rewardAT{\action}<2\gap$.

    \begin{itemize}
        \item For $\timeslot$ with $\rewardAT{\actionOptT}-\rewardAT{\action}\geq 2\gap$, by Lemma~\ref{lem:information width cost}, we have 
    \[\potentialSum{\hvecs_{\timeslot}} -\potentialSum{\hvecs_{\timeslot+1}} \geq \frac{1}{\dimension}\left(\frac{\abs{\rewardAT{\actionOptT}-\rewardAT{\actionAgent}}}{2}\gap\right)^j\geq \frac{1}{\dimension} \gap^{2j}\]
    for some $j\leq \dimension$. Since $\gap\leq \frac{1}{2}$, we have
        \[\potentialSum{\hvecs_{\timeslot}} -\potentialSum{\hvecs_{\timeslot+1}}\geq \frac{1}{\dimension}\gap^{2\dimension}.\]
        Note that $\hvecs_{\timeslot+1}\subseteq\hvecs_{\timeslot}$ holds for any $\timeslot$, $\potentialSum{\hvecs_{\timeslot}}$ does not increase as $\timeslot$ grows. Since $\potentialSum{\hvecs_1} \leq \dimension (2\pi \dimension)^{\dimension/2}$ and $\potentialSum{\hvecs_{\timeHorizon+1}}\geq 0$, there are at most $\frac{\dimension^2 (2\pi \dimension)^{\dimension/2}}{\gap^{2\dimension}}$ rounds with $\rewardAT{\actionT}<\rewardAT{\actionOptT}$. We trivially bound the immediate regret of these rounds as $1$. Then there are at most $\frac{\dimension^2(2\pi\dimension)^{\dimension/2}}{\gap^{2\dimension}}$ rounds falling into this case.
        
        \item For round $\timeslot$ with $ \rewardAT{\actionOptT}-\rewardAT{\action}<2\gap$, 
        \begin{align*}
        \regretPessimistic_\timeslot(\ALG) &=\revenuePrincipalConPessimistic_\timeslot(\hvecsPess_\timeslot) - 
        \revenuePrincipalCon_\timeslot(\contract{\timeslot})
        \\&\overset{(a)}{\leq}\revenuePrincipalConPessimistic_\timeslot(\hvecs_\timeslot) - 
        \revenuePrincipalCon_\timeslot(\contract{\timeslot})
        \\&= \left(1-\contractOptT\right)\rewardAT{\actionOptT} - \left(1-\contract{\timeslot}\right)\rewardAT{\actionT}
        \\&\overset{(b)}{\leq} \left(1-\contractOptT\right)\rewardAT{\actionOptT} - \left(1-\contractOptT-\gap\right)(\rewardAT{\actionOptT}-2\gap)
         \\&\overset{(c)}{=} \gap \rewardAT{\actionOptT} + 2\gap(1-\contractOptT-\gap)
         \\&\overset{(d)}{\leq} \gap+ 2\gap=3\gap,
    \end{align*}
    where (a) is because $\hvecsPess_\timeslot\subseteq \hvecs_\timeslot$; (b) is due to $\rewardAT{\actionAgent}\geq \rewardAT{\actionOptT}-2\gap$; (c) is by simple identical transformation; (d) is due to $\rewardAT{\actionOptT}\leq 1$ and $1-\contractOptT-\gap\leq 1$.
    \end{itemize}
    
    Overall, recall that $\gap = \min\set{(2\pi)^{\frac{\dimension}{4\dimension+2}}\dimension^{\frac{\dimension+4}{4\dimension+2}}T^{-\frac{1}{2\dimension+1}},\frac{1}{2}}$ in $\ALG$, we have
    \begin{align*}
        \regretPessimistic(\ALG) &= \sum_{\timeslot:\contract{\timeslot}=1} \regretPessimistic_\timeslot(\ALG)+\sum_{\timeslot:\rewardAT{\actionAgent}<\rewardAT{\actionOptT}-2\gap, \contract{\timeslot}<1}\regretPessimistic_\timeslot(\ALG)
        +\sum_{\timeslot:\rewardAT{\actionAgent}\geq \rewardAT{\actionOptT}-2\gap, \contract{\timeslot}<1} \regretPessimistic_\timeslot(\ALG)
        \\&\leq \timeHorizon \cdot \gap + \frac{\dimension^2(2\pi \dimension)^{\dimension/2}}{\gap^{2\dimension}}+ 3T\cdot\gap = 5 (2\pi)^{\frac{\dimension}{4\dimension+2}} \dimension^{\frac{\dimension+4}{4\dimension+2}} T^{1-\frac{1}{2\dimension+1}}. \tag*{\qed}
    \end{align*}
    \begingroup
    \renewcommand{\qedsymbol}{}
\end{proof}
\endgroup

\subsection{Lower Bound for Cost Context}\label{sec:lower-bound-cost-opt}

In this part, we now consider the \emph{pessimistic Stackelberg regret} lower bound for the contextual principal-agent game with cost contexts.
The following lower bound is established for any algorithm $\ALG$. As for the classic \emph{Stackelberg regret}, we defer the discussion in \Cref{sec:lower-cost-true}.

\begin{theorem}\label{thm:lower-bound-cost-opti}
    For contextual principal-agent game with cost contexts, let $\dimension$ be the dimension of the context, then for any algorithm $\ALG$, there exists an instance making the algorithm suffer $\Omega(T^{1-\frac{1}{d}})$ pessimistic Stackelberg regret. 

\end{theorem}

We first present our \Cref{alg: adversary_cost_optimism}, ‌analyze the important simplified instance with known actions of our construction in \Cref{sec:instance-cost}, verify the validity of adversarial construction in \Cref{sec:valid-cost-opt} and complete our lower bound analysis in \Cref{sec:regret_lower_cost_opt}. Given any algorithm $\ALG$, we construct an adversarial instance to ensure that $\ALG$ suffers high regret from the instance. The adversary makes use of the concept of \emph{spherical code} from coding theory in \Cref{def:spherical codes}.

\xhdr{Adversary.} Let $\leftGapThreshold$ be a number to be determined. For $t$-th round, we consider a four-action instance with action set $\actionsT = \set{\nullAction,\knownOptAction,\adversarialAction,\restrictingAction}$. The adversary promises that the first coefficient of the true hidden vector $\hvecTrue$ is $\frac{1}{2}$.
\begin{itemize}
    \item $\nullAction$ is the null action with zero reward and cost.
    \item $\knownOptAction$ is the action with context vector $\contextAT{\knownOptAction}=(\frac{1}{2\eulerNumber},0,\ldots,0)$ and $\rewardAT{\knownOptAction} = \frac{1}{2\eulerNumber}$. Note the context vector make its cost known to $\ALG$, which is $\costAT{\knownOptAction} = \frac{1}{4\eulerNumber}$.
    \item $\restrictingAction$ has reward $\rewardAT{\restrictingAction}=\frac{1}{2\eulerNumber}+\frac{\leftGapThreshold}{2\eulerNumber }$ with its context $\contextAT{\restrictingAction} = \left(\frac{1}{2\eulerNumber}+\frac{\leftGapThreshold}{4\eulerNumber } \cdot \frac{2+5 \leftGapThreshold}{1+ \leftGapThreshold},0,\ldots,0 \right).$
    Similarly, the adversary ensures that only the first coefficient of $\contextAT{\restrictingAction}$ is non-zero, which makes its cost known to $\ALG$.
    \item $\adversarialAction$ has reward $\rewardAT{\adversarialAction} =\frac{1}{2\eulerNumber}+\frac{\leftGapThreshold}{2\eulerNumber }$. Its context in the $\timeslot$-th round will be determined adaptively according to the behavior of $\ALG$ in the previous $\timeslot-1$ rounds. Then, only the cost of this action is unknown to $\ALG$.
\end{itemize}
The adversary's strategy at time step $\timeslot$ is determined by the contract $\contract{\timeslot}$ received from the algorithm $\ALG$. Let $$\contractThreshold \in \argmax\limits_{\contractSimple\in[0,1]}\max\limits_{\hvec\in\hvecs_\timeslot}~\revenuePrincipalCon_\timeslot(\contract{}| \hvec)$$ be the optimistic contract value given the current uncertainty set $\hvecs_{\timeslot}$. Notice that $\contractThreshold$ remains unchanged for each round by our construction of the instance. The adversary's action depends on whether $\contract{\timeslot}$ meets this threshold, which indicates if the principal is exploring.
\begin{itemize}
    \item If (No Exploration): The adversary perceives no incentive to alter the uncertainty set and thus leaves $\hvecs_{\timeslot}$ unchanged.
    \item If (Exploration): The adversary shrinks the uncertainty set $\hvecs_{\timeslot}$ by removing the spherical cap corresponding to the direction of the related codeword $\codeWord{i}$. This process is repeated in subsequent rounds for different directions. Once all codeword directions in $\sphericalCodedim{\dimension-1}$ have been exhausted, the adversary adopts a different shrinking mechanism: it directly reduces the radius by a factor of $1-\leftGapThreshold^2$.
\end{itemize}
The adversarial process described above repeats in each round. A formal specification of this constructed adversary is provided in \Cref{alg: adversary_cost_optimism}.

\begin{adversary}[t!]
\caption{\textsc{Adversary with Cost Context}}
\label[adversary]{alg: adversary_cost_optimism}
\SetAlgoLined
\KwData{$\ALG, \timeHorizon, \dimension, \leftGapThreshold$, spherical code $\sphericalCodedim{\dimension-1}$ }

Initialize $\contextAT{\knownOptAction} = (\frac{1}{2\eulerNumber},0,\ldots,0), \contextAT{\restrictingAction} = \left(\frac{1}{2\eulerNumber}+\frac{\leftGapThreshold}{4\eulerNumber } \cdot \frac{2+5 \leftGapThreshold}{1+ \leftGapThreshold},0,\ldots,0 \right), \rewardAT{\knownOptAction}=\frac{1}{2\eulerNumber}, \rewardAT{\restrictingAction}=\rewardAT{\adversarialAction}=\frac{1}{2\eulerNumber}+\frac{\leftGapThreshold}{2\eulerNumber }, \forall \timeslot\geq 1$

$\timeslot\gets 1, i\gets 1, j\gets 1$

Initialize $\radius\super{1} = \frac{\sqrt{3}}{2}$, $\hvecs_{1}=\set{\frac{1}{2}}\times \left(\radius\super{1} \cdot  \mathbb{B}_{d-1}\right)$

Arbitrarily index code words $\sphericalCodedim{\dimension-1}=\set{\codeWord{1},\codeWord{2},\ldots,\codeWord{\abs{\sphericalCodedim{\dimension-1}}}}$

 $\contractThreshold\gets \frac{2+3 \cdot \leftGapThreshold}{4(1+\leftGapThreshold)}$

\While{$\timeslot\leq \timeHorizon$}{
    Set     $\contextAT{\adversarialAction}  =  \left(\frac{1}{2\eulerNumber}+\frac{ \leftGapThreshold}{ 4\eulerNumber} \frac{2+3 \cdot  \leftGapThreshold}{1+ \leftGapThreshold}+\frac{1}{2\eulerNumber(1+ \leftGapThreshold)},   \frac{\codeWord{i}}{4\eulerNumber \left(1+\leftGapThreshold \right)\radius\super{j}} \right)$, where $\codeWord{i} \in \sphericalCodedim{\dimension-1}\subseteq \realNumbers^{\dimension-1}$.

    Show contexts to $\ALG$ and receive contract $\contract{\timeslot}$.

    \If{$\contractSimple_{\timeslot}<\contractThreshold$}{
     $\hvecs_{\timeslot+1} \gets \hvecs_{\timeslot}$

    }
    
    \If{$\contractSimple_{\timeslot}\geq\contractThreshold$}{

    $\hvecs_{\timeslot+1} \gets \hvecs_{\timeslot}\cap \set{\left(\frac{1}{2},\zeta\right) \condition \lrangle{\codeWord{i},-\zeta}\leq \radius\super{j} \left(1-\leftGapThreshold^2\right)}$

    $i \gets i+1$
    }

    \If{$i>\abs{\sphericalCodedim{\dimension-1}}$}{
    $i\gets 1$, $\radius\super{j+1} \gets \radius\super{j} \cdot \left(1-\leftGapThreshold^2\right)$

    $\hvecs_{\timeslot+1} \gets \set{\frac{1}{2}}\times  \left( \radius\super{j+1} \cdot \mathbb{B}_{d-1}\right)$

    $j\gets j+1$

    }

    $\timeslot \gets \timeslot+1$
}
Determine $\hvecTrue = \argmin_{\hvec\in\hvecs_{\timeslot}}\lrangle{\contextAT{\adversarialAction},\hvec}$
\end{adversary}

We will now show that this specific adversary forces the principal to incur a significant pessimistic Stackelberg regret. Before proceeding to the proof of \Cref{thm:lower-bound-cost-opti}, we introduce some extra notations for our analysis in the remainder of the paper.

\xhdr{Extra notations.} Let $\instance{\actions}{\theta}$ be the principal-agent game instance where the agent's action set is $\actions\subseteq\actionsT$ and the hidden vector is fixed to $\hvec$. Given a principal-agent game instance $\instanceSimple$ where the cost and reward of each action are fixed, we use $\optProfit{\instanceSimple}$ to denote the maximum principal's utility in instance $\instanceSimple$.

We also require several auxiliary lemmas. The first lemma in \Cref{sec:instance-cost} concerns a simplified instance derived from our adversarial construction, which we term the ``Instance with Known Actions''.

\subsubsection{Instance with Known Actions} \label{sec:instance-cost}
By the construction of the adversary, in each round, only the cost of $\adversarialAction$ is unknown to $\ALG$, and the action set $\actionsKnownT=\set{\nullAction,\knownOptAction,\restrictingAction}$ is totally known to $\ALG$. Moreover, $0=\rewardAT{\nullAction}<\rewardAT{\knownOptAction}<\rewardAT{\restrictingAction}$. The instance $\instanceKnown$ plays an important role in our proof. Let $0= \contractTOrder{0}\leq \contractTOrder{1} \leq \contractTOrder{2} \leq \contractTOrder{3}=1$ be the critical points of $\instanceKnown$. The following lemma provides the value of the first non-trivial critical point and a lower bound for the second. 

\begin{lemma}\label{critical-point inequ-cost}
    We have $\contractTOrder{1}=\frac{1}{2}$ and $\contractTOrder{2}\geq  \frac{1}{2}+\frac{3}{4}\leftGapThreshold-\frac{3}{4}\leftGapThreshold^2$.
\end{lemma}
\begin{proof}
    $\contractTOrder{1}$ is the solution to 
    \[\contractTOrder{1} \rewardAT{\knownOptAction}- \costAT{\knownOptAction} = 0.\]
    By simple calculation, $\contractTOrder{1}=\frac{1}{2}$. And $\contractTOrder{2}$ is the solution to 
    \[\contractTOrder{2} \rewardAT{\knownOptAction}- \costAT{\knownOptAction} = \contractTOrder{2} \rewardAT{\restrictingAction}-\costAT{\restrictingAction}.\]
    Recall that $\costAT{\knownOptAction} = \frac{1}{4\eulerNumber}$, $\costAT{\restrictingAction}=\frac{1}{4\eulerNumber}+\frac{\leftGapThreshold}{8\eulerNumber } \cdot \frac{2+5 \leftGapThreshold}{1+ \leftGapThreshold}$, $\rewardAT{\restrictingAction}=\frac{1}{2\eulerNumber}+\frac{\leftGapThreshold}{2\eulerNumber },$ we have 
    \[
        \contractTOrder{2} = \frac{2+5\leftGapThreshold}{4(1+\leftGapThreshold)} = \frac{1}{2}+\frac{3}{4}\leftGapThreshold-\frac{3}{4}\leftGapThreshold^2+O(\leftGapThreshold^3)
    \]
    by direct calculation.
\end{proof}

\begin{lemma}
\label{lem:opt-act-cost}
$\optProfitFromA{\instanceKnown}{\knownOptAction} = \frac{1}{4\eulerNumber}$, $\optProfitFromA{\instanceKnown}{\restrictingAction} = \frac{1}{4\eulerNumber}-\frac{\leftGapThreshold}{8\eulerNumber}$.
\end{lemma}
\begin{proof}
Notice that proposing $\contractTOrder{1}$ can obtain the maximal utility induced by action $\knownOptAction$. Thus, $\optProfitFromA{\instanceKnown}{\knownOptAction} =(1-\contractTOrder{1})\rewardAT{\knownOptAction}= \frac{1}{4\eulerNumber}$. And proposing $\contractTOrder{2}$ can obtain the maximal utility induced by action $\restrictingAction$. Therefore, $\optProfitFromA{\instanceKnown}{\restrictingAction} = (1-\contractTOrder{2})\rewardAT{\restrictingAction}= \frac{1}{4\eulerNumber}-\frac{\leftGapThreshold}{8\eulerNumber}$.
\end{proof}

\subsubsection{Validity of Adversary~\ref{alg: adversary_cost_optimism}}
\label{sec:valid-cost-opt}

Next, we establish two crucial properties of our adversarial construction (\Cref{alg: adversary_cost_optimism}). The following lemma demonstrates that our adversary not only successfully restricts the agent's choice to the known action set in \Cref{sec:instance-cost}, but also maintains internal consistency throughout the interaction. This latter property ensures it is a valid adversary.

\begin{lemma}\label{lem:valid adversary_cost_optimism}
    Employing the \Cref{alg: adversary_cost_optimism}, whatever the contract is, the agent will always take an action in $\actionsKnownT$. Moreover, \Cref{alg: adversary_cost_optimism} is a valid adversary, which means that $\hvecs_{\timeslot}$ is consistent with the observed feedback of $\ALG$.
\end{lemma}

\begin{proof}
Consider the $\timeslot$-th round. Let $i^*, j^*$ be the $i,j$ at the beginning of the $\timeslot$-th round.  We analyze two cases: 
\begin{itemize}
\item Case $1$: $0 \leq\contractSimple_{\timeslot} < \contractThreshold$. Regardless of the true hidden vector $\hvecTrue  \in \hvecs_{\timeslot}$, when $\contractTOrder{1} \leq\contractSimple_{\timeslot} < \contractThreshold$, the action $\knownOptAction \in \actionsKnownT$ will be induced since $$\rewardAT{\knownOptAction} \cdot \contractSimple_{\timeslot} -\lrangle{\contextAT{\knownOptAction},\hvecTrue}  \geq \max\left\{0,\rewardAT{\restrictingAction} \cdot  \contractSimple_{\timeslot} - \lrangle{\contextAT{\restrictingAction},\hvecTrue},  \rewardAT{\adversarialAction} \cdot \contractSimple_{\timeslot} - \lrangle{\contextAT{\adversarialAction},\hvecTrue}\right\}.$$
When $0 \leq\contractSimple_{\timeslot} < \contractTOrder{1}$, the agent will choose the null action $\nullAction \in \actionsKnownT$ since other actions bring negative revenue. In both subcases, the principal cannot infer new information about the true hidden vector $\hvecTrue$. Thus, $\hvecs_{\timeslot+1} = \hvecs_{\timeslot}$, which is consistent with the observed feedback from $\ALG$. 

\item Case $2$: $\contractThreshold \leq\contractSimple_{\timeslot} \leq 1$. If $i^*+1 \leq \abs{\sphericalCodedim{\dimension-1}}$, the adversary will restrict the true hidden vector to $\hvecTrue  \in \hvecs_{\timeslot}\cap \set{\left(\frac{1}{2},\zeta \right) \condition \lrangle{\codeWord{i^*},-\zeta}\leq \radius\super{j^*} \left(1-\leftGapThreshold^2\right)}$. One can verify $\costAT{\restrictingAction}  = \lrangle{\contextAT{\restrictingAction},\hvecTrue} \leq \lrangle{\contextAT{\adversarialAction},\hvecTrue} = \costAT{\adversarialAction}$ and $\rewardAT{\restrictingAction} =\rewardAT{\adversarialAction} $. Hence, action $\adversarialAction$ is fully dominated by action $\restrictingAction$. In addition, the principal can obtain the most information when offering the contract $\contractSimple_{\timeslot} = \contractTOrder{2}$. The observed feedback from $\ALG$ is in accordance with $\hvecs_{\timeslot+1} = \hvecs_{\timeslot}\cap \set{\left(\frac{1}{2},\zeta \right) \condition \lrangle{\codeWord{i^*},-\zeta}\leq \radius\super{j^*} \left(1-\leftGapThreshold^2\right)}$. If $i^*+1 > \abs{\sphericalCodedim{\dimension-1}}$, the adversary shrinks the radius by a factor of $1-\leftGapThreshold^2$ and defines $\hvecs_{\timeslot+1} = \set{\frac{1}{2}}\times \left(\radius\super{j+1} \cdot \mathbb{B}_{d-1}\right)$. Since $\radius\super{\timeslot+1}=\radius\super{\timeslot}$, it can be easily checked that
$$\set{\frac{1}{2}}\times \left(\radius\super{j+1} \cdot \mathbb{B}_{d-1} \right) \subseteq \bigcap_{1 \leq i \leq \abs{\sphericalCodedim{\dimension-1}}} \set{\left(\frac{1}{2},\zeta \right) \condition \lrangle{\codeWord{i},-\zeta}\leq \radius\super{j^*} \left(1-\leftGapThreshold^2\right)}.$$ 
Therefore, action $\adversarialAction$ remains fully dominated, and the feedback is consistent.

\end{itemize}
In each case, the principal’s observed feedback from 
$\ALG$ exactly matches the evolution of $\hvecs_{\timeslot} $ as described above, completing the proof.
\end{proof}

Our model assumption requires that all generated context vectors lie within the unit ball $\mathbb{B}_{d}$ and that their associated action costs remain non-negative. The following lemma establishes a sufficient condition on the size of the spherical code $\sphericalCodedim{\dimension-1}$ to make our adversary construction meet these foundational assumptions.

\begin{lemma}[Boundedness and Non-negativity]\label{lem:bounded context_cost_optimism}
Let $\sphericalCodedim{\dimension-1}$ be the $(\dimension-1)$-dimensional spherical code used by our adversarial construction (\Cref{alg: adversary_cost_optimism}). If $\sphericalCodedim{\dimension-1}$ satisfies the condition $\abs{\sphericalCodedim{\dimension-1}}\geq \timeHorizon \leftGapThreshold^2$, then for all rounds, the following properties hold:
\begin{enumerate}
\item[\textnormal{(i)}] All context vectors generated by the adversary are contained in the unit ball $\mathbb{B}_{d}$.
\item[\textnormal{(ii)}] The costs of all corresponding actions are non-negative.
\end{enumerate}
\end{lemma}

\begin{proof}
    By \Cref{alg: adversary_cost_optimism}, $\radius\super{j}$ will shrink at most $\frac{T}{\abs{\sphericalCodedim{\dimension-1}}}$ times. Thus, it always holds \[\radius\super{j}\geq \frac{\sqrt{3}}{2}\left(1-\leftGapThreshold^2\right)^j\geq \frac{\sqrt{3}}{2}\left(1-\leftGapThreshold^2\right)^{\frac{\timeHorizon}{\abs{\sphericalCodedim{\dimension-1}}}}\geq \frac{\sqrt{3}}{2\eulerNumber}.\]

    Recall that $\contextAT{\adversarialAction}  =  \left(\frac{1}{2\eulerNumber}+\frac{ \leftGapThreshold}{ 4\eulerNumber} \frac{2+3 \cdot  \leftGapThreshold}{1+ \leftGapThreshold}+\frac{1}{2\eulerNumber(1+ \leftGapThreshold)},   \frac{\codeWord{i}}{4\eulerNumber \left(1+\leftGapThreshold \right)\radius\super{j}} \right)$. We have
    \[
 \lim_{\leftGapThreshold \to 0} \abs{ \frac{1}{2\eulerNumber}+\frac{ \leftGapThreshold}{ 4\eulerNumber} \frac{2+3 \cdot  \leftGapThreshold}{1+ \leftGapThreshold}+\frac{1}{2\eulerNumber(1+ \leftGapThreshold)}} = \frac{1}{\eulerNumber} < \frac{1}{2},
    \]
and 
    \begin{align*}
   \lim_{\leftGapThreshold \to 0} \abs{ \frac{\codeWord{i}}{4\eulerNumber \left(1+\leftGapThreshold \right)\radius\super{j} }}
    =  \frac{\abs{\codeWord{i}}}{4\eulerNumber} \cdot \frac{1}{\radius\super{j}} \leq   \frac{1}{4 \eulerNumber } \cdot \frac{1}{\frac{\sqrt{3}}{2\eulerNumber}} = \frac{\sqrt{3}}{6}. 
    \end{align*}
Thus, $\lim_{\leftGapThreshold \to 0} \abs{\contextAT{\adversarialAction}} \leq \sqrt{\left(\frac{1}{2}\right)^2+ \left(\frac{\sqrt{3}}{6}\right)^2} = \frac{\sqrt{3}}{3} <1$. When $\leftGapThreshold$ is small enough, there holds that $\abs{\contextAT{\adversarialAction}} \leq  1 $.
 In addition, by direct calculation, we obtain
    $$\costAT{\adversarialAction} =  \lrangle{\contextAT{\adversarialAction},\hvecTrue}\geq \min_{\hvec\in \hvecs_{\timeslot}} \lrangle{\contextAT{\adversarialAction},\hvec} \geq \frac{1}{4\eulerNumber} + \frac{\leftGapThreshold}{ 8\eulerNumber} \frac{2+3 \cdot \leftGapThreshold}{1+\leftGapThreshold}  >  0 $$
which completes the proof of \Cref{lem:bounded context_cost_optimism}.
\end{proof}

\subsubsection{Lower Bound Analysis}\label{sec:regret_lower_cost_opt}
We now proceed to analyze the regret employing \Cref{alg: adversary_cost_optimism}. We first determine the exact value of the principal's optimistic utility. The following lemma shows that, under a specific geometric condition on the spherical code used by our adversary, this utility remains constant in every round.

\begin{lemma}\label{lem:optimism_utility_cost}
    Suppose the adversary (\Cref{alg: adversary_cost_optimism}) utilizes a spherical code with a minimal angle of 
 $\arccos\left(1- \leftGapThreshold^2 \right)$. Then, for any round $\timeslot$, the principal's optimistic utility, defined as the best achievable utility over the current hypothesis set $\hvecs_{\timeslot}$, is given by: $$\max_{\hvec\in\hvecs_{\timeslot}} \optProfit{\instance{\actions_{\timeslot}}{\hvec}} = \frac{1}{4\eulerNumber}+\frac{\leftGapThreshold}{8\eulerNumber}.$$
\end{lemma}
\begin{proof}
    Let $i^*, j^*$ be the $i,j$ at the beginning of the $\timeslot$-th round. We first show that $$\left(\frac{1}{2},-\radius\super{j^*} \cdot \codeWord{i^*}\right)\in \hvecs_{\timeslot}.$$
    By the design of the adversary, 
    \[ \hvecs_{\timeslot} = \set{\frac{1}{2}}\times \left( \left(\radius\super{j^*} \cdot \mathbb{B}_{d-1} \right)\bigcap \left(\bigcap_{i=1}^{i^*-1} \set{\zeta \condition \lrangle{-\zeta,  \codeWord{i}}\leq \radius\super{j^*}(1-\leftGapThreshold^2)}\right)\right).\]
    Notice that $\codeWord{i^*}\in \mathbb{B}_{\dimension-1}$, we have $\left(\frac{1}{2},-\radius\super{j^*} \cdot \codeWord{i^*}\right)\in \set{\frac{1}{2}}\times \left( \radius\super{j^*}\cdot \mathbb{B}_{d-1}\right)$. Since the minimal angle of $\sphericalCodedim{\dimension-1}$ is $\arccos (1-\leftGapThreshold^2)$, we have 
    \[ \lrangle{-(-\radius\super{j^*} \cdot \codeWord{i^*}),  \codeWord{i}}\leq \radius\super{j^*}(1-\leftGapThreshold^2) .\]
    Thus, $\left(\frac{1}{2},-\radius\super{j^*} \cdot \codeWord{i^*}\right)\in \hvecs_{\timeslot}$.
    
     Therefore, $\min_{\hvec\in\hvecs_{\timeslot}} \lrangle{\contextTA{\timeslot}{\adversarialAction}, \hvec} \leq \lrangle{\contextTA{\timeslot}{\adversarialAction}, \left(\frac{1}{2},-\radius\super{j^*} \cdot \codeWord{i^*}\right)}= \frac{1}{4\eulerNumber}+\frac{\leftGapThreshold}{8\eulerNumber} \cdot \frac{2+3\leftGapThreshold}{1+\leftGapThreshold}$. On the other hand, since $\hvecs_{\timeslot} \subseteq \set{\frac{1}{2}}\times \left( \radius\super{j^*}\cdot \mathbb{B}_{d-1}\right)$, we have $\min_{\hvec\in\hvecs_{\timeslot}} \lrangle{\contextTA{\timeslot}{\adversarialAction},\hvec}\geq \frac{1}{4\eulerNumber}+\frac{\leftGapThreshold}{8\eulerNumber} \cdot \frac{2+3\leftGapThreshold}{1+\leftGapThreshold}$. Thus, $\min_{\hvec\in\hvecs_{\timeslot}} \lrangle{\contextTA{\timeslot}{\adversarialAction},\hvec} =\frac{1}{4\eulerNumber}+\frac{\leftGapThreshold}{8\eulerNumber} \cdot \frac{2+3\leftGapThreshold}{1+\leftGapThreshold}$. Given $\hvec \in \hvecs_{\timeslot}$, we consider two cases. 
     \begin{itemize}
     \item Case $1$: if $\frac{1}{4\eulerNumber}+\frac{\leftGapThreshold}{8\eulerNumber} \cdot \frac{2+5\leftGapThreshold}{1+\leftGapThreshold} \leq \costA{\adversarialAction}_{\timeslot}=\lrangle{\contextTA{\timeslot}{\adversarialAction},\hvec} $, action $\adversarialAction$ is fully dominated by $\restrictingAction$. Therefore, $\optProfit{\instance{\actions_{\timeslot}}{\hvec}} = \optProfit{\instanceKnown} = \frac{1}{4\eulerNumber}$.
     \item Case $2$: if $\frac{1}{4\eulerNumber}+\frac{\leftGapThreshold}{8\eulerNumber} \cdot \frac{2+3\leftGapThreshold}{1+\leftGapThreshold} \leq \costA{\adversarialAction}_{\timeslot}=\lrangle{\contextTA{\timeslot}{\adversarialAction},\hvec} < \frac{1}{4\eulerNumber}+\frac{\leftGapThreshold}{8\eulerNumber} \cdot \frac{2+5\leftGapThreshold}{1+\leftGapThreshold}$, by simple calculation, we obtain 
     \begin{align*}
     \optProfit{\instance{\actions_{\timeslot}}{\hvec}} = \max \left\{\frac{1}{4\eulerNumber}, \left(1- \frac{\costA{\adversarialAction}_{\timeslot} - \frac{1}{4\eulerNumber}}{\frac{\leftGapThreshold}{2\eulerNumber}} \right) \cdot \rewardA{\adversarialAction}_{\timeslot} \right\} & \leq  \left(1- \frac{\frac{1}{4\eulerNumber}+\frac{\leftGapThreshold}{8\eulerNumber} \cdot \frac{2+3\leftGapThreshold}{1+\leftGapThreshold} - \frac{1}{4\eulerNumber}}{\frac{\leftGapThreshold}{2\eulerNumber}} \right) \cdot \rewardA{\adversarialAction}_{\timeslot} \\
     & = \frac{1}{4\eulerNumber}+\frac{\leftGapThreshold}{8\eulerNumber},
     \end{align*}
     where the first equality holds if $\costA{\adversarialAction}_{\timeslot}=\lrangle{\contextTA{\timeslot}{\adversarialAction},\hvec} = \frac{1}{4\eulerNumber}+\frac{\leftGapThreshold}{8\eulerNumber} \cdot \frac{2+3\leftGapThreshold}{1+\leftGapThreshold}$. \end{itemize}

      Thus, the best achievable utility
given current knowledge $\max_{\hvec\in\hvecs_{\timeslot}} \optProfit{\instance{\actions_{\timeslot}}{\hvec}}=\frac{1}{4\eulerNumber}+\frac{\leftGapThreshold}{8\eulerNumber}$. \qedhere
\end{proof}

Next, we demonstrate that our adversarial construction forces the principal to incur a substantial amount of regret in every single round. The following lemma formalizes this by establishing a constant lower bound on the instantaneous pessimistic Stackelberg regret.

\begin{lemma}\label{lem:optimism_regret_cost}
     In any round $\timeslot$, regardless of the contract $\contract{\timeslot}$ proposed by the principal $\ALG$, the resulting instantaneous pessimistic Stackelberg regret satisfies
     \[\regretPessimistic_\timeslot(\ALG) \geq \frac{\leftGapThreshold}{8\eulerNumber}.\]
\end{lemma}
\begin{proof}
By \Cref{lem:valid adversary_cost_optimism}, whatever  the contract $\contract{\timeslot}$ proposed by $\ALG$ is, the agent will always take an action in $\actionsKnownT$. Then combining with \Cref{lem:opt-act-cost} and \Cref{lem:optimism_utility_cost}, we obtain

    \begin{align*}
        \regretPessimistic_\timeslot(\ALG) &=\max_{\hvec\in\hvecsPess_{\timeslot}} \optProfit{\instance{\actions_{\timeslot}}{\hvec}} -\revenuePrincipalCon_\timeslot(\contract{\timeslot})  \\ &\overset{(a)}{\geq} \max_{\hvec\in\hvecs_{\timeslot}} \optProfit{\instance{\actions_{\timeslot}}{\hvec}} -\revenuePrincipalCon_\timeslot(\contract{\timeslot}) \\
        & \geq \max_{\hvec\in\hvecs_{\timeslot}} \optProfit{\instance{\actions_{\timeslot}}{\hvec}}  -\optProfit{\instanceKnown}
        \\&= \frac{1}{4\eulerNumber}+\frac{\leftGapThreshold}{8\eulerNumber}- \frac{1}{4\eulerNumber} = \frac{\leftGapThreshold}{8\eulerNumber}. 
    \end{align*}
Here, the inequality (a) is because $\hvecs_{\timeslot}$ is consistent with the observed feedback by \Cref{lem:valid adversary_cost_optimism}, leading to $\hvecs_\timeslot\subseteq \hvecsPess_\timeslot$.
\end{proof}

We are now ready to prove \Cref{thm:lower-bound-cost-opti}. The proof synthesizes the preceding lemmas to establish the final lower bound on cumulative pessimistic Stackelberg regret.

\begin{proof}[\bf{Proof of \Cref{thm:lower-bound-cost-opti}}]
The proof proceeds in two main steps. First, we verify that the conditions for our technical lemmas are met. Second, we sum the per-round regret to obtain the total regret.

First, we set $\leftGapThreshold = \min\{\frac{\sqrt{2}}{2}C^{\frac{1}{\dimension}} (\dimension-1)^{\frac{1}{2\dimension}} \timeHorizon^{-\frac{1}{\dimension}},\frac{\sqrt{2}}{2}\}$. By \Cref{lem:code lower bound}, there exists a spherical code with a minimal angle of 
 $\arccos\left(1- \leftGapThreshold^2 \right)$, whose size is lower-bounded as follows:
    $$\abs{\sphericalCodedim{\dimension-1}} \geq   C \frac{\left(1-\leftGapThreshold^2\right)\sqrt{d-1}}{\sin^{d-2} \left(\arccos{\left(1-\leftGapThreshold^2\right)}\right) }   \geq C\frac{(1-\leftGapThreshold^2)\sqrt{\dimension-1}}{(\sqrt{2}\leftGapThreshold)^{d-2}} \geq \timeHorizon \leftGapThreshold^2.$$

    Therefore, the condition of \Cref{lem:bounded context_cost_optimism}  is satisfied. This ensures that our adversarial construction is well-defined (i.e., contexts are bounded and costs are non-negative). Furthermore, by \Cref{lem:valid adversary_cost_optimism}, the adversary is guaranteed to be valid.
With the validity of our construction established, we can now compute the total regret. According to \Cref{lem:optimism_regret_cost}, the principal incurs an instantaneous regret of at least $\frac{\leftGapThreshold}{8\eulerNumber}$ in every round. Summing this per-round regret over the rounds yields the cumulative regret bound:
 \[ \regretPessimistic(\ALG)\geq \frac{\leftGapThreshold}{8\eulerNumber} \cdot \timeHorizon.\]
 Thus, we complete the proof.
 \end{proof} 
\section{Principal-Agent Games with Reward Context}
\label{sec:unknown reward}

In this section, we consider the setting where the \emph{reward is unknown} to the principal. 
For agent in each round, both the action set $\actionsT$ and the cost profile $\{\costAT{\action}\}_{\action \in \actionsT}$ are observed by the principal. However, the reward profile $\{\costAT{\action}\}_{\action \in \actionsT}$ is not directly observable. Instead, the principal observes a \emph{reward context profile} $\{\contextAT{\action}\}_{\action \in \actionsT}$, where $\contextAT{\action} \in \vecBall_\dimension$ is a $\dimension$-dimensional vector lying in the unit ball. For each action $\action \in \actionsT$, the corresponding reward satisfies the following linear structure:
\begin{align}
\label{eq:linear reward structure}
\tag{linear reward structure}
\rewardAT{\action} = \innerproduct{\contextAT{\action}, \hvecTrue},
\end{align}
where $\hvecTrue \in \vecBall_{\dimension}$ is the \emph{reward hidden vector}, an unknown vector in the $\dimension$-dimensional unit ball.

Similar to the cost-context setting, we assume the agent's actions and related data are chosen by an \emph{adaptive} adversary. We assume that the linear cost structure ensures $\rewardAT{\action} = \innerproduct{\rewardAT{\action}, \hvecTrue} \geq 0$ for all actions $\action \in \actionsT$, with $\rewardAT{0}=\costAT{0} = 0$ for the trivial action $0$. 

\xhdr{Overload notations in the cost-context setting.} Now we overload the notations used in the cost-context setting to make it fit the reward-context setting. The agent's best response action is
\begin{align*}
    \actionT\in \argmax\limits_{\action\in \actionsT}~ \contract{\timeslot}\cdot \lrangle{\contextAT{\action},\hvecTrue}-\costAT{\action}.
\end{align*}
The principal receives utility $\revenuePrincipalCon_\timeslot(\contract{\timeslot})=(1-\contract{\timeslot})\cdot \rewardAT{\actionT}$, and the agent receives utility $\utiAgent_t(\contract{\timeslot})=\contract{\timeslot}\cdot\lrangle{\contextAT{\actionAgent},\hvecTrue}-\costAT{\actionAgent}$. The principal then observes the chosen best-responding action $\actionT$ and its reward $\rewardAT{\actionAgent}$. With the overloaded definition of $\revenuePrincipalCon_\timeslot(\contract{\timeslot})$, the definition and notations of two regret notions remain the same in its form with \Cref{def:stackelberg regret} and \Cref{def:pessimistic stackelberg regret}. We also need some extra notations in this section.

\xhdr{Extra notations.} We recall and define some extra notations in this section. Let $\instance{\actions}{\theta}$ be the principal-agent game instance where the agent's action set is $\actions\subseteq\actionsT$ and the hidden vector is fixed to $\hvec$. 
Given a principal-agent game instance $\instanceSimple$ where the cost and reward of each action are fixed, we use $\contract{}(\instanceSimple)$ to denote the optimal contract in this instance. 
Given a principal-agent game instance $\instanceSimple$ where the cost and reward of each action are fixed, we use $\optProfit{\instanceSimple}$ to denote the maximum principal's utility in instance $\instanceSimple$. 
and use $\optProfitFromA{\instanceSimple}{\action}$ to denote the maximum principal's utility achieved by incentivizing action $\action$. 
we use $\contract{}(\instanceSimple)$ to denote the optimal contract in this instance. We 
use $\optProfit{\instanceSimple}$ to denote the maximum principal's utility in instance $\instanceSimple$ and 
use $\optProfitFromA{\instanceSimple}{\action}$ to denote the maximum principal's utility achieved by incentivizing action $\action$.

We make the following extra assumption in this section.
\begin{assumption}\label{assump:costLB}
    The cost of all non-trivial actions is lower bound by a constant. Namely, there exists constant $\costLB>0$ such that $\costAT{\action}\geq\costLB, \forall \action\in \actions\backslash \set{0}, \timeslot\leq \timeHorizon$. 
\end{assumption}

\subsection{Upper Bound for Reward Context}

For the contextual principal-agent game with reward context, we design Algorithm~\ref{alg: single stage} achieving the following regret upper bound guarantee.

\begin{theorem}\label{thm:reward context upper bound}
    Let $\ALG$ be Algorithm~\ref{alg: single stage} with $\leftGapThreshold = \frac{\sqrt{3\costLB} \dimension^{\frac{1}{4}+\frac{1}{\dimension}}}{\timeHorizon^{\frac{1}{2\dimension}}}$. We have the following pessimistic Stackelberg regret bound,
    \[\regretPessimistic(\ALG)\leq \frac{\sqrt{6\pi}\dimension^{\frac{1}{4}+\frac{1}{\dimension}}}{\sqrt{\costLB}} \timeHorizon^{1-\frac{1}{2\dimension}}+\dimension.\]
    Since the classic Stackelberg regret is always no larger than the pessimistic Stackelberg regret, our regret upper bound also holds for the classic Stackelberg regret.
\end{theorem}

\newcommand{\representAction}{\actionT^{-}}
\newcommand{\knownSpace}{\mathcal{S}}

We first introduce the algorithm we designed in the subsequent paragraphs. Then we analyze the regret of the proposed algorithm in \Cref{sect:reward regret analysis}.

Our algorithm maintains a convex hypothesis set $\hvecs_{\timeslot}$ of $\hvecTrue$. Each round, our algorithm updates $\hvecs_{\timeslot}$ based on the feedback, ensuring that $\hvecTrue\in \hvecsPess_\timeslot\subseteq \hvecs_{\timeslot}$ always hold. Each round, we can shrink the hypothesis set to its intersection with a hyperplane defined by $\contextAT{\actionAgent}$ and $\rewardAT{\actionAgent}$, i.e., $\hvecs_{\timeslot+1}\deq \hvecs_{\timeslot}\cap \set{\hvec \condition \lrangle{\contextAT{\actionAgent}, \hvec}=\rewardAT{\actionAgent}}$. Unless $\hvecs_{\timeslot}$ is already contained in the hyperplane, we can reduce the dimension of the hypothesis set to a strictly lower number. Thus, if we successfully incentivize the agent to take actions that have a context vector that is independent of all the actions taken by the agent before, then we only need $\dimension$ such rounds to determine $\hvecTrue$. The cumulative regret incurred in such rounds is at most $\dimension$. However, this may not always happen since the principal can not force the agent to take independent action. The idea of our algorithm is to incentivize the agent to take an independent action while controlling the immediate regret when the agent does not select an independent action.

To realize the idea, our algorithm maintains a set $\knownSpace_{\timeslot}$ which contains all the context vectors for which actions have been taken by the agent before. That is, $\knownSpace_{\timeslot}\deq \set{\context_{\tpre}\super{\action_{\tpre}}}_{\tpre\in [\timeslot]}$. Each round, we partition $\actionsT$ into $\actionsT = \actionsKnownT\cup\actionsUnknownT$. $\actionsKnownT$ is the set of all actions whose context vector is contained in $\mathrm{Span}(\knownSpace_{\timeslot})$, and $\actionsUnknownT \deq \actionsT\backslash \actionsKnownT$. By the definition of $\actionsKnownT$, the reward of any action $\action \in \actionsKnownT$ is fixed, and we can select any $\hvec \in \hvecs_{\timeslot}$ to calculate its reward $\rewardAT{\action}$ by $\lrangle{\contextAT{\action}, \hvec}$. For simplicity, we assume every action in $\actionsKnownT$ is non-degenerated. Otherwise, we just need to remove all degenerated actions and denote the remain actions as $\actionsKnownT$.

Let $\instanceKnown$ be the principal-agent game instance equipped with action set $\actionsKnownT$, notice this instance is independent from $\hvec\in \hvecs_{\timeslot}$. To maximize the knowledge gained when the agent takes action $\actionAgent\in\actionsKnownT$ while controlling the immediate regret this round, we need to select an appropriate target action $\actionOptT \in \actionsUnknownT$ to incentivize. We find it in the following way: for each $\action\in \actionsUnknownT$, let $\hvecAOptT{\action}\in \hvecs_{\timeslot}$ be the hidden vector in $\hvecs_{\timeslot}$ which maximize the reward of $\action$. Namely, $\hvecAOptT{\action} \in \argmax_{\hvec\in \hvecs_{\timeslot}}\lrangle{\contextAT{\action},\hvec}$. Consider the instance with action set $\actionsKnownT \cup \set{\action}$ and hidden vector $\hvecAOptT{\action}$. This instance is determined since the rewards of all actions are fixed. We denote this instance as $\instance{\actionsKnownT \cup \set{\action}}{\hvecAOptT{\action}}$ and calculate the maximal profit $\optProfit{\instance{\actionsKnownT \cup \set{\action}}{\hvecAOptT{\action}}}$ of the instance. We select the $\action\in \actionsT$ which maximize $\optProfitFromA{\instance{\actionsKnownT \cup \set{\action}}{\hvecAOptT{\action}}}{\action}$ as our target action $\actionOptT$, i.e., $\actionOptT \deq \argmax_{\action\in\actionsT}\optProfitFromA{\instance{\actionsKnownT \cup \set{\action}}{\hvecAOptT{\action}}}{\action}$. 
If $\actionOptT\in \actionsKnownT$, we will show there is no need to incentivize an independent action. Instead, we will suffer $0$ immediate regret by exploiting the optimal principal's utility while only considering actions in $\actionsKnownT$. 
Only if $\actionOptT\in \actionsUnknownT$, we set $\actionOptT$ as our target action to incentivize.
Denote $\hvecOptT\deq\hvecAOptT{\actionOptT}$ and $\instanceSimple_{\timeslot}\deq \instance{\actionsKnownT\cup \set{\actionOptT}}{\hvecOptT}$.

Next, we assume that $\actionOptT\in \actionsUnknownT$, which is also the more complex situation. We further divide this situation into two cases according to the \emph{left cost gap} defined as follows.

\begin{definition}[Left cost gap]
    Given an action set $\actions$, we define the \emph{left cost gap} between an action $\action^{\dagger}$ and action set $\actions$ as
    \begin{equation*}
        \leftCostGap{\action^{\dagger}}{\actions} \deq \min_{\action\in \actions, \costA{\action}<\costA{\action^{\dagger}}} \left(\costA{\action^{\dagger}}-\costA{\action} \right).
    \end{equation*}
\end{definition}

In each round, we distinguish two cases by the value of $\leftCostGap{\actionOptT}{\actionsKnownT}$. Let $\leftGapThreshold$ be a tunable parameter. Case I is when $\leftCostGap{\actionOptT}{\actionsKnownT}\geq \leftGapThreshold$, Case II is when $\leftCostGap{\actionOptT}{\actionsKnownT}<\leftGapThreshold$. 

If round $\timeslot$ falls into case I, our algorithm first calculates the minimal reward of $\action$ to make it non-degenerated. We denote this reward $\nonDeReward{\actionOptT}$, and the rigorous definition is as follows.
\begin{definition}[Non-degenerating reward]\label{def:non degenerated reward}
    Let $\contracts_{\instanceSimple,\action}$ be the set of all linear contracts that implement action $\action$ in instance $\instanceSimple$, i.e., 
    \begin{equation*}
        \contracts_{\instanceSimple,\action}= \set{\contractSimple\in [0,1]\condition \contractSimple\cdot\rewardA{\action}-\costA{\action}\geq \contractSimple\cdot\rewardA{\action^{\dagger}}-\costA{\action^{\dagger}}, \forall \action^{\dagger}\in \actions}.
    \end{equation*}
    Given $\actionsKnownT$ and action $\actionOptT\in \actionsUnknownT$, we define the \emph{non-degenerating reward} $\nonDeReward{\actionOptT}$ as 
    \begin{equation*}
        \nonDeReward{\actionOptT} \deq \inf_{\reward\in [0,1]} \set{\reward\condition \contracts_{\instance{\actionsKnown_{\timeslot}\cup \set{\actionOptT}}{\reward},\actionOptT}\neq \emptyset}.
    \end{equation*}
    Here, since only the reward of $\actionOptT$ is unknown in action set $\actionsKnownT \cup \set{\actionOptT}$, we slightly abuse the notation $\instance{\actionsKnownT\cup \set{\actionOptT}}{\reward}$ to represent the instance where $\rewardAT{\actionOptT}$ is fixed to $\reward$.
\end{definition}

\begin{figure}[t]
    \centering
    \begin{tikzpicture}[xscale=\xScale,yscale=\yScale]
    \draw[draw=none, use as bounding box](\clipX,\clipY) rectangle (9,3.75);
    \clip (\clipX,\clipY) rectangle (\clipRectX,\clipRectY);

    \draw[->,thick] (-0.5,0) -- (8.5, 0) node[right] {\scriptsize $\contractSimple$}; \draw[->,thick] (0,-7) -- (0, 4.5); \draw[->,thick] (8,-7) -- (8, 4.5) node[right] {};

\draw[-,thick] (0.1,-\cFive) -- (-0.1,-\cFive) node[left] {};

\draw[-,thick, blue] (0,-\cFour) -- (8, \RFour-\cFour); \draw[-,thick, red] (0,-\cFive) -- (8, \RFive-\cFive); 

\draw[-,thick,blue] (8.1,\RFour-\cFour) -- (7.9,\RFour-\cFour) node[anchor=west, xshift=-6pt, yshift=4pt] {~~~\scriptsize$\max_{\hvec\in \hvecs_{\timeslot}}{\lrangle{\contextAT{\actionOptT},\hvec}-\costA{\actionOptT}}$}; \draw[-,thick,red] (8.1,\RFive-\cFive) -- (7.9,\RFive-\cFive) node[anchor=west, xshift=-6pt, yshift=-3pt] {~~~\scriptsize $\nonDeReward{\actionOptT}-\costAT{\actionOptT}$}; 

    \draw[-,thick] (0*8,0.1) -- (0*8,-0.1) node[below] {\scriptsize~~~~~~~$\contractTOrder{0}$}; \draw[-,thick] (\alphaOne*8,0.1) -- (\alphaOne*8,-0.1) node[below] {\scriptsize~~~~~~~$\contractTOrder{1}$}; \draw[-,thick] (\alphaTwo*8,0.1) -- (\alphaTwo*8,-0.1) node[below] {\scriptsize~~~~~~~$\contractTOrder{2}$}; \draw[-,thick] (\alphaThree*8,0.1) -- (\alphaThree*8,-0.1) node[below] {\scriptsize${\color{red} \nonDeContract{\actionOptT}}=\contractTOrder{3}$}; \draw[-,thick] (1*8,0.1) -- (1*8,-0.1) node[below] {\scriptsize~~~~~~~~~~~~$\contractTOrder{4}=1$}; 

    \draw[ultra thick] (0*8,0) -- (\alphaOne*8,0);
    \draw[ultra thick] (\alphaOne*8,0) -- (\alphaTwo*8,\intersectionOneTwo); \draw[ultra thick] (\alphaTwo*8,\intersectionOneTwo) -- (\alphaThree*8,\intersectionTwoThree); \draw[ultra thick] (\alphaThree*8,\intersectionTwoThree) -- (1*8,\RThree-\cThree); 

    \draw[thick,black,dashed] (\alphaOne*\xxScale,\clipRectY-.9) -- (\alphaOne*\xxScale,\clipY); \node at (0.5*\alphaOne*\xxScale,0.85*\clipY) {\scriptsize $\actionTOrder{0}$}; \draw[thick,black,dashed] (\alphaTwo*\xxScale,\clipRectY-.9) -- (\alphaTwo*\xxScale,\clipY); \node at ({0.5*(\alphaOne+\alphaTwo)*\xxScale},{0.85*\clipY}) {\scriptsize  $\actionTOrder{1}$}; \draw[thick,black,dashed] (\alphaThree*\xxScale,\clipRectY-.9) -- (\alphaThree*\xxScale,\clipY); \node at ({0.5*(\alphaTwo+\alphaThree)*\xxScale},{0.85*\clipY}) {\scriptsize  $\actionTOrder{2}$}; \node at ({0.5*(\alphaThree+1)*\xxScale},{0.85*\clipY}) {\scriptsize  $\actionTOrder{3}$}; \draw[thick,black,dashed] (1*\xxScale,\clipRectY-.9) -- (1*\xxScale,\clipY); \end{tikzpicture} 	\caption{Agent's utility graph in $\instance{\actionsKnownT\cup \set{\actionOptT}}{\reward}$. The black segment represents actions in $\actionsKnownT$. The red and blue segments represent $\actionOptT$ with different rewards. The blue segment represents $\actionOptT$ with the largest possible reward over the hypothesis set $\hvecs_{\timeslot}$. The red segment represents $\actionOptT$ with the non-degenerating reward $\nonDeReward{\actionOptT}$, the non-degenerating contract $\nonDeContract{\actionOptT}$ is also marked in the graph. }\label{fig:upper-envelope-c}
\end{figure}
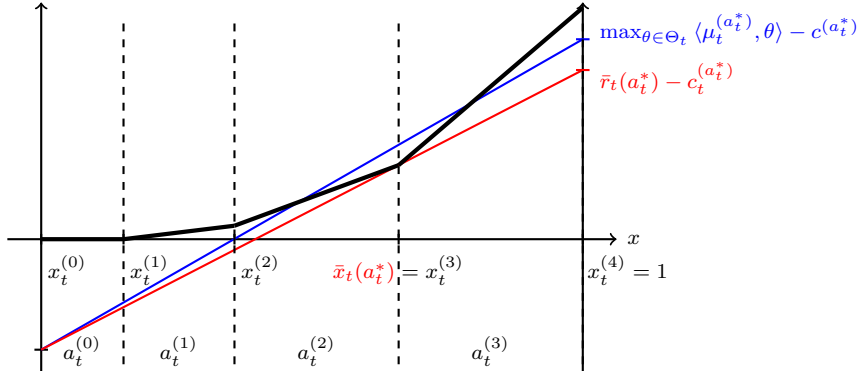

Our algorithm proposes a linear contract $\contract{\timeslot} = \nonDeContract{\actionOptT}$, where $\nonDeContract{\actionOptT}$ is the min-pay contract which incentivizes the agent to take action $\actionOptT$ in $\instance{\actionsKnownT\cup \set{\actionOptT}}{\nonDeReward{\actionOptT}}$. The rigorous definition of $\nonDeContract{\actionOptT}$ is as follows. We also provide a graph in \Cref{fig:upper-envelope-c} to help the reader understand the meaning of $\nonDeReward{\actionOptT}$ and $\nonDeContract{\actionOptT}$.

\begin{definition}[Non-degenerating contract]\label{def:non degenerated contract}
    Given $\actionsKnownT$ and action $\actionOptT\in \actionsUnknownT$, we define the \emph{non-degenerating contract} $\nonDeContract{\actionOptT}$ be the left-most contract in $\contracts_{\instance{\actionsKnown_{\timeslot}\cup \set{\actionOptT}}{\nonDeReward{\actionOptT}},\actionOptT}$. That is,
    \begin{equation*}
        \nonDeContract{\actionOptT} \deq \min_{\contractSimple\in \contracts_{\instance{\actionsKnown_{\timeslot}\cup \set{\actionOptT}}{\nonDeReward{\actionOptT}},\actionOptT}} \contractSimple.
    \end{equation*}
\end{definition}
\begin{remark}
    Note that we use $\min$ rather than $\inf$ in
    \Cref{def:non degenerated contract}. This is because that $\contracts_{\instance{\actionsKnown_{\timeslot}\cup \set{\action}}{\nonDeReward{\actionOptT}},\actionOptT}$ is a closed set by its definition. Moreover, we point out in \Cref{lem:non empty} that $\contracts_{\instance{\actionsKnown_{\timeslot}\cup \set{\actionOptT}}{\nonDeReward{\actionOptT}},\actionOptT}$ is a non-empty set. Thus, the non-degenerating contract is well-defined.
\end{remark}

The high-level intuition of proposing $\nonDeContract{\actionOptT}$ is to ensure that the hypothesis set can be shrunk as much as possible if an action in $\actionsKnownT$ is taken by the agent. Concretely, if $\actionAgent\in \actionsUnknownT$, we update the hypothesis set by $\hvecs_{\timeslot+1} = \hvecs_{\timeslot}\cap \set{\hvec \condition \lrangle{\hvec,\contextAT{\actionAgent}}=\rewardAT{\actionAgent}}$. If $\actionAgent\in \actionsKnownT$, the true reward of $\actionOptT$ is less than $\nonDeReward{\actionOptT}$. Thus, we update the hypothesis as $\hvecs_{\timeslot+1}\gets  \hvecs_{\timeslot} \cap \set{\hvec\condition \lrangle{\hvec,\contextAT{\actionOptT}}\leq \nonDeReward{\actionOptT}}$. We will show that the immediate regret when $\actionAgent\in\actionsKnownT$ is upper bounded by $O\left(\frac{1}{\leftCostGap{\actionOptT}{\actionsKnownT}}\left(\lrangle{\contextAT{\actionOptT},\hvecOptT}-\nonDeReward{\actionOptT}\right)\right)\leq O\left(\frac{1}{\leftGapThreshold}\left(\lrangle{\contextAT{\actionOptT},\hvecOptT}-\nonDeReward{\actionOptT}\right)\right)$ in Lemma~\ref{lem:contract distance case i}.

However, for rounds $\timeslot$ that fall into case II, if we still use the strategy for case I, the immediate regret when $\actionAgent\in \actionsKnownT$ can be very large compared with the knowledge about $\hvecTrue$ we obtained since $\leftCostGap{\actionOptT}{\actionsKnownT}$ can be very close to $0$. Fortunately, we show that we can find an action $\representAction\in \actionsKnownT$ to represent $\actionOptT$. That is, by proposing $\contract{\timeslot}=\optContract{\instanceKnown,\representAction}$, we can control the immediate regret suffered and make it comparable to the knowledge about $\hvecTrue$ we learned.
The $\representAction$ is the action satisfying the following two conditions:
\begin{enumerate}
    \item[(1)] Cost of $\representAction$ is in $[\costAT{\actionOptT}-\leftGapThreshold, \costAT{\actionOptT}]$;
    \item[(2)] Action $\representAction$ has the lowest cost among all $\actionsKnownT$ actions that satisfy condition (1).
\end{enumerate}
Our algorithm then propose $\contract{\timeslot}=\optContract{\instanceKnown,\representAction}$. Recall that notation $\optContract{\instanceKnown,\representAction}$ represents the min-pay contract
of $\representAction$ in $\instanceKnown$. If $\actionAgent\in \actionsUnknownT$, we update the hypothesis set $\hvecs_{\timeslot+1} = \hvecs_{\timeslot}\cap \set{\hvec \condition \lrangle{\hvec,\contextAT{\actionAgent}}=\rewardAT{\actionAgent}}$. If $\actionAgent\in \actionsKnownT$, we update $\hvecs_{\timeslot+1}= \hvecs_{\timeslot} \cap \set{\hvec\condition \lrangle{\hvec,\contextAT{\actionOptT}}\leq \frac{\costAT{\actionOptT}-\costAT{\representAction}}{\contract{\timeslot}}+\rewardAT{\representAction}}$ since the utility of $\actionOptT$ is lower than $\representAction$.

\begin{algorithm}[t!]
\caption{\textsc{Maximal Learning}}
\label{alg: single stage}
\SetAlgoLined
\KwData{$\dimension, \timeHorizon, \leftGapThreshold$}
$\timeslot\gets 1$, $\hvecs_1 \gets \mathbb{B}_{\dimension}$

$\knownSpace_0 \gets \emptyset$

\While{$\timeslot \leq \timeHorizon$}{

Receive $\contextAT{\action}, \costAT{\action}$ for each $\action\in \actionsT$

$\actionsKnownT \gets \set{\action\in \actionsT \condition \contextAT{\action}\in \mathrm{Span}(\knownSpace_{\timeslot-1})}$

Arbitrarily select $\hvec \in \hvecs_{\timeslot}$ and calculate $\rewardAT{\action} = \lrangle{\contextAT{\action},\hvec}, \forall \action\in\actionsKnownT$

$\actionsUnknownT \gets \actionsT \backslash \actionsKnownT$

$\hvecAOptT{\action} \gets \argmax_{\hvec \in \hvecs_{\timeslot}} \lrangle{ \contextAT{\action}, \hvec}$ for each $\action\in \actionsT$

$\actionOptT\gets \argmax_{\action\in \actionsT} \optProfit{\instance{\actionsKnownT\cup\set{\action}}{\hvec_{\timeslot}\super{\action}}, \action}$

$\hvecOptT\gets \hvecAOptT{\actionOptT}$, $\instanceSimple_{\timeslot} \gets \instance{\actionsKnownT\cup \set{\actionOptT}}{\hvecOptT}$

\If{$\actionOptT\in \actionsKnownT$}{
    propose contract $\contract{\timeslot} = \optContract{\instanceSimple_{\timeslot}}$, observe the action $\actionAgent$ and reward $\rewardAT{\actionAgent}$.

    Update hypothesis set: $\hvecs_{\timeslot+1}\gets  \hvecs_{\timeslot} \cap \set{\hvec\condition \lrangle{\hvec,\contextAT{\actionOptT}} = \rewardAT{\actionAgent}}$ if $\actionAgent\in \actionsUnknownT$; $\hvecs_{\timeslot+1}\gets  \hvecs_{\timeslot}$ if $\actionAgent\in \actionsKnownT$.
 }
 \Else{
\If{$\leftCostGap{\actionOptT}{\actionsKnownT}\geq \leftGapThreshold$}{
        Calculate $\nonDeReward{\actionOptT}$ and $\nonDeContract{\actionOptT}$ as in \Cref{def:non degenerated reward,def:non degenerated contract}

        Propose contract $\contract{\timeslot}=\nonDeContract{\actionOptT}$, observe the action $\actionAgent$ and reward $\rewardAT{\actionAgent}$

        Update hypothesis set: $\hvecs_{\timeslot+1}\gets  \hvecs_{\timeslot} \cap \set{\hvec\condition \lrangle{\hvec,\contextAT{\actionOptT}} = \rewardAT{\actionAgent}}$ if $\actionAgent\in \actionsUnknownT$; $\hvecs_{\timeslot+1}\gets  \hvecs_{\timeslot} \cap \set{\hvec\condition \lrangle{\hvec,\contextAT{\actionOptT}}\leq \nonDeReward{\actionOptT}}$ if $\actionAgent\in \actionsKnownT$.
    }
    \Else{
        Let $\representAction\in \actionsKnownT$ be the action satisfying the following two conditions: (1) cost of $\representAction$ is in $[\costAT{\actionOptT}-\leftGapThreshold, \costAT{\actionOptT}]$; (2) cost of $\representAction$ is the lowest among all $\actionsKnownT$ actions that satisfy condition (1)

        Propose contract $\contract{\timeslot}\gets \optContract{\instanceKnown,\representAction}$, observe the action $\actionAgent$ and reward $\rewardAT{\actionAgent}$

        Update hypothesis set: $\hvecs_{\timeslot+1}\gets  \hvecs_{\timeslot} \cap \set{\hvec\condition \lrangle{\hvec,\contextAT{\actionOptT}} = \rewardAT{\actionAgent}}$ if $\actionAgent\in \actionsUnknownT$; $\hvecs_{\timeslot+1}\gets  \hvecs_{\timeslot} \cap \set{\hvec\condition \lrangle{\hvec,\contextAT{\actionOptT}}\leq \frac{\costAT{\actionOptT}-\costAT{\representAction}}{\contract{\timeslot}}+\rewardAT{\representAction}}$ if $\actionAgent\in \actionsKnownT$.
    }
    $\knownSpace_{\timeslot}\gets \knownSpace_{\timeslot-1}\cup \set{\contextAT{\actionAgent}}$

    $\timeslot \gets \timeslot+1$
}
}

\end{algorithm}

\subsubsection{Regret Analysis of Algorithm~\ref{alg: single stage}}\label{sect:reward regret analysis}

Our proof is divided into three parts. 
\begin{itemize}
    \item The first part is relaxing the original regret benchmark to a simpler benchmark that simplifies our analysis. We accomplish this goal by \Cref{lem:restricted instance}, \Cref{lem:monotone reward profit}, and \Cref{lem:benchmark relaxation}. \Cref{lem:restricted instance} and \Cref{lem:monotone reward profit} are the supporting technical lemmas for proving \Cref{lem:benchmark relaxation}, which is the formal statement of the benchmark relaxation. 
    \item The second part aims to build the connection between immediate regret $\regretImmediate$ with information width, which is defined in \Cref{def:information  width}. The proof is further divided into several cases according to whether the cost gap is large.
\item In the final part, we utilize the connection between information width and potential decay we built in \Cref{sec:information width and potential}. 
    It helps us to lower bound $\regretImmediate$ by the potential decay in \Cref{lem:immediate regret single stage}. 
    Together with the result in the second part, we derive the regret upper bound.
\end{itemize}

\noindent\emph{\underline{[Part (i)] Benchmark relaxing}}\label{sect:part 1}

\begin{definition}[Restricted instance]
    If the action set of $\instanceSimple$ contains the action set of $\instanceSimple'$ (with the same cost and reward), we say $\instanceSimple'$ is a \emph{restricted instance} of $\instanceSimple$, denoted by $\instanceSimple'\prec \instanceSimple$.
\end{definition}

\begin{lemma}\label{lem:restricted instance}
    Let $\actions$ and $\actions'$ be the action set of $\instanceSimple$ and $\instanceSimple'$, respectively.
    If $\instanceSimple'\prec \instanceSimple$, then $\optProfitFromA{\instanceSimple'}{\action}\geq \optProfitFromA{\instanceSimple}{\action}\ \forall \action\in \actions'$.
\end{lemma}
\begin{proof}
    For any action $\action \in \actions$, recall $\costA{\action}$ is its cost and $\rewardA{\action}$ is its reward. Let $\contracts_{\instanceSimple,\action}$ be the set of all linear contracts that action $\action$ is induced in instance $\instanceSimple$. Thus, we have
    \begin{equation*}
        \contracts_{\instanceSimple,\action}= \set{\contractSimple\in [0,1]\condition \contractSimple\cdot\rewardA{\action}-\costA{\action}\geq \contractSimple\cdot \rewardA{\action'}-\costA{\action'}, \forall \action'\in \actions},
    \end{equation*}
    and
    \begin{equation*}
        \contracts_{\instanceSimple',\action}= \set{\contractSimple\in [0,1]\condition \contractSimple\cdot \rewardA{\action}-\costA{\action}\geq \contractSimple\cdot\rewardA{\action'}-\costA{\action'}, \forall \action' \in \actions'}.
    \end{equation*}
    Since $\actions'\subseteq \actions$, we have $\contracts_{\instanceSimple,\action}\subseteq\contracts_{\instanceSimple',\action}$ for any $\action\in \actions'$.

    Note that 
    \begin{equation*}
        \optProfitFromA{\instanceSimple}{\action} = \max_{\contractSimple\in\contracts_{\instanceSimple,\action}} (1-\contractSimple)\rewardA{\action}\mbox{  and  } \optProfitFromA{\instanceSimple'}{\action} = \max_{\contractSimple\in\contracts_{\instanceSimple',\action}} (1-\contractSimple)\rewardA{\action}.
    \end{equation*}
    Since $\contracts_{\instanceSimple,\action}\subseteq\contracts_{\instanceSimple',\action}$, we have $\optProfitFromA{\instanceSimple'}{\action}\geq \optProfitFromA{\instanceSimple}{\action}$.
\end{proof}

\begin{lemma}\label{lem:monotone reward profit}
    Let $\actions=\bar{\actions}\cup\set{\action},\actions'= \bar{\actions}\cup\set{\action'}$ be the action set of $\instanceSimple$ and $\instanceSimple'$ respectively. Assume that $\costA{\action}=\costA{\action'}$ and $\rewardA{\action'}\geq \rewardA{\action}$, than $\optProfitFromA{\instanceSimple'}{\action'}\geq \optProfitFromA{\instanceSimple}{\action}$.
\end{lemma}
\begin{proof}
    Let $\contracts_{\instanceSimple,\action}$ be defined as in the proof of Lemma~\ref{lem:restricted instance}. Specifically,
    \begin{equation*}
        \contracts_{\instanceSimple,\action}= \set{\contractSimple\in [0,1]\condition \contractSimple\cdot \rewardA{\action}-\costA{\action}\geq \contractSimple\cdot\rewardA{\bar{\action}}-\costA{\bar{\action}}, \forall \bar{\action}\in \bar{\actions}}
    \end{equation*}
    and 
    \begin{equation*}
        \contracts_{\instanceSimple',\action'}= \set{\contractSimple\in [0,1]\condition \contractSimple\cdot \rewardA{\action'}-\costA{\action'}\geq \contractSimple\cdot\rewardA{\bar{\action}}-\costA{\bar{\action}}, \forall \bar{\action}\in \bar{\actions}}.
    \end{equation*}
    Since $\costA{\action}=\costA{\action'}$ and $\rewardA{\action'}\geq \rewardA{\action}$ we have $\contracts_{\instanceSimple,\action}\subseteq \contracts_{\instanceSimple',\action'}$. Then
    \begin{equation*}
        \optProfitFromA{\instanceSimple'}{\action'} 
        = \max_{\contractSimple\in\contracts_{\instanceSimple',\action'}} (1-\contractSimple)\rewardA{\action'}
        \overset{(a)}{\geq} 
        \max_{\contractSimple\in\contracts_{\instanceSimple',\action'}} (1-\contractSimple)\rewardA{\action}
        \overset{(b)}{\geq}
        \max_{\contractSimple\in\contracts_{\instanceSimple,\action}} (1-\contractSimple)\rewardA{\action}
        =
        \optProfitFromA{\instanceSimple}{\action},
    \end{equation*}
    where inequality (a) is due to $\rewardA{\action'}\geq \rewardA{\action}$ and inequality (b) is due to $\contracts_{\instanceSimple,\action}\subseteq \contracts_{\instanceSimple',\action'}$.
\end{proof}

\newcommand{\hvecOptimism}{\hvec^{\dagger}_{\timeslot}}
\newcommand{\actionOptimism}{\action^{\dagger}_{\timeslot}}

\begin{lemma}\label{lem:hypothesis valid reward}
    During the running of Algorithm~\ref{alg: single stage}, $\hvecTrue\in \hvecsPess_\timeslot\subseteq \hvecs_{\timeslot}$ holds for all $\timeslot\in [\timeHorizon]$.
\end{lemma}
\begin{proof}
    It is enough to prove $\hvecsPess_\timeslot\subseteq \hvecs_{\timeslot}, \forall \timeslot\in [\timeHorizon]$. We prove this by induction. For $\timeslot=1$, $\hvecsPess_1=\hvecs_1=\vecBall$, the claim holds. Suppose $\hvecsPess_\timeslot\subseteq \hvecs_{\timeslot}$. If the agent takes an independent action $\actionAgent\in \actionsUnknownT$, then
    \[\hvecsPess_{\timeslot+1} \subseteq \hvecsPess_\timeslot \cap \set{\hvec\condition \lrangle{\hvec,\contextAT{\actionOptT}}=\rewardAT{\actionAgent}}\subseteq \hvecs_\timeslot \cap  \set{\hvec\condition \lrangle{\hvec,\contextAT{\actionOptT}}=\rewardAT{\actionAgent}} = \hvecs_{\timeslot+1}.\]
    If $\actionAgent\in \actionsKnownT$ and $\actionOptT\in \actionsKnownT$, $\hvecs_{\timeslot+1} = \hvecs_\timeslot$. Thus, $\hvecsPess_{\timeslot+1}\subseteq \hvecsPess_{\timeslot}\subseteq \hvecs_\timeslot\subseteq\hvecs_{\timeslot+1}$.
    If $\actionAgent \in \actionsKnownT$, $\actionOptT\in \actionsUnknownT$, and
    $\leftCostGap{\actionOptT}{\actionsKnownT}\geq \leftGapThreshold$. The agent's utility of taking action $\actionOptT$ is no larger than taking $\actionAgent$. By the definition of non-degenerating contract $\nonDeContract{\actionOptT}$, the reward of $\actionOptT$ is no larger than $\nonDeReward{\actionOptT}$. Then, 
    \[\hvecsPess_{\timeslot+1} \subseteq \hvecsPess_\timeslot \cap \set{\hvec\condition \lrangle{\hvec,\contextAT{\actionOptT}}\leq \nonDeReward{\actionOptT}}\subseteq \hvecs_\timeslot \cap  \set{\hvec\condition \lrangle{\hvec,\contextAT{\actionOptT}}\leq \nonDeReward{\actionOptT}}= \hvecs_{\timeslot+1}.\]
    If $\actionAgent \in \actionsKnownT$, $\actionOptT\in \actionsUnknownT$, and
    $\leftCostGap{\actionOptT}{\actionsKnownT}\geq \leftGapThreshold$. The agent's utility of taking action $\representAction$ is no larger than taking $\actionAgent$. So
    \begin{align*}
        \hvecsPess_{\timeslot+1}& \subseteq \hvecsPess_\timeslot \cap \set{\hvec\condition \lrangle{\hvec,\contextAT{\actionOptT}}\leq \frac{\costAT{\actionOptT}-\costAT{\representAction}}{\contract{\timeslot}}+\rewardAT{\representAction}}
        \\&\subseteq \hvecs_\timeslot \cap  \set{\hvec\condition \lrangle{\hvec,\contextAT{\actionOptT}}\leq \frac{\costAT{\actionOptT}-\costAT{\representAction}}{\contract{\timeslot}}+\rewardAT{\representAction}}= \hvecs_{\timeslot+1}.
    \end{align*}
    Therefore, our claim holds for all $\timeslot$ by induction.
\end{proof}

\begin{lemma}[Benchmark relaxation]\label{lem:benchmark relaxation}
    Let $\ALG$ be Algorithm~\ref{alg: single stage} with any inputting parameter, we have
    $\regretPessimistic_\timeslot(\ALG)\leq \optProfitFromA{\instance{\actionsKnown_{\timeslot}\cup \set{\actionOptT}}{\hvecOptT}}{\actionOptT}-\revenuePrincipalCon_\timeslot(\contract{\timeslot})$.
\end{lemma}
\begin{proof}
    Recall that $\regretPessimistic_\timeslot(\ALG) = \revenuePrincipalConPessimistic_\timeslot(\hvecsPess_\timeslot) - \revenuePrincipalCon_\timeslot(\contract{\timeslot})= \max_{\hvec\in \hvecsPess_{\timeslot}}\optProfit{\instance{\actionsT}{\hvec}}- \revenuePrincipalCon_\timeslot(\contract{\timeslot})$. Since $\hvecsPess\subseteq\hvecs_{\timeslot}$ by \Cref{lem:hypothesis valid reward}, we have $\max_{\hvec\in\hvecsPess_\timeslot}\optProfit{\instance{\actionsT}{\hvec}} \leq \max_{\hvec\in \hvecs_{\timeslot}}\optProfit{\instance{\actionsT}{\hvec}}$.
    
    Let $\hvecOptimism\in \argmax_{\hvec \in \hvecs_{\timeslot}}\optProfit{\instance{\actionsT}{\hvec}}$ and $\actionOptimism$ be the agent's action when the principal proposed the optimal contract in $\instance{\actionsT}{\hvecOptimism}$.  
    By \Cref{lem:restricted instance}, we have 
    \[\optProfitFromA{\instance{\actions_{\timeslot}}{\hvecOptimism}}{\actionOptimism} \leq \optProfitFromA{\instance{\actionsKnown_{\timeslot}\cup \set{\actionOptimism}}{\hvecOptimism}}{\actionOptimism},\]
    which is due to $ \instance{\actionsKnown_{\timeslot}\cup \set{\actionOptimism}}{\hvecOptimism}\prec \instance{\actions_{\timeslot}}{\hvecOptimism}$. 
    
    Furthermore, we can consider $\optProfitFromA{\instance{\actionsKnown_{\timeslot}\cup \set{\actionOptimism}}{\hvec_{\timeslot}\super{\actionOptimism}}}{\actionOptimism}$ where $\hvecAOptT{\actionOptimism}$ is the possible hidden vector that makes $\actionOptimism$ has the largest reward. We have 
    \[\optProfitFromA{\instance{\actionsKnown_{\timeslot}\cup \set{\actionOptimism}}{\hvecOptimism}}{\actionOptimism} 
    \overset{(a)}{\leq} 
    \optProfitFromA{\instance{\actionsKnown_{\timeslot}\cup \set{\actionOptimism}}{\hvecAOptT{\actionOptimism}}}{\actionOptimism}
    \overset{(b)}{\leq}
    \optProfitFromA{\instance{\actionsKnown_{\timeslot}\cup \set{\actionOptT}}{\hvecOptT}}{\actionOptT},
    \]
    where inequality (a) is due to Lemma~\ref{lem:monotone reward profit} and the fact that the reward of $\actionOptimism$ in $\instance{\actionsKnown_{\timeslot}\cup \set{\actionOptimism}}{\hvecOptimism}$ is no larger than which in $\instance{\actionsKnown_{\timeslot}\cup \set{\actionOptimism}}{\hvecAOptT{\actionOptimism}}$ and  
    inequality (b) is due to the property of the max operator and the definition of $\actionOptT, \hvecOptT$.

    Combining all inequalities above, we have 
    \[\regretPessimistic_\timeslot(\ALG)\leq \optProfitFromA{\instance{\actionsKnown_{\timeslot}\cup \set{\actionOptT}}{\hvecOptT}}{\actionOptT.} - \revenuePrincipalCon_\timeslot(\contract{\timeslot}). \]
    which completes the proof of \Cref{lem:benchmark relaxation}.
\end{proof}

\noindent\emph{\underline{[Part (ii)] Building connection between $\regretImmediate$ and information width}}\label{sect:part 2}

Consider the instance $\instance{\actionsKnown_{\timeslot}\cup \set{\actionOptT}}{\hvec}$, as $\hvec$ is fixed, the reward of $\actionOptT$ is determined, and the instance is determined by the reward of $\actionOptT$. By the abuse of the notation, we further let $\instance{\actionsKnown_{\timeslot}\cup \set{\actionOptT}}{\reward}$ be the instance where the reward of $\actionOptT$ is set to $\reward$.

\noindent \textbf{At each round, our analysis considers four cases and analyzes the immediate regret respectively:}
\begin{enumerate}
    \item $\actionAgent \in \actionsUnknownT$;
    \item $\actionOptT \in \actionsKnownT$ and $\actionAgent \in \actionsKnownT$;
    \item $\actionOptT \in \actionsUnknownT$, $\actionAgent \in \actionsKnownT$ and $\leftCostGap{\actionOptT}{\actionsKnownT}\geq \leftGapThreshold$;
    \item $\actionOptT \in \actionsUnknownT$, $\actionAgent \in \actionsKnownT$ and $\leftCostGap{\actionOptT}{\actionsKnownT}< \leftGapThreshold$.
\end{enumerate}

\xhdr{Case I: $\actionAgent\in \actionsUnknownT$.} 

In this case, our algorithm adds a context vector $\contextAT{\actionOptT}$ into $\knownSpace$ and makes the dimension of $\mathrm{Span}(\knownSpace)$ added by one. Since the dimension of context vectors is $\dimension$, this case happens at most $\dimension$ times. Therefore, the cumulative regret incurred by this case is at most $\dimension$.

\xhdr{Case II: $\actionAgent \in \actionsKnownT$ and $\actionOptT \in \actionsKnownT$.} 
In this case, the principal's utility is $\revenuePrincipalCon_{\timeslot}(\contract{\timeslot}) = \optProfit{\instanceKnown}$. We show that the immediate regret is $0$ in this case. 
\begin{lemma}\label{lem:case II regret 0}
    Let $\ALG$ denote Algorithm~\ref{alg: single stage}.
    If round $\timeslot$ of $\ALG$ falls into Case II, we have 
    \[ \regretPessimistic_{\timeslot} (\ALG) = 0.\]
\end{lemma}
\begin{proof}
    By \Cref{lem:benchmark relaxation}, 
    $\regretPessimistic_{\timeslot}(\ALG)\leq \optProfitFromA{\instance{\actionsKnown_{\timeslot}\cup \set{\actionOptT}}{\hvecOptT}}{\actionOptT} - \revenuePrincipalCon_{\timeslot}(\contract{\timeslot})$. 
    By the condition of Case II, we have $\optProfitFromA{\instance{\actionsKnown_{\timeslot}\cup \set{\actionOptT}}{\hvecOptT}}{\actionOptT} \leq \max_{\action\in \actionsKnownT}  \optProfitFromA{\instance{\actionsKnown_{\timeslot}\cup \set{\action}}{\hvec_{\timeslot}\super{\action}}}{\action} $. Since action in $\actionsKnownT$ is independent of the hidden vector $\hvec\in \hvecs_{\timeslot}$, we have 
    \[\max_{\action\in \actionsKnownT} \optProfitFromA{\instance{\actionsKnown_{\timeslot}\cup \set{\action}}{\hvec_{\timeslot}\super{\action}}}{\action} = \max_{\action\in \actionsKnownT}  \optProfitFromA{\instanceKnown}{\action} = \optProfit{\instanceKnown}.\]
    Overall, we have 
    \[  \optProfitFromA{\instance{\actionsKnown_{\timeslot}\cup \set{\actionOptT}}{\hvecOptT}}{\actionOptT} = \optProfit{\instanceKnown}.\]
    Therefore, $\regretPessimistic_{\timeslot}(\ALG)=0.$ since $\revenuePrincipalCon_{\timeslot}(\contract{\timeslot}) = \optProfit{\instanceKnown}$.
\end{proof}

\xhdr{Case III: $\actionOptT \in \actionsUnknownT$, $\actionAgent \in \actionsKnownT$ and $\leftCostGap{\actionOptT}{\actionsKnownT}\geq \leftGapThreshold$.} 
In this case, Algorithm~\ref{alg: single stage} proposes non-degenerating contract $\contract{\timeslot} \deq \nonDeContract{\actionOptT}$~(\Cref{def:non degenerated contract}).

Let $\instanceKnown$ be the instance equipped with action set $\actionsKnownT$, then this instance is independent of $\hvec$. Let the linear contract $\contractSimple$ vary from $0$ to $1$ in $\instanceKnown$, then different actions in $\actionsKnownT$ will be induced. We temporarily index the actions in $\actionsKnownT$ by this order. That is, we index the actions as $\actionsKnownT\deq\set{\actionTOrder{0},\actionTOrder{1},\ldots,\actionTOrder{m-1}}$ where $m\deq\abs{\actionsKnownT}$. The index ensures that there exists $0=\contractTOrder{0}\leq \contractTOrder{1}\leq \ldots \leq \contractTOrder{m}=1$ such that $\actionTOrder{j}$ is induced if the principal issues contract $\contractSimple\in [\contractTOrder{j},\contractTOrder{j+1})$. Moreover, $0=\costAT{\actionTOrder{0}}\leq \costAT{\actionTOrder{1}}\leq \ldots\leq \costAT{\actionTOrder{m-1}}$ and $0=\rewardAT{\actionTOrder{0}}\leq \rewardAT{\actionTOrder{1}}\leq \ldots\leq \rewardAT{\actionTOrder{m-1}}$. Here, the action $\actionTOrder{0}$ is the null action.

\begin{lemma}\label{lem:contract distance case i}
    If round $\timeslot$ in Algorithm~\ref{alg: single stage} falls into Case III, then 
    \[\nonDeContract{\actionOptT}-\optContract{\instanceSimple_{\timeslot}}\leq \frac{1}{\Delta}\informationWidth{\contextAT{\actionOptT}}{\nonDeReward{\actionOptT}}{\hvecs_{\timeslot}}.\]
\end{lemma}
\begin{proof}
    There exists $j\geq 1$ such that $\nonDeContract{\actionOptT}=\contractTOrder{j}$. 
    Moreover, there exists $k \leq j-1$ such that $\optContract{\instanceSimple_{\timeslot}}\in [\contractTOrder{k},\contractTOrder{k+1}]$ since $\optContract{\instanceSimple_{\timeslot}}\in [0,\nonDeContract{\actionOptT}]$. The above claim is easy to see on the \Cref{fig:upper-envelope-c}.

    Let $\utiAgent_{\instanceKnown}(\contractSimple)$ denote the agent's utility in instance $\instanceKnown$ when proposing the contract $\contractSimple$. Namely,
    \[\utiAgent_{\instanceKnown}(\contractSimple) = \contractSimple \cdot \rewardAT{\actionTOrder{i}} - \costAT{\actionTOrder{i}}, \mbox{ if } \contractSimple\in [\contractTOrder{i},\contractTOrder{i+1}].\]
    
    Let $\unknownRewardFunc{\contractSimple}$ be the reward of $\actionOptT$ such that the min-pay contract that implements $\actionOptT$ is $\contractSimple$. That is, it is the solution to
    \[\contractSimple\cdot \unknownRewardFunc{\contractSimple} - \costAT{\actionOptT} = \utiAgent_{\instanceKnown}(\contractSimple), \quad \contractSimple\in (0, \nonDeContract{\actionOptT}].\]
    By identical transformation,
    \[\unknownRewardFunc{\contractSimple} = \frac{\utiAgent_{\instanceKnown}(\contractSimple)+\costAT{\actionOptT}}{\contractSimple}, \quad  \contractSimple\in (0, \nonDeContract{\actionOptT}].\]
    Note that $\utiAgent_{\instanceKnown}(\contractSimple)$ is a continuous piecewise linear function. Thus, $\unknownRewardFunc{\contractSimple}$ can also be written as a piecewise function. Moreover, $\unknownRewardFunc{\contractSimple}$ is differentiable on $[\contractTOrder{i},\contractTOrder{i+1}]$ and its derivative is
    
    \[\unknownRewardFuncDerivative{\contractSimple} = \frac{\rewardAT{\actionTOrder{i}}}{\contractSimple} -\frac{\contractSimple\cdot \rewardAT{\actionTOrder{i}}+\costAT{\actionOptT}-\costAT{\actionTOrder{i}}}{\contractSimple^2}=-\frac{\costAT{\actionOptT}-\costAT{\actionTOrder{i}}}{\contractSimple^2}, \quad \forall \contractSimple\in [\contractTOrder{i},\contractTOrder{i+1}]\] by simply applying derivative rules. 

    Note that $\costAT{\actionOptT}-\costAT{\actionTOrder{i}}\geq \leftGapThreshold$ by the condition of Case I and $\contractSimple\leq 1$, we have $\unknownRewardFuncDerivative{\contractSimple}\leq - \leftGapThreshold$ almost everywhere on $[\optContract{\instanceSimple_{\timeslot}},\nonDeContract{\actionOptT}]$. Since $\unknownRewardFunc{\optContract{\instanceSimple_{\timeslot}}}=\max_{\hvec\in\hvecs_{\timeslot}}\lrangle{\contextAT{\actionOptT},\hvec}$ and $\unknownRewardFunc{\nonDeContract{\actionOptT}} =\nonDeReward{\actionOptT}$, we have
    \[\nonDeReward{\actionOptT}-\max_{\hvec\in\hvecs_{\timeslot}}\lrangle{\contextAT{\actionOptT},\hvec}\leq -\leftGapThreshold\left(\nonDeContract{\actionOptT}-\optContract{\instanceSimple_{\timeslot}}\right).\]
    By identical transformation and the definition of information width, we have
    \[\nonDeContract{\actionOptT}-\optContract{\instanceSimple_{\timeslot}}\leq \frac{1}{\leftGapThreshold} \informationWidth{\contextAT{\actionOptT}}{\nonDeReward{\actionOptT}}{\hvecs_{\timeslot}}.\]
    This completes the proof of \Cref{lem:contract distance case i}.

    \let\unknownRewardFunc\undefined
    \let\unknownRewardFuncDerivative\undefined
    \let\agentRevenueFunc\undefined

\end{proof}

\begin{lemma}\label{lem:case i}
    Let $\ALG$ denote Algorithm~\ref{alg: single stage}. If round $\timeslot$ of $\ALG$ falls into Case III, then the immediate pessimistic Stackelberg regret suffered by $\ALG$ in $\timeslot$-th round is
    \[\regretPessimistic_\timeslot(\ALG) \leq \frac{2}{\leftGapThreshold}\informationWidth{\contextAT{\actionOptT}}{\nonDeReward{\actionOptT}}{\hvecs_{\timeslot}}.\]
\end{lemma}
\begin{proof}
    By the definition of our notations, $\optProfit{\instanceSimple_{\timeslot},\actionOptT}= \left(1-\optContract{\instanceSimple_{\timeslot}}\right)\max_{\hvec\in \hvecs_{\timeslot}}\lrangle{\contextAT{\actionOptT},\hvec}$. Recall that there exist $j\geq 1$ such that $\nonDeContract{\actionOptT}=\contractTOrder{j}$.

    If $j\leq m-1$, then $\actionTOrder{j}$ is induced, we have $\revenuePrincipalCon_{\timeslot}(\contract{\timeslot}) = \left(1-\nonDeContract{\actionOptT}\right)\rewardAT{\actionTOrder{j}}\geq \left(1-\nonDeContract{\actionOptT}\right)\nonDeReward{\actionOptT}$.

    If $j=m$, then $\nonDeContract{\actionOptT}=\contractTOrder{m}=1$. Thus, $\revenuePrincipalCon_{\timeslot}(\contract{\timeslot}) = 0 $. We can also write $\revenuePrincipalCon_{\timeslot}(\contract{\timeslot}) = (1-\nonDeContract{\actionOptT})\nonDeReward{\actionOptT}$.
    
    Therefore, whatever $j$ is, we have
    \begin{align*}
        \regretPessimistic_\timeslot (\ALG)&\overset{(a)}{\leq} \optProfitFromA{\instanceSimple_{\timeslot}}{\actionOptT}-\revenuePrincipalCon_{\timeslot}(\contract{\timeslot})
        \\ &\overset{(b)}{\leq} \left(1-\optContract{\instanceSimple_{\timeslot}}\right)\max_{\hvec\in \hvecs_{\timeslot}}\lrangle{\contextAT{\actionOptT},\hvec} - \left(1-\nonDeContract{\actionOptT}\right)\nonDeReward{\actionOptT}
        \\ &= \left(1-\optContract{\instanceSimple_{\timeslot}}\right)\max_{\hvec\in \hvecs_{\timeslot}}\lrangle{\contextAT{\actionOptT},\hvec}- \left(1-\nonDeContract{\actionOptT}\right)\max_{\hvec\in \hvecs_{\timeslot}}\lrangle{\contextAT{\actionOptT},\hvec}
        \\&\quad + \left(1-\nonDeContract{\actionOptT}\right)\max_{\hvec\in \hvecs_{\timeslot}}\lrangle{\contextAT{\actionOptT},\hvec}- \left(1-\nonDeContract{\actionOptT}\right)\nonDeReward{\actionOptT}
        \\&=\left(\nonDeContract{\actionOptT}-\optContract{\instanceSimple_{\timeslot}}\right)\max_{\hvec\in \hvecs_{\timeslot}}\lrangle{\contextAT{\actionOptT},\hvec}+\left(1-\nonDeContract{\actionOptT}\right)\left(\max_{\hvec\in \hvecs_{\timeslot}}\lrangle{\contextAT{\actionOptT},\hvec}-\nonDeReward{\actionOptT}\right)
        \\&\overset{(c)}{\leq} \frac{1}{\leftGapThreshold}\informationWidth{\contextAT{\actionOptT}}{\nonDeReward{\actionOptT}}{\hvecs_{\timeslot}} + \frac{1}{\leftGapThreshold}\informationWidth{\contextAT{\actionOptT}}{\nonDeReward{\actionOptT}}{\hvecs_{\timeslot}}
        \\ &\leq \frac{2}{\leftGapThreshold}\informationWidth{\contextAT{\actionOptT}}{\nonDeReward{\actionOptT}}{\hvecs_{\timeslot}},
    \end{align*}
    where inequality (a) is because of \Cref{lem:benchmark relaxation}, inequality (b) is by $ \revenuePrincipalCon_{\timeslot}(\contract{\timeslot}) \geq \left(1-\nonDeContract{\actionOptT}\right)\nonDeReward{\actionOptT}$ we have just proved, and inequality (c) is by \Cref{lem:contract distance case i}, $\max_{\hvec\in \hvecs_{\timeslot}}\lrangle{\contextAT{\actionOptT},\hvec}\leq 1$ and the definition of information width.
\end{proof}

\xhdr{Case IV: $\actionOptT \in \actionsUnknownT$, $\actionAgent \in \actionsKnownT$ and $\leftCostGap{\actionOptT}{\actionsKnownT}< \leftGapThreshold$.} 
When $\leftCostGap{\actionOptT}{\actionsKnownT}<\leftGapThreshold$, recall that Algorithm~\ref{alg: single stage} proposes contract $\contract{\timeslot} = \optContract{\instanceKnown,\representAction}$. Since an $\actionsKnownT$ action is induced, then $\representAction$ is induced by the linear contract we proposed.

\begin{lemma}\label{lem:reward gap case ii}
    If round $\timeslot$ in Algorithm~\ref{alg: single stage} falls into Case IV, then 
    \[ \max_{\hvec\in\hvecs_{\timeslot+1}}\lrangle{\hvec,\contextAT{\actionOptT}}-\rewardAT{\representAction}\leq \frac{\leftGapThreshold}{\costLB}.\]
\end{lemma}
\begin{proof}
    If $\representAction$ is induced, the hidden vector space is reduced to 
    \[\hvecs_{\timeslot+1} = \hvecs_{\timeslot} \cap \set{\hvec\condition \optContract{\instanceKnown,\representAction}\cdot\rewardAT{\representAction}-\costAT{\representAction}\geq \optContract{\instanceKnown,\representAction}\left(\lrangle{\hvec,\contextAT{\actionOptT}}\right)-\costAT{\actionOptT}}.\]
    By a simple identical transformation,
    \begin{equation}\label{eq:caseII new hvecs}
        \hvecs_{\timeslot+1} = \hvecs_{\timeslot} \cap \set{\hvec\condition \lrangle{\hvec,\contextAT{\actionOptT}}\leq \frac{\costAT{\actionOptT}-\costAT{\representAction}}{\optContract{\instanceKnown,\representAction}}+\rewardAT{\representAction}}.
    \end{equation}
    By \eqref{eq:caseII new hvecs}, 
   \[
        \max_{\hvec\in\hvecs_{\timeslot+1}}\lrangle{\hvec,\contextAT{\actionOptT}}-\rewardAT{\representAction}\leq \frac{\costAT{\actionOptT}-\costAT{\representAction}}{\optContract{\instanceKnown,\representAction}}\leq \frac{\leftGapThreshold}{\costLB}
    \]
    which completes the proof of \Cref{lem:reward gap case ii}.
\end{proof}

\begin{lemma} \label{lem:contract distance case ii}
    If round $\timeslot$ in Algorithm~\ref{alg: single stage} falls into Case IV, then
    \[\contract{\timeslot}-\optContract{\instanceSimple_{\timeslot}}\leq \frac{1}{\leftGapThreshold}\informationWidth{\contextAT{\actionOptT}}{\frac{\costAT{\actionOptT}-\costAT{\representAction}}{\contract{\timeslot}}+\rewardAT{\representAction}}{\hvecs_{\timeslot}}.\]
\end{lemma}
\begin{proof}
    There exists $m-1\geq j\geq 1$ such that $\representAction = \actionTOrder{j}$, then $\contract{\timeslot}=\optContract{\instanceKnown,\representAction} = \contractTOrder{j}$. Moreover, there exists $k$ such that $\optContract{\instanceSimple_{\timeslot}}\in [\contractTOrder{k},\contractTOrder{k+1}]$. If $k\geq j$, then 
    \[\contract{\timeslot}-\optContract{\instanceSimple_{\timeslot}}\leq 0 \leq  \frac{1}{\leftGapThreshold}\informationWidth{\contextAT{\actionOptT}}{\frac{\costAT{\actionOptT}-\costAT{\representAction}}{\contract{\timeslot}}+\rewardAT{\representAction}}{\hvecs_{\timeslot}},\] 
    since the information width is non-negative.
    Therefore, we assume $k\leq j-1$ in the following proof.
    
    Recall that $\utiAgent_{\instanceKnown}(\contractSimple)$ is the agent's utility in instance $\instanceKnown$ when proposing the contract $\contractSimple$. Namely,
    \[\utiAgent_{\instanceKnown}(\contractSimple) = \contractSimple \cdot \rewardAT{\actionTOrder{i}} - \costAT{\actionTOrder{i}}, \mbox{ if } \contractSimple\in [\contractTOrder{i},\contractTOrder{i+1}].\]
    
    Let $\unknownRewardFunc{\contractSimple}$ be the reward of $\actionOptT$ such that the min-pay contract that implements $\actionOptT$ is $\contractSimple$. That is, it is the solution to
    \[\contractSimple\cdot \unknownRewardFunc{\contractSimple} - \costAT{\actionOptT} = \utiAgent_{\instanceKnown}(\contractSimple), \quad \contractSimple\in (0, \nonDeContract{\actionOptT}].\]

    By identical transformation,
    \[\unknownRewardFunc{\contractSimple} = \frac{\utiAgent_{\instanceKnown}(\contractSimple)+\costAT{\actionOptT}}{\contractSimple}, \quad  \contractSimple\in (0, \nonDeContract{\actionOptT}].\]
    Note that $\utiAgent_{\instanceKnown}(\contractSimple)$ is a continuous piecewise linear function. Thus, $\unknownRewardFunc{\contractSimple}$ can also be written as a piecewise function. Moreover, $\unknownRewardFunc{\contractSimple}$ is differentiable on $[\contractTOrder{i},\contractTOrder{i+1}]$ and its derivative is
    
    \[\unknownRewardFuncDerivative{\contractSimple} = \frac{\rewardAT{\actionTOrder{i}}}{\contractSimple} -\frac{\contractSimple\cdot \rewardAT{\actionTOrder{i}}+\costAT{\actionOptT}-\costAT{\actionTOrder{i}}}{\contractSimple^2}=-\frac{\costAT{\actionOptT}-\costAT{\actionTOrder{i}}}{\contractSimple^2}, \quad \forall \contractSimple\in [\contractTOrder{i},\contractTOrder{i+1}]\] by simply applying derivative rules. 

    Note that $\costAT{\actionOptT}-\costAT{\actionTOrder{i}}\geq \leftGapThreshold, \forall i<j$ by the condition of Case II, then $\unknownRewardFuncDerivative{\contractSimple}\leq -\leftGapThreshold$ almost everywhere on $[0, \contractTOrder{j})$. Since $\unknownRewardFunc{\optContract{\instanceSimple_{\timeslot}}}=\max_{\hvec\in\hvecs_{\timeslot}}\lrangle{\contextAT{\actionOptT},\hvec}$ 
    and 
    $\unknownRewardFunc{\contractTOrder{j}} = \frac{\costAT{\actionOptT}-\costAT{\representAction}}{\contract{\timeslot}}+\rewardAT{\representAction}$, 
    we have
    \[\frac{\costAT{\actionOptT}-\costAT{\representAction}}{\contract{\timeslot}}+\rewardAT{\representAction}-\max_{\hvec\in\hvecs_{\timeslot}}\lrangle{\contextAT{\actionOptT},\hvec}= \unknownRewardFunc{\contractTOrder{j}}-\optContract{\instanceSimple_{\timeslot}} \leq -\leftGapThreshold\left(\contractTOrder{j}-\optContract{\instanceSimple_{\timeslot}}\right)=-\leftGapThreshold\left(\contract{\timeslot}-\optContract{\instanceSimple_{\timeslot}}\right).\]

    By identical transformation,
    \[\contract{\timeslot}-\optContract{\instanceSimple_{\timeslot}}\leq \frac{1}{\leftGapThreshold}\informationWidth{\contextAT{\actionOptT}}{\frac{\costAT{\actionOptT}-\costAT{\representAction}}{\contract{\timeslot}}+\rewardAT{\representAction}}{\hvecs_{\timeslot}}\]
    which completes the proof of \Cref{lem:contract distance case ii}.
\end{proof}

\begin{lemma}\label{lem:case ii}
     Let $\ALG$ denote Algorithm~\ref{alg: single stage}. If round $\timeslot$ of $\ALG$ falls into Case IV, then the immediate pessimistic Stackelberg regret suffered by $\ALG$ in $\timeslot$-th round is
    \[\regretPessimistic_\timeslot (\ALG) \leq \frac{\leftGapThreshold}{\costLB}+\left(\frac{2}{\leftGapThreshold}+1\right)\informationWidth{\contextAT{\actionOptT}}{\frac{\costAT{\actionOptT}-\costAT{\representAction}}{\contract{\timeslot}}+\rewardAT{\representAction}}{\hvecs_{\timeslot}}.\]
\end{lemma}
\begin{proof}
    By the definition of our notations, $\optProfitFromA{\instanceSimple_{\timeslot}}{\actionOptT}= \left(1-\optContract{\instanceSimple_{\timeslot}}\right)\max_{\hvec\in \hvecs_{\timeslot}}\lrangle{\contextAT{\actionOptT},\hvec}$. Since the proposed contract is $\contract{\timeslot}$ and the action $\representAction=\actionTOrder{j}$ is induced, we have $\revenuePrincipalCon_{\timeslot}(\contract{\timeslot}) = \left(1-\contract{\timeslot}\right)\rewardAT{\representAction}$.

    Therefore,
    \begin{align*}
        \regretPessimistic_\timeslot (\ALG)&\leq \optProfitFromA{\instanceSimple_{\timeslot}}{\actionOptT}-\revenuePrincipalCon_{\timeslot}(\contract{\timeslot})
        \\ &\leq \left(1-\optContract{\instanceSimple_{\timeslot}}\right)\max_{\hvec\in \hvecs_{\timeslot}}\lrangle{\contextAT{\actionOptT},\hvec} - \left(1-\contract{\timeslot}\right)\rewardAT{\representAction}
        \\ &= \left(1-\optContract{\instanceSimple_{\timeslot}}\right)\max_{\hvec\in \hvecs_{\timeslot}}\lrangle{\contextAT{\actionOptT},\hvec}- \left(1-\contract{\timeslot}\right)\max_{\hvec\in \hvecs_{\timeslot}}\lrangle{\contextAT{\actionOptT},\hvec}
        \\&\quad + \left(1-\contract{\timeslot}\right)\left(\max_{\hvec\in \hvecs_{\timeslot}}\lrangle{\contextAT{\actionOptT},\hvec}-\max_{\hvec\in \hvecs_{\timeslot+1}}\lrangle{\contextAT{\actionOptT},\hvec}\right) +(1-\contract{\timeslot}) \max_{\hvec\in \hvecs_{\timeslot+1}}\lrangle{\contextAT{\actionOptT},\hvec}- \left(1-\contract{\timeslot}\right)\rewardAT{\representAction}
        \\&=\left(\contract{\timeslot}-\optContract{\instanceSimple_{\timeslot}}\right)\max_{\hvec\in \hvecs_{\timeslot}}\lrangle{\contextAT{\actionOptT},\hvec} + \left(1-\contract{\timeslot}\right)\left(\max_{\hvec\in \hvecs_{\timeslot}}\lrangle{\contextAT{\actionOptT},\hvec}-\max_{\hvec\in \hvecs_{\timeslot+1}}\lrangle{\contextAT{\actionOptT},\hvec}\right) 
        \\&\quad +\left(1-\contract{\timeslot}\right)\left(\max_{\hvec\in \hvecs_{\timeslot+1}}\lrangle{\contextAT{\actionOptT},\hvec}-\rewardAT{\representAction}\right)
        \\&\overset{(a)}{\leq} \left(\frac{2}{\leftGapThreshold}+1\right)\informationWidth{\contextAT{\actionOptT}}{\frac{\costAT{\actionOptT}-\costAT{\representAction}}{\contract{\timeslot}}+\rewardAT{\representAction}}{\hvecs_{\timeslot}} +\frac{\leftGapThreshold}{\costLB}.
    \end{align*}
    The inequality (a) is because of \Cref{lem:reward gap case ii}, \Cref{lem:contract distance case ii}, and the definition of information width.
\end{proof}

\noindent \emph{\underline{[Part (iii)] Final regret analysis}}

\begin{lemma}\label{lem:immediate regret single stage}
    Let $\ALG$ denote Algorithm~\ref{alg: single stage}. If round $\timeslot$ of $\ALG$ falls into Case III or Case IV, then the immediate pessimistic Stackelberg regret suffered by $\ALG$ in $\timeslot$-th round is
    \begin{equation}\label{eq:noname4}
        \regretPessimistic_\timeslot (\ALG)\leq \frac{\leftGapThreshold}{\costLB} + \left(\frac{2}{\leftGapThreshold} +1\right)\dimension^{1/\dimension}\left(\potentialSum{\hvecs_{\timeslot}}-\potentialSum{\hvecs_{\timeslot+1}}\right)^{1/\dimension}.
    \end{equation}
    Here $\potentialSum{\cdot}$ is the potential function we defined in \Cref{def:potential function}.
\end{lemma}
\begin{proof}
    In case III, $\hvecs_{\timeslot+1} = \hvecs_{\timeslot}\cap \set{\hvec\condition \lrangle{\hvec,\contextAT{\actionOptT}}\leq \nonDeReward{\actionOptT}}$. Let $\convexBody = \hvecs_{\timeslot}$ and $\convexBodyNext=\hvecs_{\timeslot+1}$ in the \Cref{lem:potential sum information width}, then there exist $j\leq \dimension$ such that
    \[\potentialSum{\hvecs_{\timeslot}}-\potentialSum{\hvecs_{\timeslot+1}} \geq \frac{1}{\dimension}\left(\frac{1}{\norm{\contextAT{\actionOptT}}}\informationWidth{\contextAT{\actionOptT}}{\nonDeReward{\actionOptT}}{\hvecs_{\timeslot}}\right)^j.\]
    Here, since $\nonDeReward{\actionOptT}\geq 0$ and $\hvecs_{\timeslot}\subseteq \mathbb{B}_{\dimension}$, we have $\informationWidth{\contextAT{\actionOptT}}{\nonDeReward{\actionOptT}}{\hvecs_{\timeslot}}\leq \norm{\contextAT{\actionOptT}}$. Therefore, $\frac{1}{\norm{\contextAT{\actionOptT}}}\informationWidth{\contextAT{\actionOptT}}{\nonDeReward{\actionOptT}}{\hvecs_{\timeslot}}\leq 1$ and 
    \[\potentialSum{\hvecs_{\timeslot}}-\potentialSum{\hvecs_{\timeslot+1}} \geq \frac{1}{\dimension}\left(\frac{1}{\norm{\contextAT{\actionOptT}}}\informationWidth{\contextAT{\actionOptT}}{\nonDeReward{\actionOptT}}{\hvecs_{\timeslot}}\right)^d.\]

    By \Cref{lem:case i}, we have
    \[\regretPessimistic_\timeslot (\ALG) \leq \frac{2}{\leftGapThreshold} \dimension^{1/\dimension} \left(\potentialSum{\hvecs_{\timeslot}}-\potentialSum{\hvecs_{\timeslot+1}} \right)^{1/\dimension}.\]
    
    In case IV, $\hvecs_{\timeslot+1} = \hvecs_{\timeslot}\cap \set{\hvec\condition \lrangle{\hvec,\contextAT{\actionOptT}}\leq \frac{\costAT{\actionOptT}-\costAT{\representAction}}{\contract{\timeslot}}+\rewardAT{\representAction}}$.  Let $\convexBody = \hvecs_{\timeslot}$ and $\convexBodyNext=\hvecs_{\timeslot+1}$ in the \Cref{lem:potential sum information width}, using the similar argument from the above, we have
    \[\potentialSum{\hvecs_{\timeslot}}-\potentialSum{\hvecs_{\timeslot+1}} \geq \frac{1}{\dimension}\left(\informationWidth{\contextAT{\actionOptT}}{\frac{\costAT{\actionOptT}-\costAT{\representAction}}{\contract{\timeslot}}+\rewardAT{\representAction}}{\hvecs_{\timeslot}}\right)^d .\]
    By \Cref{lem:case ii}, we have
    \[\regretPessimistic_\timeslot (\ALG) \leq \left(\frac{2}{\leftGapThreshold}+1\right) \dimension^{1/\dimension} \left(\potentialSum{\hvecs_{\timeslot}}-\potentialSum{\hvecs_{\timeslot+1}} \right)^{1/\dimension} + \frac{\leftGapThreshold}{\costLB}.\]
    Overall, the immediate pessimistic regret in Case III and IV can be upper bounded as \eqref{eq:noname4}
\end{proof}

\begin{proof}[\bf{Proof of \Cref{thm:reward context upper bound}}]
    Let $\mathcal{T}_{1}$, $\mathcal{T}_2$ be the set of rounds that fall into Case I and Case II, respectively. Let $\mathcal{T}_{3,4}$ be the set of rounds that fall into Case III or IV. Then $\abs{\mathcal{T}_1}\leq \dimension$.
    
    By \Cref{lem:immediate regret single stage},
    \begin{align*}
        \regretPessimistic(\ALG) &= \sum_{\timeslot\in\mathcal{T}_{3,4}} \regretPessimistic_\timeslot(\ALG)+\sum_{\timeslot\in\mathcal{T}_1} \regretPessimistic_\timeslot(\ALG) + \sum_{\timeslot\in\mathcal{T}_2} \regretPessimistic_\timeslot(\ALG)
        \\ &\overset{(a)}{\leq} \sum_{\timeslot\in\mathcal{T}_{3,4}} \left(\left(\frac{2}{\leftGapThreshold}+1\right) \dimension^{1/\dimension} \left(\potentialSum{\hvecs_{\timeslot}}-\potentialSum{\hvecs_{\timeslot+1}} \right)^{1/\dimension} + \frac{\leftGapThreshold}{\costLB}\right) + \abs{\mathcal{T}_1}
        \\ &\leq \sum_{\timeslot\in [\timeHorizon]}\left(\left(\frac{2}{\leftGapThreshold}+1\right) \dimension^{1/\dimension} \left(\potentialSum{\hvecs_{\timeslot}}-\potentialSum{\hvecs_{\timeslot+1}} \right)^{1/\dimension} + \frac{\leftGapThreshold}{\costLB}\right) + \abs{\mathcal{T}_1}
        \\ &\leq \frac{\leftGapThreshold}{\costLB} \timeHorizon + \left(\frac{2}{\leftGapThreshold}+1\right)\dimension^{1/\dimension} \sum_{\timeslot\in [\timeHorizon]}\left(\potentialSum{\hvecs_{\timeslot}}-\potentialSum{\hvecs_{\timeslot+1}} \right)^{1/\dimension} +\dimension
        \\&\overset{(b)}{\leq} \frac{\leftGapThreshold}{\costLB} \timeHorizon + \left(\frac{2}{\leftGapThreshold}+1\right)\dimension^{1/\dimension} \timeHorizon^{1-1/\dimension} \left(\sum_{t=1}^{\timeHorizon} \left(\potentialSum{\hvecs_{\timeslot}}-\potentialSum{\hvecs_{\timeslot+1}} \right)\right)^{1/\dimension} + \dimension
        \\&\leq\frac{\leftGapThreshold}{\costLB} \timeHorizon + \frac{3}{\leftGapThreshold}\dimension^{1/\dimension} \timeHorizon^{1-1/\dimension} \left(\potentialSum{\hvecs_1}-\potentialSum{\hvecs_{\timeHorizon+1}}\right)^{1/\dimension}+\dimension
        \\&\overset{(c)}{\leq} \frac{\leftGapThreshold}{\costLB} \timeHorizon + \frac{3}{\leftGapThreshold}\dimension^{2/\dimension} \timeHorizon^{1-1/\dimension} \sqrt{2\pi \dimension}+\dimension
        \\&\overset{(d)}{\leq} \frac{\sqrt{6\pi}\dimension^{\frac{1}{4}+\frac{1}{\dimension}}}{\sqrt{\costLB}} \timeHorizon^{1-\frac{1}{2\dimension}}+\dimension.
    \end{align*}
    The inequality (a) is because of \Cref{lem:immediate regret single stage}, \Cref{lem:case II regret 0}.
    The inequality (b) is by Holder's inequality. 
    The inequality (c) is by \Cref{lem:initial potential}. 
    The inequality (d) is due to $\leftGapThreshold = \frac{\sqrt{3\costLB} \dimension^{\frac{1}{4}+\frac{1}{\dimension}}}{\timeHorizon^{\frac{1}{2\dimension}}}$.
\end{proof}

\subsection{Lower Bound for Reward Context}\label{sec:lower-bound-reward-opt}
To establish a \emph{pessimistic Stackelberg regret} lower bound about any algorithm $\ALG$
 for the reward-context setting, we construct an adversarial instance analogous to our previous cost-context analysis. Our construction forces the algorithm to incur significant pessimistic Stackelberg regret. This guarantee is formally stated in the theorem below, which provides a regret lower bound for the reward-context setting. We also defer the discussion about the classic \emph{Stackelberg regret} in \Cref{sec:lower-reward-true}.

\begin{theorem}\label{thm:lower-bound-reward-opti}
    For contextual principal-agent game with reward contexts, let $\dimension$ be the dimension of the context, for any algorithm $\ALG$, there is an instance making the algorithm suffer $\Omega(T^{1-\frac{1}{d}})$ pessimistic Stackelberg regret.
\end{theorem}

Following a similar line of reasoning as in the cost-context setting, we first present our \Cref{alg: adversary_optimism}, ‌analyze the important simplified instance with known actions of our construction in \Cref{sec:instance-known-reward}, verify the validity of adversarial construction in \Cref{sec:valid-reward-opt} and complete our lower bound analysis in \Cref{sec:analysis-reward-opt}. Given any algorithm $\ALG$, a key component of adversarial strategy also relies on the use of \emph{spherical code} from coding theory with minimal angle in \Cref{def:spherical codes} to adaptively define the context vectors.

\xhdr{Adversary.} Let $\leftGapThreshold$ be a number to be determined. We consider a four-action instance with action set $\actionsT = \set{\nullAction,\knownOptAction,\adversarialAction,\restrictingAction}$. Similarly, the first coefficient of the true hidden vector $\hvecTrue$ is fixed as $\frac{1}{2}$.
\begin{itemize}
    \item $\nullAction$ is the null action with zero reward and cost.
    \item $\knownOptAction$ is the action with context vector $\contextAT{\knownOptAction}=(\frac{1}{\eulerNumber},0,\ldots,0)$ and $\costAT{\knownOptAction} = \frac{1}{4\eulerNumber}$. Note the context vector make its reward known to $\ALG$, which is $\rewardAT{\knownOptAction} = \frac{1}{2\eulerNumber}$.
    \item $\restrictingAction$ has cost $\costAT{\restrictingAction}=\frac{1}{4\eulerNumber}+\frac{\leftGapThreshold}{\eulerNumber}$ with context vector $\contextAT{\restrictingAction}=(2\equalRevenueReward,0,...,0)$. Here, the adversary also ensures that only the first coefficient of $\contextAT{\restrictingAction}$ is non-zero, which makes its reward known to $\ALG$, which is $\rewardAT{\restrictingAction}=\equalRevenueReward$.
    \item $\adversarialAction$ has cost $\costAT{\adversarialAction} = \frac{1}{4\eulerNumber}+\frac{\leftGapThreshold}{\eulerNumber}$. Its context in the $\timeslot$-th round will be determined adaptively according to the behavior of $\ALG$ in the previous $\timeslot-1$ rounds. Only the reward of this action is unknown to $\ALG$.
\end{itemize}
The adversary's strategy for updating its hypothesis set mirrors that of the cost-context setting. Specifically, it evaluates the principal's proposed contract $\contract{\timeslot}$ against an optimistic threshold derived from the current hypothesis set $\hvecs_{\timeslot}$. A formal description of this adaptive procedure is provided in \Cref{alg: adversary_optimism}.

\begin{adversary}[t!]
\caption{\textsc{Adversary with Reward Context}}
\label[adversary]{alg: adversary_optimism}
\SetAlgoLined
\KwData{$\ALG, \timeHorizon, \dimension, \leftGapThreshold$, spherical code $\sphericalCodedim{\dimension-1}$ }

Initialize $\contextAT{\knownOptAction} = (\frac{1}{\eulerNumber},0,\ldots,0), \costAT{\knownOptAction}=\frac{1}{4\eulerNumber}, \contextAT{\restrictingAction} = \left(2\equalRevenueReward,0,\ldots,0\right), \costAT{\restrictingAction}=\costAT{\adversarialAction}=\frac{1}{4\eulerNumber}+\frac{1}{\eulerNumber}\leftGapThreshold, \forall \timeslot\geq 1$

$\timeslot\gets 1, i\gets 1, j\gets 1$

Initialize $\radius\super{1} = \frac{\sqrt{3}}{2}$, $\hvecs_{1}=\set{\frac{1}{2}}\times \left(\radius\super{1} \cdot \mathbb{B}_{d-1}\right)$ 

Arbitrarily index code words $\sphericalCodedim{\dimension-1}=\set{\codeWord{1},\codeWord{2},\ldots,\codeWord{\abs{\sphericalCodedim{\dimension-1}}}}$

Set $\rewardLargestPossible$ so that it satisfies
    $\left(1-\frac{\frac{1}{\eulerNumber}\leftGapThreshold}{\rewardLargestPossible-\frac{1}{2\eulerNumber}}\right)\rewardLargestPossible=\frac{1}{4\eulerNumber}+\frac{\leftGapThreshold}{2\eulerNumber}$. That is, 
    \[\rewardLargestPossible = \frac{3}{8\eulerNumber}(1+2\leftGapThreshold) + \frac{1}{8\eulerNumber}\sqrt{36\leftGapThreshold^2+20\leftGapThreshold+1}.\]

Set $\equalRevenueReward$ so that it satisfies $\left(1-\frac{\frac{1}{\eulerNumber}\leftGapThreshold}{\equalRevenueReward-\frac{1}{2\eulerNumber}}\right)\equalRevenueReward=\frac{1}{4\eulerNumber}-\frac{\leftGapThreshold}{2\eulerNumber}$. That is
    \[\equalRevenueReward = \frac{3}{8\eulerNumber}+\frac{\leftGapThreshold}{4\eulerNumber} + \frac{1}{8\eulerNumber}\sqrt{4\leftGapThreshold^2+28\leftGapThreshold+1}.\]

$\contractThreshold\gets \frac{\leftGapThreshold}{\eulerNumber\rewardLargestPossible-\frac{1}{2}}$

\While{$\timeslot\leq \timeHorizon$}{
    Set $\contextAT{\adversarialAction} = \left(\rewardLargestPossible, \frac{\rewardLargestPossible}{2\radius\super{j}}\cdot\codeWord{i}\right)$ where $\codeWord{i} \in \sphericalCodedim{\dimension-1}\subseteq \realNumbers^{\dimension-1}$.

    Show contexts to $\ALG$ and receive contract $\contract{\timeslot}$.

    \If{$\contractSimple_{\timeslot}<\contractThreshold$}{
     $\hvecs_{\timeslot+1} \gets \hvecs_{\timeslot}$

    }
    
    \If{$\contractSimple_{\timeslot}\geq\contractThreshold$}{
    $i \gets i+1$

    $\hvecs_{\timeslot+1} \gets \hvecs_{\timeslot}\cap \set{\hvec = \left( \frac{1}{2}, \zeta\right) \condition \lrangle{\contextAT{\adversarialAction},\hvec}\leq \equalRevenueReward}$

    }

    \If{$i>\abs{\sphericalCodedim{\dimension-1}}$}{
    $i\gets 1$, $\radius\super{j+1} \gets \radius\super{j} \cdot \left( \frac{2\equalRevenueReward}{\rewardLargestPossible}-1\right)$

    $\hvecs_{\timeslot+1} \gets \set{\frac{1}{2}}\times \left(\radius\super{j+1}\cdot \mathbb{B}_{d-1}\right)$

    $j\gets j+1$

    }

    $\timeslot \gets \timeslot+1$
}
Determine $\hvecTrue = \argmax_{\hvec\in\hvecs_{\timeslot}}\lrangle{\contextAT{\adversarialAction},\hvec}$
\end{adversary}

We now prove that our adversarial construction compels any algorithm $\ALG$ to accumulate significant pessimistic Stackelberg regret. In parallel with the preceding analysis in \Cref{sec:lower-bound-cost-opt}, the proof of \Cref{thm:lower-bound-reward-opti} also relies on several auxiliary lemmas. We begin by characterizing the simplified instance including known actions that the principal faces in each round.

\subsubsection{Instance with Known Actions} 
\label{sec:instance-known-reward}
By the construction of the adversary. In each round, only the reward of $\adversarialAction$ is unknown to $\ALG$, and the action set $\actionsKnownT=\set{\nullAction,\knownOptAction,\restrictingAction}$ is totally known to $\ALG$. Moreover, $0=\rewardAT{\nullAction}<\rewardAT{\knownOptAction}<\rewardAT{\restrictingAction}$. The instance $\instanceKnown$ plays an important role in our proof. Let $0= \contractTOrder{0}\leq \contractTOrder{1} \leq \contractTOrder{2} \leq \contractTOrder{3}=1$ be the critical points of $\instanceKnown$. The following lemma provides the value of the first non-trivial critical point and a lower bound for the second.

\begin{lemma}
    $\contractTOrder{1}=\frac{1}{2}, \contractTOrder{2}\geq \frac{1}{2}+3\leftGapThreshold-24\leftGapThreshold^2$.
\end{lemma}
\begin{proof}
    $\contractTOrder{1}$ is the solution to 
    \[\contractTOrder{1} \rewardAT{\knownOptAction}- \costAT{\knownOptAction} = 0.\]
    By simple calculation, $\contractTOrder{1}=\frac{1}{2}$. And $\contractTOrder{2}$ is the solution to 
    \[\contractTOrder{2} \rewardAT{\knownOptAction}- \costAT{\knownOptAction} = \contractTOrder{2} \rewardAT{\restrictingAction}-\costAT{\restrictingAction}.\]
    Recall that $\rewardAT{\knownOptAction} = \frac{1}{2\eulerNumber}$, $\rewardAT{\restrictingAction}=\frac{3}{8\eulerNumber}+\frac{\leftGapThreshold}{4\eulerNumber} + \frac{1}{8\eulerNumber}\sqrt{4\leftGapThreshold^2+28\leftGapThreshold+1}$, $\costAT{\knownOptAction}=\frac{1}{4\eulerNumber}$, $\costAT{\restrictingAction}=\frac{1}{4\eulerNumber}+\frac{\leftGapThreshold}{\eulerNumber}$, we have 
    \begin{align*}
        \contractTOrder{2} &= \frac{\leftGapThreshold}{\eulerNumber}\left(1/ (\frac{\leftGapThreshold}{4\eulerNumber}+\frac{1}{8\eulerNumber}\sqrt{4\leftGapThreshold^2+28\leftGapThreshold+1}-\frac{1}{8\eulerNumber})\right)
        \\ &= \frac{8\leftGapThreshold}{2\leftGapThreshold+ \frac{4\leftGapThreshold^2+28\leftGapThreshold}{1+\sqrt{4\leftGapThreshold^2+28\leftGapThreshold +1}}}
        \\ &= \frac{1}{2}+3\leftGapThreshold -24\leftGapThreshold^2 +O(\leftGapThreshold^3)
    \end{align*}
    by simple calculation.
\end{proof}

\begin{lemma}
\label{lem:opt-act-reward}
$\optProfitFromA{\instanceKnown}{\knownOptAction} = \frac{1}{4\eulerNumber}$, $\optProfitFromA{\instanceKnown}{\restrictingAction} = \frac{1}{4\eulerNumber}-\frac{\leftGapThreshold}{2\eulerNumber}$.
\end{lemma}
\begin{proof}
Notice that proposing $\contractTOrder{1}$ can obtain the maximal utility induced by action $\knownOptAction$. Thus, $\optProfitFromA{\instanceKnown}{\knownOptAction} =(1-\contractTOrder{1})\rewardAT{\knownOptAction}= \frac{1}{4\eulerNumber}$. And proposing $\contractTOrder{2}$ can obtain the maximal utility induced by action $\restrictingAction$. Therefore, $\optProfitFromA{\instanceKnown}{\restrictingAction} = (1-\contractTOrder{2})\rewardAT{\restrictingAction}= \frac{1}{4\eulerNumber}-\frac{\leftGapThreshold}{2\eulerNumber}$.
\end{proof}

\subsubsection{Validity of Adversary~\ref{alg: adversary_optimism}}
\label{sec:valid-reward-opt}
We must also establish that our adversarial construction (\Cref{alg: adversary_optimism}) functions as intended. This involves showing that the agent is restricted to the known action set in \Cref{sec:instance-known-reward} and that the adversary maintains a valid representation of the hypothesis set.

\begin{lemma}\label{lem:valid adversary_optimism}
    Employing the \Cref{alg: adversary_optimism}, whatever the contract is, the agent will always take an action in $\actionsKnownT$. Moreover, \Cref{alg: adversary_optimism} is a valid adversary, which means that $\hvecs_{\timeslot}$ is consistent with the observed feedback of $\ALG$.
\end{lemma}

\begin{proof}
Consider the $\timeslot$-th round. Let $i^*, j^*$ be the $i,j$ at the beginning of the $\timeslot$-th round.  We analyze two cases: 
\begin{itemize}
\item Case $1$: $0 \leq\contractSimple_{\timeslot} < \contractThreshold$. One can easily verify that $\contractThreshold > \frac{1}{2} = \contractTOrder{1}$. Regardless of the true hidden vector $\hvecTrue  \in \hvecs_{\timeslot}$, when $\contractTOrder{1} \leq\contractSimple_{\timeslot} < \contractThreshold$, the action $\knownOptAction \in \actionsKnownT$ will be induced since $$  \lrangle{\contextAT{\knownOptAction},\hvecTrue} \cdot \contractSimple_{\timeslot} -\costAT{\knownOptAction}   \geq \max\left\{0,  \lrangle{\contextAT{\restrictingAction},\hvecTrue} \cdot  \contractSimple_{\timeslot} - \costAT{\restrictingAction} ,   \lrangle{\contextAT{\adversarialAction},\hvecTrue} \cdot \contractSimple_{\timeslot} - \costAT{\adversarialAction} \right\}.$$
When $0 \leq\contractSimple_{\timeslot} < \contractTOrder{1}$, the agent will choose the null action $\nullAction \in \actionsKnownT$ since other actions bring negative revenue. In both subcases, the principal cannot infer new information about the true hidden vector $\hvecTrue$. Thus, $\hvecs_{\timeslot+1} = \hvecs_{\timeslot}$, which aligns with the observed feedback from $\ALG$. 

\item Case $2$: $\contractThreshold \leq\contractSimple_{\timeslot} \leq 1$. If $i^*+1 \leq \abs{\sphericalCode}$, the adversary will restrict the true vector to $\hvecTrue  \in \hvecs_{\timeslot}\cap \set{\hvec = \left( \frac{1}{2}, \zeta\right) \condition \lrangle{\contextAT{\adversarialAction},\hvec}\leq \equalRevenueReward}$. One can verify $\rewardAT{\restrictingAction}  = \lrangle{\contextAT{\restrictingAction},\hvecTrue}= \equalRevenueReward \geq \lrangle{\contextAT{\adversarialAction},\hvecTrue} = \rewardAT{\adversarialAction}$ and $\costAT{\restrictingAction} =\costAT{\adversarialAction} $. Hence, action $\adversarialAction$ is fully dominated by action $\restrictingAction$. In addition, the principal can obtain the most information when offering the contract $\contractSimple_{\timeslot} = \contractTOrder{2}$. The observed feedback from $\ALG$ is in accordance with $\hvecs_{\timeslot+1} = \hvecs_{\timeslot}\cap \set{\hvec = \left( \frac{1}{2}, \zeta\right) \condition \lrangle{\contextAT{\adversarialAction},\hvec}\leq \equalRevenueReward}$. If $i^*+1 > \abs{\sphericalCode}$, then $i^*=\abs{\sphericalCode}$ and index code word $\codeWord{i}$ ranges over all elements of $\sphericalCode$. The adversary instead shrinks the radius $\radius\super{j^*}$ by a factor of $\frac{2\equalRevenueReward}{\rewardLargestPossible}-1$. Concretely, $\hvecs_{\timeslot+1} = \set{\frac{1}{2}}\times \left(\radius\super{j^*+1}\cdot \mathbb{B}_{d-1}\right)$.  Because for every given $1 \leq i \leq \abs{\sphericalCode}$, the constraint $\lrangle{\contextAT{\adversarialAction},\hvec}\leq \equalRevenueReward$ is equivalent to $\lrangle{ \codeWord{i},\zeta} \leq \left(\frac{2\equalRevenueReward}{\rewardLargestPossible}-1\right) \cdot \radius\super{j^*} = \radius\super{j^*+1}$. We have $\set{\frac{1}{2}}\times \left(\radius\super{j^*+1}\cdot \mathbb{B}_{d-1}\right) \subseteq \bigcap_{1 \leq i \leq \abs{\sphericalCode}} \set{\hvec = \left(\frac{1}{2},\zeta \right) \condition \lrangle{\codeWord{i},\zeta}\leq \radius\super{j^*+1} }$. Thus again action $\adversarialAction$ is fully dominated and the feedback is consistent. 

\end{itemize}
In each case, the principal’s observed feedback from 
$\ALG$ exactly matches the evolution of $\hvecs_{\timeslot} $ as described above, completing the proof.
\end{proof}

Just as in the cost-context setting, our construction must adhere to our modeling assumptions. Specifically, all generated context vectors must lie within the unit ball $\mathbb{B}_{d}$, and their corresponding rewards must be non-negative. The following lemma provides a sufficient condition on the size of the spherical code to ensure these properties hold.

\begin{lemma}[Boundedness and non-negativity]\label{lem:bounded context_optimism}
    If the spherical code $\sphericalCodedim{\dimension-1}$ used by \Cref{alg: adversary_optimism} satisfies $\abs{\sphericalCodedim{\dimension-1}}\geq 2 \cdot\timeHorizon \left(1-\frac{\equalRevenueReward}{\rewardLargestPossible}\right)$, then for all rounds, the following properties hold:
\begin{enumerate}
\item[\textnormal{(i)}] All context vectors generated by the adversary are contained in the unit ball $\mathbb{B}_{d}$.
\item[\textnormal{(ii)}] The rewards of all corresponding actions are non-negative.
\end{enumerate}
\end{lemma}
\begin{proof}
    By \Cref{alg: adversary_optimism}, $\radius\super{j}$ will shrink at most $\frac{T}{\abs{\sphericalCodedim{\dimension-1}}}$ times. Thus, it always holds \[\radius\super{j}\geq \frac{\sqrt{3}}{2}\left(\frac{2\equalRevenueReward}{\rewardLargestPossible}-1\right)^j\geq \frac{\sqrt{3}}{2}\left(1-2 \cdot \frac{\rewardLargestPossible-\equalRevenueReward}{\rewardLargestPossible}\right)^{\frac{\timeHorizon}{\abs{\sphericalCodedim{\dimension-1}}}}\geq \frac{\sqrt{3}}{2\eulerNumber}.\]

    Therefore, we have $ \lim_{ \leftGapThreshold \to 0} \abs{\contextAT{\adversarialAction{1}}} \leq \lim_{ \leftGapThreshold \to 0} \abs{\left(\rewardLargestPossible, \frac{\rewardLargestPossible}{2 \cdot \frac{\sqrt{3}}{2\eulerNumber}} \right)} <1 $. This implies, when $\leftGapThreshold$ is small enough, we have $\abs{\contextAT{\adversarialAction{1}}} < 1$.
    In addition, one can check that $$\rewardAT{\adversarialAction} = \lrangle{\contextAT{\adversarialAction},\hvecTrue} \geq \min_{\hvec\in \hvecs_{\timeslot}} \lrangle{\contextAT{\adversarialAction},\hvec} \geq \frac{\beta}{2} - \frac{\beta \radius\super{j_\timeslot}}{2 \radius\super{j_\timeslot}} = 0,$$
where $j_\timeslot$ is the value of $j$ in $\timeslot$-th round.
\end{proof}

\subsubsection{Lower Bound Analysis}
\label{sec:analysis-reward-opt}

For the reward-context setting, we can also determine the exact value of the principal's optimistic utility. The following lemma shows that, under a specific geometric condition on the spherical code used by our adversary, this utility remains constant in every round.

\begin{lemma}\label{lem:optimism_utility}
    Assuming the input spherical code is with minimal angle $\arccos \left( \frac{2\equalRevenueReward}{\rewardLargestPossible}-1\right)$. For every $\timeslot$-th round, the best achievable principal's utility given the current knowledge is
$$\max_{\hvec\in\hvecs_{\timeslot}} \optProfit{\instance{\actions_{\timeslot}}{\hvec}} = \frac{1}{4\eulerNumber}+\frac{\leftGapThreshold}{2\eulerNumber}.$$
\end{lemma}
\begin{proof}
    Let $i^*, j^*$ be the $i,j$ at the beginning of the $\timeslot$-th round. We first show that $\left(\frac{1}{2},\radius\super{j^*} \cdot \codeWord{i^*}\right)\in \hvecs_{\timeslot}$. By the design of the adversary, 
    \[ \hvecs_{\timeslot} = \set{\frac{1}{2}}\times \left(\radius\super{j^*}\mathbb{B}_{\dimension-1}\bigcap \left(\bigcap_{i=1}^{i^*-1} \set{\zeta \condition \frac{\rewardLargestPossible}{2}+ \lrangle{\zeta, \frac{\rewardLargestPossible}{2\radius\super{j^*}}\cdot \codeWord{i}}\leq \equalRevenueReward}\right)\right).\]
    Since $\codeWord{i^*}\in \mathbb{B}_{\dimension-1}$, we have $\left(\frac{1}{2},\radius\super{j^*} \cdot \codeWord{i^*}\right)\in \set{\frac{1}{2}}\times \radius\super{j^*}\mathbb{B}_{d-1}$. Since the minimal angle of $\sphericalCodedim{\dimension-1}$ is $\arccos \left(\frac{2\equalRevenueReward}{\rewardLargestPossible}-1\right)$, we have 
    \[ \frac{\rewardLargestPossible}{2}+\lrangle{\radius\super{j^*}\codeWord{i^*},\frac{\rewardLargestPossible}{2\radius\super{j^*}}\codeWord{i}} \leq \frac{\rewardLargestPossible}{2}+\frac{\rewardLargestPossible}{2}\cdot \left( \frac{2\equalRevenueReward}{\rewardLargestPossible} -1\right) \leq \equalRevenueReward.\]
    Thus, $\left(\frac{1}{2},\radius\super{j^*} \cdot \codeWord{i^*}\right)\in \hvecs_{\timeslot}$.
    
     Therefore, $ \max_{\hvec\in\hvecs_{\timeslot}} \lrangle{\contextTA{\timeslot}{\adversarialAction}, \hvec}\geq \lrangle{\contextTA{\timeslot}{\adversarialAction}, \left(\frac{1}{2},\radius\super{j^*} \cdot \codeWord{i^*}\right)}= \rewardLargestPossible$. On the other hand, since $\hvecs_{\timeslot} \subseteq \set{\frac{1}{2}}\times \radius\super{j^*}\mathbb{B}_{\dimension-1}$, we have $ \max_{\hvec\in\hvecs_{\timeslot}} \lrangle{\contextTA{\timeslot}{\adversarialAction},\hvec}\leq \rewardLargestPossible$. Thus, $\max_{\hvec\in\hvecs_{\timeslot}} \lrangle{\contextTA{\timeslot}{\adversarialAction},\hvec}=\rewardLargestPossible$. Given $\hvec \in \hvecs_{\timeslot}$, we consider two cases. 
     \begin{itemize}
     \item Case $1$: if $\rewardA{\adversarialAction}_{\timeslot}=\lrangle{\contextTA{\timeslot}{\adversarialAction},\hvec} \leq \equalRevenueReward$, action $\adversarialAction$ is fully dominated by $\restrictingAction$. Therefore, $\optProfit{\instance{\actions_{\timeslot}}{\hvec}} = \optProfit{\instanceKnown} = \frac{1}{4 \eulerNumber}$.
     \item Case $2$: if $\equalRevenueReward < \rewardA{\adversarialAction}_{\timeslot}=\lrangle{\contextTA{\timeslot}{\adversarialAction},\hvec} \leq \rewardLargestPossible$, by simple calculation, we have $$\optProfit{\instance{\actions_{\timeslot}}{\hvec}} = \max \left\{\frac{1}{4\eulerNumber},\left(1- \frac{\frac{\leftGapThreshold}{\eulerNumber}}{\rewardA{\adversarialAction}_{\timeslot} - \frac{1}{2 \eulerNumber}} \right) \cdot \rewardA{\adversarialAction}_{\timeslot} \right\} \overset{(a)}{\leq} \left(1- \frac{\frac{\leftGapThreshold}{\eulerNumber}}{\rewardLargestPossible - \frac{1}{2 \eulerNumber}} \right) \cdot \rewardLargestPossible \overset{(b)}{=} \frac{1}{4\eulerNumber}+\frac{\leftGapThreshold}{2\eulerNumber},$$
     where (a) becomes equality if $\rewardA{\adversarialAction}_{\timeslot}=\lrangle{\contextTA{\timeslot}{\adversarialAction},\hvec} = \rewardLargestPossible$, and (b) is by the definition of $\beta$.
     \end{itemize}
      Thus, the best achievable utility
given current knowledge $\max_{\hvec\in\hvecs_{\timeslot}} \optProfit{\instance{\actions_{\timeslot}}{\hvec}}=\frac{1}{4\eulerNumber}+\frac{\leftGapThreshold}{2\eulerNumber}$. 

\end{proof}

The strategic design of our reward-unknown adversary has a direct and costly consequence for the principal. We will now show that, much like in the cost-unknown scenario, the principal is compelled to suffer a non-trivial amount of regret in each and every round. This per-round performance penalty is quantified in the lemma below.

\begin{lemma}\label{lem:optimism_regret_reward}
     In any round $\timeslot$, regardless of the contract $\contract{\timeslot}$ proposed by the principal ($\ALG$), the resulting instantaneous pessimistic Stackelberg regret satisfies:
     \[\regretPessimistic_\timeslot(\ALG)\geq \frac{1}{2\eulerNumber}\leftGapThreshold.\]
\end{lemma}
\begin{proof}
By \Cref{lem:valid adversary_optimism}, whatever  the contract $\contract{\timeslot}$ proposed by $\ALG$ is, the agent will always take an action in $\actionsKnownT$. Then combining with \Cref{lem:opt-act-reward} and \Cref{lem:optimism_utility}, we obtain

    \begin{align*}
        \regretPessimistic_\timeslot(\ALG) &=\max_{\hvec\in\hvecsPess_{\timeslot}} \optProfit{\instance{\actions_{\timeslot}}{\hvec}} -\revenuePrincipalCon_\timeslot(\contract{\timeslot})  \\ &\overset{(a)}{\geq} \max_{\hvec\in\hvecs_{\timeslot}} \optProfit{\instance{\actions_{\timeslot}}{\hvec}} -\revenuePrincipalCon_\timeslot(\contract{\timeslot}) \\
        & \geq \max_{\hvec\in\hvecs_{\timeslot}} \optProfit{\instance{\actions_{\timeslot}}{\hvec}}  -\optProfit{\instanceKnown}
        \\&= \frac{1}{4\eulerNumber}+\frac{\leftGapThreshold}{2\eulerNumber}- \frac{1}{4\eulerNumber} = \frac{\leftGapThreshold}{2\eulerNumber},
    \end{align*}
    where inequality (a) is because $\hvecs_{\timeslot}$ is consistent with the observed feedback by \Cref{lem:valid adversary_optimism}, leading to $\hvecs_\timeslot\subseteq \hvecsPess_\timeslot$.
\end{proof}

With all the necessary components in place, we can now proceed with the proof of \Cref{thm:lower-bound-reward-opti}. The argument integrates the results from the preceding lemmas to derive the final lower bound on cumulative regret.

\begin{proof}[\bf{Proof of \Cref{thm:lower-bound-reward-opti}}]
First, we set $\leftGapThreshold = \min\{\frac{1}{8}C^{\frac{1}{\dimension}} (\dimension-1)^{\frac{1}{2\dimension}} \timeHorizon^{-\frac{1}{\dimension}},\frac{1}{8}\}$. Notice that $1-\left(\frac{2\equalRevenueReward}{\rewardLargestPossible}-1 \right)=2- 2 \cdot \frac{3+2\leftGapThreshold+\sqrt{4\leftGapThreshold^2+28\leftGapThreshold+1}}{3+6\leftGapThreshold+\sqrt{36\leftGapThreshold^2+20\leftGapThreshold+1}} = 32\leftGapThreshold^2-640\leftGapThreshold^3+O(\leftGapThreshold^4)$. By \Cref{lem:code lower bound}, we have
    $$\abs{\sphericalCodedim{\dimension-1}} \geq C \frac{\left(\frac{2\equalRevenueReward}{\rewardLargestPossible} -1\right) \sqrt{d-1}}{ \sin^{d-2} \left(\arccos{\left(\frac{2\equalRevenueReward}{\rewardLargestPossible} -1\right)}\right)}   \geq C\frac{(1-32\leftGapThreshold^2)\sqrt{\dimension-1}}{(8\leftGapThreshold)^{d-2}} \geq 32\timeHorizon \leftGapThreshold^2 \geq 2 \cdot \timeHorizon \left(1- \frac{\equalRevenueReward}{\rewardLargestPossible}\right).$$
    Therefore, the conditions of \Cref{lem:bounded context_optimism} are satisfied and our construction of adversary is valid.
By \Cref{lem:optimism_regret_reward}, \Cref{alg: adversary_optimism} goes through $T$ rounds and the cumulative regret 
 \[ \regretPessimistic(\ALG)\geq \frac{\leftGapThreshold}{2\eulerNumber} \cdot \timeHorizon.\]
 Thus, we complete the proof.

\end{proof}

\section{Conclusion and Future Directions}
\label{sec:conclusion}
This paper pioneers the study of contextual learning in the general principal-agent game. Our central finding is that the degeneracy of the agent's action set fundamentally increases the complexity of the learning problem, rendering $O(\log\log \timeHorizon)$-type regret bounds impossible. This work opens up several promising avenues for future research:

\xhdr{Optimal classic Stackelberg regret bound.} Though our algorithm is asymptotically optimal in the sense of pessimistic Stackelberg regret, there still exists a gap in the classic Stackelberg regret bound. It is not clear whether one could prove a classic Stackelberg regret lower bound similar to the pessimistic regret lower bound in this paper. 
On the other hand, since the regret bound in most of the previous works is essentially the pessimistic Stackelberg regret bound, achieving an $O(\sqrt{T})$ upper bound for the classic regret will likely require novel algorithmic and analytical approaches that do not rely on immediate, round-by-round regret estimation. In addition, it would be interesting to explore whether our analysis framework with information width can be applied to the classic contextual pricing model and obtain the optimal regret of $O_{\dimension}(\log\log T)$.

\xhdr{Stochastic contexts.} We suggest an inquiry into stochastic contexts. Our current lower bound construction heavily relies on adversarially selecting the context each round. Notably, to the best of our knowledge, the adversarial context is considered in all prior work in the contextual pricing literature \citep[e.g.,][]{LLV-18,LS-18,LLS-21,KLPS-21}. A compelling open problem is to determine if assuming an i.i.d.\ context sampled from an unknown distribution would circumvent this hardness result, potentially permitting the achievement of more favorable $O(\log \timeHorizon)$ or $O(\log\log \timeHorizon)$ regret bounds.

\xhdr{Stochastic outcomes and hidden action model.} Our work follows the line of works on contextual pricing, assuming deterministic feedback. Another significant direction for future work is to consider the contextual learning problem of more general principal-agent models incorporating stochastic outcomes and hidden actions, as seen in settings like contract design. As this model is a superset of our own, our lower bound holds directly, highlighting the inherent challenges in this broader class of problems.

\section*{Acknowledgments}

We are grateful to the anonymous reviewers of SODA 2026 for their time and effort in carefully reading this paper. 
\bibliographystyle{plainnat}
\bibliography{mybibfile.bib}

\appendix

\section{Discussion on Practical Applications}
\label{apx:practical application}
Our model provides a unified foundation for studying sequential decision-making and learning in principal–agent interactions under contextual uncertainty. The cost-context setting captures scenarios in which the principal faces uncertainty about agents' effort costs while action rewards are known. This includes crowdsourcing markets: the platform repeatedly offers payment contracts over multiple tasks to sequentially arriving workers characterized by cost-related context features; each worker chooses a task based on private, task-specific effort cost. The platform observes only the chosen task and the realized outcome, and learns over time how the observed context predicts hidden costs.

The reward-context setting, by contrast, captures situations in which the principal can estimate the cost for an agent to perform each action—such as required time, resources, or effort—yet remains uncertain about the resulting benefit. For example, in product development or R\&D, the sponsor (a government agency, firm, or university) issues repeated funding calls across multiple projects and offers outcome-contingent contracts. Teams arrive over time with contextual features on execution capacity, risk management, and impact potential; each submits a proposal and bids for a project based on private, project-specific success assessments, and the sponsor learns from outcomes how context predicts rewards. 
\section{Lower Bound against Optimum-in-Hindsight Benchmark}
\label{apx:classic lower bound}

In this section, we shift our focus from the optimistic benchmark to the more challenging optimum-in-hindsight benchmark. To establish a standard Stackelberg regret lower bound in the cost-context setting and reward-context setting, we adapt the adversarial construction from the previous section. The key modifications are  
\begin{itemize}
    \item Introduction of an Exploration-Detection Threshold ($\determineThreshold$): We introduce a parameter that allows the adversary to detect rounds in which the principal engages in exploration (i.e., plays more aggressively than myopic optimization would suggest).
    \item Inclusion of Early Halting Conditions: We equip the adversary with specific conditions under which it can terminate the interaction prematurely. This prevents a patient principal from eventually uncovering the true model through prolonged exploration.
\end{itemize}
Our analysis proceeds in two parts. We first establish the Stackelberg regret lower bound for the cost-context setting. Subsequently, we will address the reward-context setting.
\subsection{Lower Bound with Unknown Cost}
\label{sec:lower-cost-true}

To establish a standard Stackelberg regret lower bound, we now adapt our adversarial strategy for the cost-context setting in \Cref{sec:lower-bound-cost-opt}. Crucially, we reuse the same underlying instance as in the analysis of the pessimistic Stackelberg regret. The key change lies not in the instance itself, but in the adversary's algorithm, which we modify to be effective against a principal optimizing for the true outcome. The details of this revised adversarial procedure are presented below.

\xhdr{Adversary.} The adversary retains the same underlying instance and notations as defined in \Cref{sec:lower-bound-cost-opt}. However, its strategy is now enhanced with a mechanism to counter principals that may act cautiously. The core of this strategy is a dichotomy based on the principal's exploratory behavior, controlled by a counter 
$k$ and a threshold $\determineThreshold$:
\begin{itemize}
    \item Case 1: Passive Principal (Insufficient Exploration). The adversary uses the counter $k$ to track the number of consecutive rounds in which the principal avoids aggressive exploration. If this count reaches the threshold (i.e., $\determineThreshold$), the adversary executes its endgame:
    \begin{itemize}
        \item It halts the interaction prematurely.
        \item It finalizes its choice of the true hidden vector $\hvecTrue$. This choice is made strategically to ensure that the actions taken by the principal during these last "passive" rounds were demonstrably suboptimal, thereby crystallizing their regret.
    \end{itemize}
    \item Case 2: Active Principal (Sufficient Exploration). Conversely, if the principal explores aggressively enough to prevent the halting condition from being met, they are funneled into the adversary's primary regret-inducing mechanism (inherited from the optimistic benchmark construction). As we have shown, this mechanism is itself designed to generate significant cumulative regret.
\end{itemize}
In either case, the principal is forced to incur a large regret. The formal construction is presented in \Cref{alg: adversary_cost}.

\begin{adversary}[t!]
\caption{\textsc{Adversary with Cost Context}}
\label[adversary]{alg: adversary_cost}
\SetAlgoLined
\KwData{$\ALG, \timeHorizon, \dimension, \leftGapThreshold, \determineThreshold$, spherical code $\sphericalCodedim{\dimension-1}$ }

Initialize $\contextAT{\knownOptAction} = (\frac{1}{2\eulerNumber},0,\ldots,0), \contextAT{\restrictingAction} = \left(\frac{1}{2\eulerNumber}+\frac{\leftGapThreshold}{4\eulerNumber } \cdot \frac{2+5 \leftGapThreshold}{1+ \leftGapThreshold},0,\ldots,0 \right), \rewardAT{\knownOptAction}=\frac{1}{2\eulerNumber}, \rewardAT{\restrictingAction}=\rewardAT{\adversarialAction}=\frac{1}{2\eulerNumber}+\frac{\leftGapThreshold}{2\eulerNumber}, \forall \timeslot\geq 1$

$\timeslot\gets 1, i\gets 1, j\gets 1, k\gets 0$

Initialize $\radius\super{1} = \frac{\sqrt{3}}{2}$, $\hvecs_{1}=\set{\frac{1}{2}}\times  \left( \radius\super{1} \cdot \mathbb{B}_{d-1} \right)$

Arbitrarily index code words $\sphericalCodedim{\dimension-1}=\set{\codeWord{1},\codeWord{2},\ldots,\codeWord{\abs{\sphericalCodedim{\dimension-1}}}}$

$\contractThreshold\gets \frac{2+3 \cdot  \leftGapThreshold}{4(1+ \leftGapThreshold)}$

\While{$\timeslot\leq \timeHorizon$}{
    Set     $\contextAT{\adversarialAction}  =  \left(\frac{1}{2\eulerNumber}+\frac{ \leftGapThreshold}{ 4\eulerNumber} \frac{2+3 \cdot  \leftGapThreshold}{1+ \leftGapThreshold}+\frac{1}{2\eulerNumber(1+ \leftGapThreshold)},   \frac{\codeWord{i}}{4\eulerNumber \left(1+\leftGapThreshold \right)\radius\super{j}} \right)$
    
    Show contexts to $\ALG$ and receive contract $\contract{\timeslot}$.

    \If{$\contractSimple_{\timeslot}<\contractThreshold$}{
     $\hvecs_{\timeslot+1} \gets \hvecs_{\timeslot}$
     
     $k\gets k+1$
     
     \If{$k=\determineThreshold$}{
        Determine $\hvecTrue = \argmin_{\hvec\in\hvecs_{\timeslot}}\lrangle{\contextAT{\adversarialAction},\hvec}$ and \textbf{halt}
     }
    }
    
    \If{$\contractSimple_{\timeslot}\geq\contractThreshold$}{

    $\hvecs_{\timeslot+1} \gets \hvecs_{\timeslot}\cap \set{\left(\frac{1}{2},\zeta\right) \condition \lrangle{\codeWord{i},-\zeta}\leq \radius\super{j} \left(1-\leftGapThreshold^2\right)}$

    $i \gets i+1$, $k\gets 0$
    }

    \If{$i>\abs{\sphericalCodedim{\dimension-1}}$}{
    $i\gets 1$, $\radius\super{j+1} \gets \radius\super{j} \cdot \left(1-\leftGapThreshold^2\right)$

    $\hvecs_{\timeslot+1} \gets \set{\frac{1}{2}}\times  \left(\radius\super{j+1} \cdot \mathbb{B}_{d-1} \right)$

    $j\gets j+1$

    \If{$j\geq \frac{\timeHorizon}{\abs{\sphericalCodedim{\dimension-1}}\determineThreshold}$}{
            Arbitrarily determine a $\hvecTrue\in \hvecs_{\timeslot+1}$ and \textbf{halt}
        }
    }

    $\timeslot \gets \timeslot+1$
}
\end{adversary}

Next, we prove that our adversarial construction guarantees a significant Stackelberg regret. This is formally stated in the theorem below.

\begin{theorem}\label{thm:lower-bound-true}
    For contextual principal-agent game with cost contexts, let $d$ be the dimension of the context, for any algorithm $\ALG$, there is an instance making the algorithm suffer $\Omega(T^{\frac{1}{2}-\frac{1}{2d}})$ Stackelberg regret.
\end{theorem}

Before proceeding to the proof of \Cref{thm:lower-bound-true}, we first present the necessary auxiliary lemmas. The foundational step, consistent with our previous analysis, is to confirm the well-behaved nature of our new adversary (\Cref{alg: adversary_cost}). The first lemma accomplishes this by establishing its validity and action-restricting properties.

\begin{lemma}\label{lem:valid adversary_cost}
    Facing the \Cref{alg: adversary_cost}, whatever the contract is, the agent will always take an action in $\actionsKnownT$. Moreover, \Cref{alg: adversary_cost} is a valid adversary, which means that $\hvecs_{\timeslot}$ is consistent with the observed feedback of $\ALG$.
\end{lemma}

\begin{proof}
    The core mechanics of the adversary in \Cref{alg: adversary_cost}—specifically, how it constructs uncertainty sets and determines outcomes for given contracts—are identical to those in \Cref{alg: adversary_cost_optimism}. The only modification is the addition of early halting conditions.

These halting conditions do not affect the validity of the adversary within any given round, as they only determine whether the interaction continues to the next round. Therefore, the arguments proving that the agent's action is confined to 
 and that the uncertainty set remains consistent are the same as those in the proof of \Cref{lem:valid adversary_cost_optimism}. The original proof thus applies directly.
\end{proof}

As a foundational check, we must also ensure that our adversarial construction remains physically plausible throughout the interaction. This requires confirming that all generated context vectors lie within the unit ball $\mathbb{B}_{d}$ and that their associated costs are always non-negative. The following lemma establishes a sufficient condition on the spherical code's size to guarantee these properties hold even with the new halting mechanism.

\begin{lemma}\label{lem:bounded context_cost}
    Let $\sphericalCodedim{\dimension-1}$ be the spherical code used by our adversarial construction (\Cref{alg: adversary_cost}). If $\sphericalCodedim{\dimension-1}$ satisfies the condition $\abs{\sphericalCodedim{\dimension-1}}\geq \frac{\timeHorizon}{\determineThreshold}\leftGapThreshold^2$, then before \Cref{alg: adversary_cost} halts, the following properties hold:
\begin{enumerate}
\item[\textnormal{(i)}] All context vectors generated by the adversary are contained in the unit ball $\mathbb{B}_{d}$.
\item[\textnormal{(ii)}] The costs of all corresponding actions are non-negative.
\end{enumerate}
\end{lemma}
\begin{proof}
The proof structure mirrors that of \Cref{lem:bounded context_cost_optimism}. The key difference lies in establishing the lower bound for the radius under the new halting condition. By the halting mechanism in \Cref{alg: adversary_cost}, $\radius\super{j}$ will shrink at most $\frac{T}{\abs{\sphericalCodedim{\dimension-1}}\determineThreshold}$ times. Thus, before \Cref{alg: adversary_cost} halts, it always holds \[\radius\super{j}\geq \frac{\sqrt{3}}{2}\left(1-\leftGapThreshold^2\right)^j\geq \frac{\sqrt{3}}{2}\left(1-\leftGapThreshold^2\right)^{\frac{\timeHorizon}{\abs{\sphericalCodedim{\dimension-1}}\determineThreshold}}\geq \frac{\sqrt{3}}{2\eulerNumber}.\]
 This confirms that the radius 
 maintains the same lower bound of 
 as in the previous analysis. Since the lower bound on the radius remains unchanged, and the definitions of the context vectors and costs as functions of this radius are also identical to those in the previous setting, the remainder of the proof follows directly from the arguments in \Cref{lem:bounded context_cost_optimism}. Specifically, the same logic guarantees that context vectors are bounded within $\mathbb{B}_{d}$ and costs are non-negative. This completes the proof.
\end{proof}

To facilitate the subsequent regret analysis, a more precise estimate of the optimistic contract threshold, $\contractThreshold$, is required. The following lemma establishes a tight lower bound on the difference between $\contractThreshold$ and the baseline value of $\contractTOrder{1}=\frac{1}{2}$.

\begin{lemma}
    $\contractThreshold-\frac{1}{2}\geq \frac{\leftGapThreshold}{4}-\frac{\leftGapThreshold^2}{4}$.
\end{lemma}
\begin{proof}
    By definition, we have
        $\contractThreshold-\frac{1}{2} = \frac{\leftGapThreshold}{4(1+\leftGapThreshold)} = \frac{\leftGapThreshold}{4}-\frac{\leftGapThreshold^2}{4} +O(\leftGapThreshold^3).$
\end{proof}

Our Stackelberg regret analysis requires a more nuanced approach than the one used for the pessimistic Stackelberg regret. Specifically, we must analyze the principal's regret by conditioning on their chosen contract, $\contract{\timeslot}$. We therefore proceed with a case analysis, dividing the outcome based on whether $\contract{\timeslot}$ exceeds the optimistic threshold $\contractThreshold$.

The following lemma establishes the regret lower bound for the first case, where the principal acts aggressively.

\begin{lemma}\label{lem:case 2_cost}
     At any $\timeslot$-th round, if $\ALG$ propose $\contract{\timeslot}\geq \contractThreshold$, then the immediate regret of this round satisfies
     \[\regretImmediate(\ALG)\geq \frac{\leftGapThreshold}{8\eulerNumber} - \frac{\leftGapThreshold^2}{8\eulerNumber}.\]
\end{lemma}
\begin{proof}
    If $\contract{\timeslot}\ge \contractThreshold$, the adversary will shrink the possible hidden vector such that $\lrangle{\codeWord{i},-\hvecTrue} \leq \radius\super{j} \cdot \left(1-\leftGapThreshold^2 \right)$, that is $\costAT{\adversarialAction}=\lrangle{\contextAT{\adversarialAction},\hvecTrue}\geq \frac{1}{4\eulerNumber}+\frac{\leftGapThreshold}{8\eulerNumber } \cdot \frac{2+5 \leftGapThreshold}{1+ \leftGapThreshold}$ by \Cref{alg: adversary_cost}. Note that $\costAT{\restrictingAction} = \frac{1}{4\eulerNumber}+\frac{\leftGapThreshold}{8\eulerNumber } \cdot \frac{2+5 \leftGapThreshold}{1+ \leftGapThreshold}$ and $\rewardAT{\restrictingAction} = \rewardAT{\adversarialAction}$. Thus, action $\adversarialAction$ is fully degenerated, i.e., it can not be induced whatever $\contract{\timeslot}$ is. Then, the instance is equivalent to $\instanceKnown$ when $\contract{\timeslot}\geq \contractThreshold$.

    If $\contract{\timeslot}\in [\contractTOrder{1},\contractTOrder{2})$, then $\knownOptAction$ is induced and 
    \begin{align*}
        \regretImmediate(\ALG) &\geq \optProfit{\instanceKnown}-\revenuePrincipal 
        \\&= \frac{1}{4\eulerNumber}- (1-\contract{\timeslot}) \rewardAT{\knownOptAction} 
        \\& \geq \frac{1}{4\eulerNumber}-(1-\contractThreshold)\rewardAT{\knownOptAction}
        \\&\geq  \frac{1}{4\eulerNumber}- (\frac{1}{2}-\frac{\leftGapThreshold}{4}+\frac{\leftGapThreshold^2}{4})\frac{1}{2\eulerNumber}
        \\&= \frac{\leftGapThreshold}{8\eulerNumber} - \frac{\leftGapThreshold^2}{8\eulerNumber}.
    \end{align*}

    If $\contract{\timeslot}\in [\contractTOrder{2},1)$, then $\restrictingAction$ is induced and 
    \begin{align*}
        \regretImmediate(\ALG) &\geq  \optProfit{\instanceKnown}-\revenuePrincipal 
        \\& \geq \frac{1}{4\eulerNumber} -\optProfitFromA{\instanceKnown}{\restrictingAction}
        \\& = \frac{1}{4\eulerNumber} - \left(\frac{1}{4\eulerNumber}-\frac{\leftGapThreshold}{8\eulerNumber}\right) = \frac{\leftGapThreshold}{8\eulerNumber}.
    \end{align*}
\end{proof}

We now turn to the second case in our analysis: when the principal acts passively for a sustained period, triggering the adversary's halting mechanism. In this scenario, we will show that the adversary strategically ends the game to crystallize a significant amount of regret for the principal.

\begin{lemma}\label{lem:case 1_cost}
    Assuming the input spherical code is with minimal angle $\arccos\left(1- \leftGapThreshold^2 \right)$. Then there exists $\timeslot^*$ such that $k$ reaches $\determineThreshold$ at $\timeslot$-th round. Then the cumulative regret until $\timeslot^*$-th round satisfies
    \[\sum_{\timeslot=\timeslot^*-\determineThreshold+1}^{\timeslot^*}\regretImmediate(\ALG) \geq \frac{\leftGapThreshold}{8\eulerNumber}\determineThreshold.\]
\end{lemma}
\begin{proof}
    By the design of \Cref{alg: adversary_cost}, there are consecutive $\determineThreshold$ rounds where $\ALG$ propose $\contract{\timeslot}<\contractThreshold$. That is because proposing $\contract{\timeslot}\geq \contractThreshold$ makes the adversary reset $k$. 

    Let $i^*, j^*$ be the $i,j$ at the beginning of the $\timeslot^*$-th round. We first show that $\left(\frac{1}{2},-\radius\super{j^*} \cdot \codeWord{i^*}\right)\in \hvecs_{\timeslot^*}$. By the design of the adversary, 
    \[ \hvecs_{\timeslot^*} = \set{\frac{1}{2}}\times \left(\left(\radius\super{j^*} \cdot \mathbb{B}_{d-1} \right) \bigcap \left(\bigcap_{i=1}^{i^*-1} \set{\zeta \condition \lrangle{-\zeta,  \codeWord{i}}\leq \radius\super{j^*}(1-\leftGapThreshold^2)}\right)\right).\]
    Notice that $\codeWord{i^*}\in \mathbb{B}_{\dimension-1}$, there holds $\left(\frac{1}{2},-\radius\super{j^*} \cdot \codeWord{i^*}\right)\in \set{\frac{1}{2}}\times \left( \radius\super{j^*} \cdot \mathbb{B}_{d-1}\right)$. Since the minimal angle of $\sphericalCodedim{\dimension-1}$ is $\arccos (1-\leftGapThreshold^2)$, we have 
    \[ \lrangle{-(-\radius\super{j^*} \cdot \codeWord{i^*}),  \codeWord{i}}\leq \radius\super{j^*}(1-\leftGapThreshold^2) .\]
    Thus, $\left(\frac{1}{2},-\radius\super{j^*} \cdot \codeWord{i^*}\right)\in \hvecs_{\timeslot^*}$.
    
     Therefore, $\lrangle{\contextTA{\timeslot^*}{\adversarialAction}, \hvecTrue}\leq \lrangle{\contextTA{\timeslot^*}{\adversarialAction}, \left(\frac{1}{2},-\radius\super{j^*} \cdot \codeWord{i^*}\right)}= \frac{1}{4\eulerNumber}+\frac{\leftGapThreshold}{8\eulerNumber} \cdot \frac{2+3\leftGapThreshold}{1+\leftGapThreshold}$. On the other hand, since $\hvecTrue\in \set{\frac{1}{2}}\times \left(\radius\super{j^*} \cdot \mathbb{B}_{d-1} \right)$, we have $\lrangle{\contextTA{\timeslot^*}{\adversarialAction},\hvecTrue}\geq \frac{1}{4\eulerNumber}+\frac{\leftGapThreshold}{8\eulerNumber} \cdot \frac{2+3\leftGapThreshold}{1+\leftGapThreshold}$. Thus, $\costA{\adversarialAction}_{\timeslot^*}=\lrangle{\contextTA{\timeslot^*}{\adversarialAction},\hvecTrue}=\frac{1}{4\eulerNumber}+\frac{\leftGapThreshold}{8\eulerNumber} \cdot \frac{2+3\leftGapThreshold}{1+\leftGapThreshold}$. It's easy to check that $\optProfit{\instance{\actions_{\timeslot^*}}{\hvecTrue}}=\frac{1}{4\eulerNumber}+\frac{\leftGapThreshold}{8\eulerNumber}$. For any $\timeslot\in [\timeslot^*-\determineThreshold+1,\timeslot^*]$, the context $\contextTA{\timeslot^*}{\adversarialAction}$ is the same as $\contextAT{\adversarialAction}$, thus 
 \[\optProfit{\instance{\actions_{\timeslot}}{\hvecTrue}}=\frac{1}{4\eulerNumber}+\frac{\leftGapThreshold}{8\eulerNumber}, \quad \forall \timeslot \in [\timeslot^*-\determineThreshold, \timeslot^*].\] For any $\timeslot\in [\timeslot^*-\determineThreshold, \timeslot^*]$, since the action will always take an action in $\actionsKnownT$ by \Cref{lem:valid adversary_cost}. Then 
     \[\revenuePrincipal \leq \max_{\action\in\actionsKnownT}\optProfitFromA{\instance{\actions}{\hvecTrue}}{\action} \leq \max_{\action\in\actionsKnownT}\optProfitFromA{\instanceKnown}{\action}= \optProfit{\instanceKnown} = \frac{1}{4\eulerNumber}. \]
    Therefore, 
    \[\regretImmediate(\ALG) =\optProfit{\instance{\actions_{\timeslot}}{\hvecTrue}} -\revenuePrincipal\geq \frac{\leftGapThreshold}{8\eulerNumber}, \forall \timeslot\in  [\timeslot^*-\determineThreshold, \timeslot^*].\]
    Simply sum up $\regretImmediate(\ALG)$ over rounds $\timeslot\in  [\timeslot^*-\determineThreshold, \timeslot^*]$, we have 
    \[\sum_{\timeslot=\timeslot^*-\determineThreshold+1}^{\timeslot^*}\regretImmediate(\ALG) \geq \frac{\leftGapThreshold}{8\eulerNumber}\determineThreshold.\]
\end{proof}

Having established the necessary auxiliary lemmas, we are now equipped to present the proof of \Cref{thm:lower-bound-true}. The argument synthesizes our findings from the case analysis to derive the final regret lower bound.

\begin{proof}[\bf{Proof of \Cref{thm:lower-bound-true}}]
    First, we set $\leftGapThreshold = \min\{\frac{\sqrt{2}}{2}C^{\frac{1}{\dimension}} (\dimension-1)^{\frac{1}{2\dimension}} \timeHorizon^{-\frac{1}{2\dimension}},\frac{\sqrt{2}}{2}\}$ and $\determineThreshold=\sqrt{\timeHorizon}$. By \Cref{lem:code lower bound}, we have
     $\arccos\left(1- \leftGapThreshold^2 \right)$, whose size is lower-bounded as follows:
    $$\abs{\sphericalCodedim{\dimension-1}} \geq   C \frac{\left(1-\leftGapThreshold^2\right)\sqrt{d-1}}{\sin^{d-2} \left(\arccos{\left(1-\leftGapThreshold^2\right)}\right) }   \geq C\frac{(1-\leftGapThreshold^2)\sqrt{\dimension-1}}{(\sqrt{2}\leftGapThreshold)^{d-2}} \geq \sqrt{\timeHorizon} \leftGapThreshold^2=\frac{\timeHorizon}{\determineThreshold} \leftGapThreshold^2.$$
    Therefore, the conditions of \Cref{lem:bounded context_cost} are satisfied and our construction of adversary is valid.
    We claim that one of the following cases will happen:
    \begin{enumerate}
        \item There are at least $\frac{\timeHorizon}{\determineThreshold}$ rounds that $\ALG$ proposed $\contract{\timeslot}\geq \contractThreshold$ before the adversary halts.
        \item There exist some $\timeslot^*$ such that the counter $k$ reaches $\determineThreshold$.
    \end{enumerate}
    That is because, if $k$ never reaches $\determineThreshold$, then $\ALG$ never make contract which is less than $\contractThreshold$ for consecutive $\determineThreshold$ rounds, this directly imply that there are at least $\frac{\timeHorizon}{\determineThreshold}$ rounds for which $\ALG$ propose $\contract{\timeslot}\geq \contractThreshold$.

    If we fall into case $1$, the cumulative regret
    \[ \regret(\ALG)\geq \left(\frac{\leftGapThreshold}{8\eulerNumber} - \frac{\leftGapThreshold^2}{8\eulerNumber}\right)\frac{\timeHorizon}{\determineThreshold}  \]
    by \Cref{lem:case 2_cost}.

    If we fall into case $2$, the cumulative regret 
    \[\regret(\ALG)\geq \sum_{\timeslot=\timeslot^*-\determineThreshold+1}^{\timeslot^*}\regretImmediate(\ALG) \geq \frac{\leftGapThreshold}{8\eulerNumber}\determineThreshold  \]
    by \Cref{lem:case 1_cost}.\\
    Combining the results of both two cases, we complete the proof.

\end{proof}

\subsection{Lower Bound with Unknown Reward}
\label{sec:lower-reward-true}

Similarly, our adversarial construction against the optimum-in-hindsight benchmark in the reward-context setting also relies on the same underlying instance in \Cref{sec:lower-bound-reward-opt}. The proof structure parallels with that of the cost-context setting.

\xhdr{Adversary.} 
For the reward-context setting, we again adapt the adversary from the optimistic benchmark analysis in \Cref{sec:lower-bound-reward-opt}. The constructed instance is identical, but the algorithm incorporates a new halting rule. This rule is also governed by a counter $k$, which tracks consecutive non-exploratory rounds, and a threshold $\determineThreshold$.
The adversary's logic is straightforward and similar as that of the cost-context setting: if $k$ reaches $\determineThreshold$, the game terminates, and a large regret is forced over the last rounds. Otherwise, if the principal explores consistently, they incur regret through the baseline mechanism inherited from the optimistic setting.
The behavior of our constructed adversary is depicted in \Cref{alg: adversary}.

\begin{adversary}[t!]
\caption{\textsc{Adversary with Reward Context}}
\label[adversary]{alg: adversary}
\SetAlgoLined
\KwData{$\ALG, \timeHorizon, \dimension, \leftGapThreshold, \determineThreshold$, spherical code $\sphericalCodedim{\dimension-1}$ }

Initialize $\contextAT{\knownOptAction} = (\frac{1}{\eulerNumber},0,\ldots,0), \costAT{\knownOptAction}=\frac{1}{4\eulerNumber}, \contextAT{\restrictingAction} = \left(2\equalRevenueReward,0,\ldots,0\right), \costAT{\restrictingAction}=\costAT{\adversarialAction}=\frac{1}{4\eulerNumber}+\frac{1}{\eulerNumber}\leftGapThreshold, \forall \timeslot\geq 1$

$\timeslot\gets 1, i\gets 1, j\gets 1, k\gets 0$

Initialize $\radius\super{1} = \frac{\sqrt{3}}{2}$, $\hvecs_{1}=\set{\frac{1}{2}}\times \left(\radius\super{1} \cdot \mathbb{B}_{d-1}\right)$ 

Arbitrarily index code words $\sphericalCodedim{\dimension-1}=\set{\codeWord{1},\codeWord{2},\ldots,\codeWord{\abs{\sphericalCodedim{\dimension-1}}}}$

Set $\rewardLargestPossible$ so that it satisfies
    $\left(1-\frac{\frac{1}{\eulerNumber}\leftGapThreshold}{\rewardLargestPossible-\frac{1}{2\eulerNumber}}\right)\rewardLargestPossible=\frac{1}{4\eulerNumber}+\frac{\leftGapThreshold}{2\eulerNumber}$. That is, 
    \[\rewardLargestPossible = \frac{3}{8\eulerNumber}(1+2\leftGapThreshold) + \frac{1}{8\eulerNumber}\sqrt{36\leftGapThreshold^2+20\leftGapThreshold+1}.\]

Set $\equalRevenueReward$ so that it satisfies $\left(1-\frac{\frac{1}{\eulerNumber}\leftGapThreshold}{\equalRevenueReward-\frac{1}{2\eulerNumber}}\right)\equalRevenueReward=\frac{1}{4\eulerNumber}-\frac{\leftGapThreshold}{2\eulerNumber}$. That is
    \[\equalRevenueReward = \frac{3}{8\eulerNumber}+\frac{\leftGapThreshold}{4\eulerNumber} + \frac{1}{8\eulerNumber}\sqrt{4\leftGapThreshold^2+28\leftGapThreshold+1}.\]

$\contractThreshold\gets \frac{\leftGapThreshold}{\eulerNumber\rewardLargestPossible-\frac{1}{2}}$

\While{$\timeslot\leq \timeHorizon$}{
    Set $\contextAT{\adversarialAction} = \left(\rewardLargestPossible, \frac{\rewardLargestPossible}{2\radius}\cdot\codeWord{i}\right)$, $\contextAT{\restrictingAction} = \left(2\equalRevenueReward,0,\ldots,0\right)$.

    Show contexts to $\ALG$ and receive contract $\contract{\timeslot}$.

    \If{$\contractSimple_{\timeslot}<\contractThreshold$}{
     $\hvecs_{\timeslot+1} \gets \hvecs_{\timeslot}$
     
     $k\gets k+1$
     
     \If{$k=\determineThreshold$}{
        Determine $\hvecTrue = \argmax_{\hvec\in\hvecs_{\timeslot}}\lrangle{\contextAT{\adversarialAction},\hvec}$ and \textbf{halt}
     }
    }
    
    \If{$\contractSimple_{\timeslot}\geq\contractThreshold$}{
    $i \gets i+1$, $k\gets 0$

    $\hvecs_{\timeslot+1} \gets \hvecs_{\timeslot}\cap \set{\hvec = \left( \frac{1}{2}, \zeta\right) \condition \lrangle{\contextAT{\adversarialAction},\hvec}\leq \equalRevenueReward}$

    }

    \If{$i>\abs{\sphericalCodedim{\dimension-1}}$}{
    $i\gets 1$, $\radius\super{j+1} \gets \radius\super{j} \cdot \left( \frac{2\equalRevenueReward}{\rewardLargestPossible}-1 \right)$

    $\hvecs_{\timeslot+1} \gets \set{\frac{1}{2}}\times \left(\radius\super{j+1}\cdot \mathbb{B}_{d-1}\right)$

    $j\gets j+1$

    \If{$j\geq \frac{\timeHorizon}{\abs{\sphericalCodedim{\dimension-1}}\determineThreshold}$}{
            Arbitrarily determine a $\hvecTrue\in \hvecs_{\timeslot+1}$ and \textbf{halt}
        }
    }

    $\timeslot \gets \timeslot+1$
}
\end{adversary}

Here is our result about the Stackelberg regret.
\begin{theorem}\label{thm:lower-bound-reward-true}
    For contextual principal-agent game with reward contexts, let $\dimension$ be the dimension of the context, for any algorithm $\ALG$, there is an instance making the algorithm suffer $\Omega(T^{\frac{1}{2}-\frac{1}{2d}})$ Stackelberg regret.
\end{theorem}

We defer the proof of \Cref{thm:lower-bound-reward-true} at the end of this section, after establishing the necessary auxiliary lemmas. We begin with a foundational lemma confirming that our new adversary for the reward-context setting (\Cref{alg: adversary}) is also well-behaved.

\begin{lemma}\label{lem:valid adversary}
    Facing the \Cref{alg: adversary}, whatever the contract is, the agent will always take an action in $\actionsKnownT$. Moreover, \Cref{alg: adversary} is a valid adversary, which means that $\hvecs_{\timeslot}$ is consistent with the observed feedback of $\ALG$.
\end{lemma}

\begin{proof}
    The proof is analogous to that of \Cref{lem:valid adversary_cost} for the cost-context setting. The adversary in \Cref{alg: adversary} differs from its optimistic-benchmark counterpart (\Cref{alg: adversary_optimism}) only by the inclusion of early halting conditions.

As argued previously, these halting conditions only determine whether the interaction proceeds to the next round; they do not alter the adversary's behavior within any given round. Consequently, the logic establishing that the agent's action is confined to 
 and that the uncertainty set remains consistent is identical to that in the proof of \Cref{lem:valid adversary_optimism}. The original proof therefore applies without modification.
\end{proof}

Consistent with our analysis for the cost-context setting, we must also establish that the adversary for the reward-context case is well-behaved. This requires ensuring that all generated context vectors remain within the unit ball $\mathbb{B}_{d}$ and that their associated rewards are non-negative. The following lemma provides the sufficient condition for these properties to hold.

\begin{lemma}\label{lem:bounded context}

    Let $\sphericalCodedim{\dimension-1}$ be the spherical code used by our adversarial construction (\Cref{alg: adversary}). If $\sphericalCodedim{\dimension-1}$ satisfies the condition $\abs{\sphericalCodedim{\dimension-1}}\geq \frac{2\timeHorizon}{\determineThreshold}\left(1-\frac{\equalRevenueReward}{\rewardLargestPossible}\right)$, then before \Cref{alg: adversary} halts, the following properties hold:
\begin{enumerate}
\item[\textnormal{(i)}] All context vectors generated by the adversary are contained in the unit ball $\mathbb{B}_{d}$.
\item[\textnormal{(ii)}] The rewards of all corresponding actions are non-negative.
\end{enumerate}
\end{lemma}
\begin{proof}
    Similarly, we can obtain the same lower bound of the radius. By the halting condition in \Cref{alg: adversary}, $\radius\super{j}$ will shrink at most $\frac{T}{\abs{\sphericalCodedim{\dimension-1}}\determineThreshold}$ times. Thus, before \Cref{alg: adversary} halts, it always holds \[\radius\super{j}\geq \frac{\sqrt{3}}{2}\left(\frac{2\equalRevenueReward}{\rewardLargestPossible}-1\right)^j\geq \frac{\sqrt{3}}{2}\left(1-2 \cdot \frac{\rewardLargestPossible-\equalRevenueReward}{\rewardLargestPossible}\right)^{\frac{\timeHorizon}{\abs{\sphericalCodedim{\dimension-1}}\determineThreshold}}\geq \frac{\sqrt{3}}{2\eulerNumber}.\]

    Therefore, we have $ \lim_{ \leftGapThreshold \to 0} \abs{\contextAT{\adversarialAction{1}}} \leq \lim_{ \leftGapThreshold \to 0} \abs{\left(\rewardLargestPossible, \frac{\rewardLargestPossible}{2 \cdot \frac{\sqrt{3}}{2\eulerNumber}} \right)} <1 $. This implies when $\timeHorizon$ is large enough ($\leftGapThreshold$ is small enough), we have $\abs{\contextAT{\adversarialAction{1}}} < 1$.

\end{proof}

In the reward-context setting, we also require a more precise estimate of the optimistic contract threshold $\contractThreshold$.

\begin{lemma}
    $\contractThreshold-\frac{1}{2}\geq \leftGapThreshold -\frac{1}{8}\leftGapThreshold^2$.
\end{lemma}
\begin{proof}
    By the definition, 
    \begin{align*}
        \contractThreshold-\frac{1}{2} &= \frac{\leftGapThreshold}{\frac{3\leftGapThreshold}{4}+\frac{1}{8}\sqrt{36\leftGapThreshold^2+20\leftGapThreshold+1}-\frac{1}{8}} -\frac{1}{2}
        \\& = \leftGapThreshold-\frac{1}{8}\leftGapThreshold^2 +O(\leftGapThreshold^3).
    \end{align*}
\end{proof}

Here, we mirror our case analysis for the cost-context setting. First, we provide the regret lower bound when the principal acts aggressively.

\begin{lemma}\label{lem:case 2}
     At any $\timeslot$-th round, if $\ALG$ propose $\contract{\timeslot}\geq \contractThreshold$, then the immediate regret of this round satisfies
     \[\regretImmediate(\ALG)\geq \frac{1}{2\eulerNumber}\leftGapThreshold - \frac{1}{16\eulerNumber}\leftGapThreshold^2.\]
\end{lemma}
\begin{proof}
    If $\contract{\timeslot}\ge \contractThreshold$, the adversary will shrink the possible hidden vector such that $\rewardAT{\adversarialAction}=\lrangle{\contextAT{\adversarialAction},\hvecTrue}\leq \equalRevenueReward$ by \Cref{alg: adversary}. Note that $\rewardAT{\restrictingAction} = \equalRevenueReward$ and $\costAT{\restrictingAction} = \costAT{\adversarialAction}$. Thus, action $\adversarialAction$ is fully degenerated, i.e., it can not be induced whatever $\contract{\timeslot}$ is. Then, the instance is equivalent to $\instanceKnown$ when $\contract{\timeslot}\geq \contractThreshold$.

    If $\contract{\timeslot}\in [\contractTOrder{1},\contractTOrder{2})$, then $\knownOptAction$ is induced and 
    \begin{align*}
        \regretImmediate(\ALG) &\geq \optProfit{\instanceKnown}-\revenuePrincipal 
        \\&= \frac{1}{4\eulerNumber}- (1-\contract{\timeslot}) \rewardAT{\knownOptAction} 
        \\& \geq \frac{1}{4\eulerNumber}-(1-\contractThreshold)\rewardAT{\knownOptAction}
        \\&\geq  \frac{1}{4\eulerNumber}- (\frac{1}{2}-\leftGapThreshold+\frac{1}{8}\leftGapThreshold^2)\frac{1}{2\eulerNumber}
        \\&= \frac{1}{2\eulerNumber}\leftGapThreshold - \frac{1}{16\eulerNumber}\leftGapThreshold^2.
    \end{align*}

    If $\contract{\timeslot}\in [\contractTOrder{2},1)$, then $\restrictingAction$ is induced and 
    \begin{align*}
        \regretImmediate(\ALG) &\geq  \optProfit{\instanceKnown}-\revenuePrincipal 
        \\& \geq \frac{1}{4\eulerNumber} -\optProfitFromA{\instanceKnown}{\restrictingAction}
        \\& = \frac{1}{4\eulerNumber} - \left(\frac{1}{4\eulerNumber}-\frac{\leftGapThreshold}{2\eulerNumber}\right) = \frac{\leftGapThreshold}{2\eulerNumber}.
    \end{align*}
\end{proof}

Next, we analyze the scenario where the principal's cautious strategy leads to the adversary's early termination. This case demonstrates that prolonged passivity is, in itself, a costly strategy for the principal, as formalized in the lemma below.

\begin{lemma}\label{lem:case 1}
    Assuming the input spherical code is with minimal angle $\arccos \left(\frac{2\equalRevenueReward}{\rewardLargestPossible}-1 \right)$. Then there exists $\timeslot^*$ such that $k$ reaches $\determineThreshold$ at $\timeslot$-th round. Then the cumulative regret until $\timeslot^*$-th round satisfies
    \[\sum_{\timeslot=\timeslot^*-\determineThreshold+1}^{\timeslot^*}\regretImmediate(\ALG) \geq \frac{\leftGapThreshold}{2\eulerNumber}\determineThreshold.\]
\end{lemma}
\begin{proof}
    By the design of \Cref{alg: adversary}, there are consecutive $\determineThreshold$ rounds where $\ALG$ propose $\contract{\timeslot}<\contractThreshold$. That is because proposing $\contract{\timeslot}\geq \contractThreshold$ makes the adversary reset $k$. 

    Let $i^*, j^*$ be the $i,j$ at the beginning of the $\timeslot^*$-th round. We first show that $\left(\frac{1}{2},\radius\super{j^*} \cdot \codeWord{i^*}\right)\in \hvecs_{\timeslot^*}$. By the design of the adversary, 
    \[ \hvecs_{\timeslot^*} = \set{\frac{1}{2}}\times \left(\radius\super{j^*}\mathbb{B}_{\dimension-1}\bigcap \left(\bigcap_{i=1}^{i^*-1} \set{\zeta \condition \frac{\rewardLargestPossible}{2}+ \lrangle{\zeta, \frac{\rewardLargestPossible}{2\radius\super{j^*}}\cdot \codeWord{i}}\leq \equalRevenueReward}\right)\right).\]
    Since $\codeWord{i^*}\in \mathbb{B}_{\dimension-1}$, we have $\left(\frac{1}{2},\radius\super{j^*} \cdot \codeWord{i^*}\right)\in \set{\frac{1}{2}}\times \radius\super{j^*}\mathbb{B}_{d-1}$. Since the minimal angle of $\sphericalCodedim{\dimension-1}$ is $\arccos \left(\frac{2\equalRevenueReward}{\rewardLargestPossible}-1\right)$, we have 
    \[ \frac{\rewardLargestPossible}{2}+\lrangle{\radius\super{j^*}\codeWord{i^*},\frac{\rewardLargestPossible}{2\radius\super{j^*}}\codeWord{i}} \leq \frac{\rewardLargestPossible}{2}+\frac{\rewardLargestPossible}{2}\cdot \left( \frac{2\equalRevenueReward}{\rewardLargestPossible} -1\right) \leq \equalRevenueReward.\]
    Thus, $\left(\frac{1}{2},\radius\super{j^*} \cdot \codeWord{i^*}\right)\in \hvecs_{\timeslot^*}$.
    
     Therefore, $\lrangle{\contextTA{\timeslot^*}{\adversarialAction}, \hvecTrue}\geq \lrangle{\contextTA{\timeslot^*}{\adversarialAction}, \left(\frac{1}{2},\radius\super{j^*} \cdot \codeWord{i^*}\right)}= \rewardLargestPossible$. On the other hand, since $\hvecTrue\in \set{\frac{1}{2}}\times \radius\super{j^*}\mathbb{B}_{\dimension-1}$, we have $\lrangle{\contextTA{\timeslot^*}{\adversarialAction},\hvecTrue}\leq \rewardLargestPossible$. Thus, $\rewardA{\adversarialAction}_{\timeslot^*}=\lrangle{\contextTA{\timeslot^*}{\adversarialAction},\hvecTrue}=\rewardLargestPossible$. It's easy to check that $$\optProfit{\instance{\actions_{\timeslot^*}}{\hvecTrue}}=\frac{1}{4\eulerNumber}+\frac{\leftGapThreshold}{2\eulerNumber}.$$
     For any $\timeslot\in [\timeslot^*-\determineThreshold+1,\timeslot^*]$, the context $\contextTA{\timeslot^*}{\adversarialAction}$ is the same as $\contextAT{\adversarialAction}$, thus \[\optProfit{\instance{\actions_{\timeslot}}{\hvecTrue}}=\frac{1}{4\eulerNumber}+\frac{\leftGapThreshold}{2\eulerNumber}, \quad \forall \timeslot \in [\timeslot^*-\determineThreshold, \timeslot^*].\] For any $\timeslot\in [\timeslot^*-\determineThreshold, \timeslot^*]$, since the action will always take an action in $\actionsKnownT$ by \Cref{lem:valid adversary}. Then 
     \[\revenuePrincipal \leq \max_{\action\in\actionsKnownT}\optProfitFromA{\instance{\actions}{\hvecTrue}}{\action} \leq \max_{\action\in\actionsKnownT}\optProfitFromA{\instanceKnown}{\action}= \optProfit{\instanceKnown} = \frac{1}{4\eulerNumber}. \]
    Therefore, 
    \[\regretImmediate(\ALG) =\optProfit{\instance{\actions_{\timeslot}}{\hvecTrue}} -\revenuePrincipal\geq \frac{\leftGapThreshold}{2\eulerNumber}, \forall \timeslot\in  [\timeslot^*-\determineThreshold, \timeslot^*].\]
    Simply sum up $\regretImmediate(\ALG)$ over rounds $\timeslot\in  [\timeslot^*-\determineThreshold, \timeslot^*]$, we have 
    \[\sum_{\timeslot=\timeslot^*-\determineThreshold+1}^{\timeslot^*}\regretImmediate(\ALG) \geq \frac{\leftGapThreshold}{2\eulerNumber}\determineThreshold.\]
\end{proof}

We are now in a position to prove \Cref{thm:lower-bound-reward-true}.

\begin{proof}[\bf{Proof of \Cref{thm:lower-bound-reward-true}}]
First, we set $\leftGapThreshold = \min\{\frac{1}{8}C^{\frac{1}{\dimension}} (\dimension-1)^{\frac{1}{2\dimension}} \timeHorizon^{-\frac{1}{2\dimension}},\frac{1}{8}\}$ and $\determineThreshold=\sqrt{\timeHorizon}$. Notice that $1-\left(\frac{2\equalRevenueReward}{\rewardLargestPossible}-1 \right)=2- 2 \cdot \frac{3+2\leftGapThreshold+\sqrt{4\leftGapThreshold^2+28\leftGapThreshold+1}}{3+6\leftGapThreshold+\sqrt{36\leftGapThreshold^2+20\leftGapThreshold+1}} = 32\leftGapThreshold^2-640\leftGapThreshold^3+O(\leftGapThreshold^4)$. By \Cref{lem:code lower bound}, we have
    $$\abs{\sphericalCodedim{\dimension-1}} \geq C \frac{\left(\frac{2\equalRevenueReward}{\rewardLargestPossible} -1\right) \sqrt{d-1}}{ \sin^{d-2} \left(\arccos{\left(\frac{2\equalRevenueReward}{\rewardLargestPossible} -1\right)}\right)}   \geq C\frac{(1-32\leftGapThreshold^2)\sqrt{\dimension-1}}{(8\leftGapThreshold)^{d-2}} \geq 32\sqrt{\timeHorizon} \leftGapThreshold^2 \geq \frac{2\timeHorizon}{\determineThreshold} \left(1- \frac{\equalRevenueReward}{\rewardLargestPossible}\right).$$
    Therefore, the conditions of \Cref{lem:bounded context} are satisfied and our construction of adversary is valid.

    We claim that one of the following cases will happen:
    \begin{enumerate}
        \item There are at least $\frac{\timeHorizon}{\determineThreshold}$ rounds that $\ALG$ proposed $\contract{\timeslot}\geq \contractThreshold$ before the adversary halts.
        \item There exist some $\timeslot^*$ such that the counter $k$ reaches $\determineThreshold$.
    \end{enumerate}
    That is because, if $k$ never reaches $\determineThreshold$, then $\ALG$ never make contract which is less than $\contractThreshold$ for consecutive $\determineThreshold$ rounds, this directly imply that there are at least $\frac{\timeHorizon}{\determineThreshold}$ rounds for which $\ALG$ propose $\contract{\timeslot}\geq \contractThreshold$.

    If we fall into case $1$, the cumulative regret
    \[ \regret(\ALG)\geq \left(\frac{1}{2\eulerNumber}\leftGapThreshold - \frac{1}{16\eulerNumber}\leftGapThreshold^2\right)\frac{\timeHorizon}{\determineThreshold}\]
    by \Cref{lem:case 2}.

    If we fall into case $2$, the cumulative regret 
    \[\regret(\ALG)\geq \sum_{\timeslot=\timeslot^*-\determineThreshold+1}^{\timeslot^*}\regretImmediate(\ALG) \geq \frac{\leftGapThreshold}{2\eulerNumber}\determineThreshold\]
    by \Cref{lem:case 1}.\\
    Combining the results of both two cases, we complete the proof.
\end{proof}

\section{Missing Proofs}
\label{apx:proofs}

\subsection{Proof of Theorem~\ref{thm:warm-up:pricing connection}}
\label{apx:pricing connection proof}
\thmTwoActionRegret*
\begin{proof}
Since we consider the single non-trivial action case, we simplify the notations as follows.
We use $\context_{\timeslot}$ to denote the context of the non-trivial action, and use $\context_{\timeslot}^{\dagger}$ to denote the context of the valuation of the buyer in contextual pricing. 
We use $\cost_{\timeslot}$ to denote the cost of the non-trivial action, and use $v_{\timeslot}$ to denote the valuation of the buyer.
Since there is only $1$ non-trivial action, its reward can be normalized to $1$.

Suppose we have an algorithm $\ALG^{\dagger}$ for the contextual pricing problem. We construct an algorithm $\ALG$ for the contextual principal-agent game as follows:
\begin{itemize}
    \item $\ALG$ receive context $\context_{\timeslot}\in \mathbb{R}^{\dimension}$ and feed a valuation context $\context_{\timeslot}^{\dagger} \deq (\sqrt{2},-\sqrt{2}\context_{\timeslot})\in \mathbb{R}^{\dimension+1}$ to $\ALG^{\dagger}$. 
    \item $\ALG$ receive a price $p_{\timeslot}$ from $\ALG^{\dagger}$ and propose a linear contract $\contract{\timeslot} = 1-p_{\timeslot}$.
    \item $\ALG$ observe the feedback. If the agent takes non-trivial action, tell $\ALG^{\dagger}$ that the buyer makes a purchase. Else, tell $\ALG^{\dagger}$ there is no purchase happens.
\end{itemize}
Algorithm $\ALG$ repeats the above procedure each round. If the contextual principal-agent game has a hidden vector $\hvecTrue\in \mathbb{R}^{\dimension}$, then the feedback received by algorithm $\ALG^{\dagger}$ is consistent with a $(\dimension+1)$-dimensional contextual pricing with hidden vector $\hvecTrue^{\dagger}=(\frac{\sqrt{2}}{2}, \frac{\sqrt{2}}{2}\hvecTrue)\in \mathbb{B}_{\dimension+1}$. Note that $\lrangle{\context_{\timeslot}^{\dagger}, \hvecTrue^{\dagger}} = 1 - \lrangle{\context_{\timeslot},\hvecTrue}$ and $\ALG$ propose $\contract{\timeslot}=1-p_{\timeslot}$. By the correspondence relationship between the principal-agent game and pricing we described in \Cref{sec:contextual pricing connection}, the feedback to $\ALG^{\dagger}$ is the same as the corresponding pricing problem. Also, the regret of $\ALG$ and $\ALG^{\dagger}$ are the same, which is 
\[O((\dimension+1)\log\log \timeHorizon+(\dimension+1)\log (\dimension+1)) = O(\dimension \log\log \timeHorizon + \dimension\log\dimension)\]
if we employ the algorithm in \citep{LLS-21} as $\ALG^{\dagger}$.
\end{proof}

\begin{remark}
    Using the same idea, we can also reduce contextual pricing to the contextual principal-agent game with an additional context dimension.
\end{remark}

\section{Action-Dependent Contract and Contextual Multi-item Dynamic Pricing}
\label{apx:act-dep contract and pricing}

In this section, we consider a variant setting where the principal is allowed to specify action-dependent payments, the model becomes (almost) equivalent to contextual multi-item dynamic pricing with unit-demand buyers. This setting is similar to the multi-item pricing problem, in which we show that a regret with a double-logarithmic dependence on $T$ is achievable. We then introduce a reduction to the principal-agent setting with action-dependent payments.

\xhdr{Contextual Multi-item Dynamic Pricing Model.}
There are $n$ items, each represented by a vector $\hvecTrue^{(i)} \in \mathbb{R}^{\dimension}$ for $i \in [n]$.\footnote{When all items share the same hidden vector, we can further improve the regret in \Cref{thm:n loglog T upper bound} by dropping the dependence on $n$.} Each timestep $\timeslot \in [\timeHorizon]$, a seller observes  a buyer with context profile $\{\mu_{t}^{(i)}\}_{i\in[n]}$ and posts prices $p_{t}^{(1)}, \ldots, p_t^{(n)} \in [0,1]$ for the items. The buyer chooses the utility-maximizing item, i.e the item such that $\langle \mu_{t}^{(i)}, \hvecTrue^{(i)} \rangle - p_t^{(i)}$ is maximized. If no such item exists, the buyer does not make a purchase. Lastly, the seller observes the purchased item and obtains revenue $p_t^{(i)}$ if item $i$ is chosen by the buyer.

The seller's goal is to maximize cumulative revenue. We can define regret with respect to the best choice of prices. Let $j_t$ denote the item chosen by the buyer at timestep $t$ i.e $j_t = \argmax_{i \in [n]} \langle \mu_{t}^{(i)}, \hvecTrue^{(i)} \rangle - p_t^{(i)}$. The \emph{Stackelberg regret} of the seller is 
\begin{align*}
\REG \triangleq  \sum\nolimits_{t\in[T]} \max_{i \in [n]} \langle \mu_{t}^{(i)}, \hvecTrue^{(i)} \rangle - p_t^{(j_t)}.
\end{align*}

\begin{lemma}
\label{thm:n loglog T upper bound}
    There exists an online algorithm with Stackelberg regret~\footnote{Note that our proof is a reduction to the single-item pricing algorithm in \cite{LLS-21}, which has the essentially same bound for pessimistic Stackelberg regret. Thus, our lemma also holds for pessimistic Stackelberg regret.} of 
    $$O(\dimension \log\log T + \dimension\log\log\dimension).$$
\end{lemma}
\begin{proof}
We first introduce the algorithm and then analyze its regret.

\xhdr{Algorithm construction.} We use Algorithm 2 in \citet{LLS-21} as a subroutine of our algorithm, which we refer to as \LLS. Specifically, for each item $i\in[n]$, we create a sub-algorithm $\LLS_i$ based on \LLS\ (imagining selling a single item with unknown item-dependent feature vector $\hvecTrue^{(i)}$). We specify the interaction to all these sub-algorithms $\{\LLS_i\}_{i\in[n]}$ below.

For each buyer $t\in[T]$ and each item $i\in[n]$, query $\LLS_i$ with input $\mu_{t}^{(i)}$ as a new buyer and let $y_{t}^{(i)}$ be the price computed in $\LLS_i$. Let 
\begin{align*}
    i^* \triangleq \argmax_{i\in[n]} y_{t}^{(i)}.
\end{align*}
Post uniform price $y_t^{(i^*)}$ for all items to buyer $t$. Consider the following two cases:
\begin{itemize}
    \item Suppose buyer $t$ purchases nothing. Respond to $\LLS_{i^*}$ that the buyer with context $\mu_{t}^{(i)}$ does not purchase item $i^*$ given price $y_{t}^{(i^*)}$. For each other item $i'\not= i^*$, respond nothing to $\LLS_{i'}$ (as if no buyer with context $\mu_t^{(i')}$ appears to $\LLS_{i'}$).
    \item Suppose buyer $t$ purchases item $i$ for some $i\in[n]$. Respond to $\LLS_{i}$ that the buyer with context $\mu_{t}^{(i)}$ purchases item $i$ given price $y_{t}^{(i)}$.\footnote{Actually, the buyer with context $\mu_{t}^{(i)}$ purchases item $i$ even given higher price $y_{t}^{(i^*)} > y_{t}^{(i)}$. Nonetheless, telling sub-algorithm that the buyer with context $\mu_{t}^{(i)}$ purchases item $i$ given price $y_{t}^{(i)}$ is enough for our regret upper bound.} For each other item $i'\not= i$, respond nothing to $\LLS_{i'}$ (as if no buyer with context $\mu_t^{(i')}$ appears to $\LLS_{i'}$).
\end{itemize}

\xhdr{Regret analysis.} In the algorithm construction, for each buyer $t$, the algorithm responds to sug-algorithm $\LLS_{i}$ for a unique item $i\in[n]$. Let $\timeset_i$ be the subset of buyers (time periods) that the algorithm responds to sub-algorithm $\LLS_{i}$. We compare the regret of the constructed algorithm for all buyers $t\in[T]$ with the total regret from $\LLS_{i}$, each facing a sequence of buyers specified by $\timeset_i$. 

Specifically, let $\REG$ be the regret of the constructed algorithm for all buyers $t\in[T]$. For each item $i\in[n]$, let $\REG_i$ be the regret of running LLS for selling a single item with item-dependent feature $\hvecTrue^{(i)}$ for $|\timeset_i|$ number of buyers, each with context $\{\mu_{t}^{(i)}\}_{t\in\timeset_i}$. Recall \citet{LLS-21} shows that 
\begin{align*}
    \REG_i = O(\dimension\log\log|\timeset_i| + \dimension\log\log\dimension).
\end{align*}
which is further upper bounded by $O(\dimension\log\log T + \dimension\log\log\dimension)$ since $|\timeset_i|\leq T$.

To prove the theorem statement, it suffices to show 
\begin{align*}
    \REG \leq \sum\nolimits_{i\in[n]}\REG_i.
\end{align*}
We prove this inequality by comparing the left-hand side and right-hand side for each buyer $t\in[T]$ individually.
\begin{itemize}
    \item Suppose buyer $t$ purchases nothing. In this case, $\REG$ increases by one, while $\REG_{i^*}$ increases by one as well.\footnote{Here we do not consider the actual regret, but an upper bound of the actual regret by relaxing the benchmark to be one, instead of $\max_{i\in[n]}\langle\hvecTrue^{(i)},\mu_{t}^{(i)}\rangle$ and $\langle\theta^{(i^*)},\mu_t^{(i^*)}\rangle$ for the algorithm and $\LLS_{i^*}$, respectively. This is valid, since the regret upper bound of $\LLS_{i^*}$ in \citet{LLS-21} considers the same relaxation.}
    \item Suppose buyer $t$ purchases item $i$ for some $i\in[n]$. In this case, due to the uniform price, we must have $i = \argmax_{i'\in[n]}\langle\hvecTrue^{(i')},\mu_t^{(i')}\rangle$. Hence, $\REG$ increases by $\max_{i'\in[n]}\langle\hvecTrue^{(i')},\mu_t^{(i')}\rangle - y_{t}^{(i^*)} = \langle\hvecTrue^{(i)},\mu_{t}^{(i)}\rangle - y_{t}^{(i^*)}$, while $\REG_i$ increases by $\langle\hvecTrue^{(i)},\mu_{t}^{(i)}\rangle - y_{t}^{(i)}$. The definition of index $i^*$ ensures that $y_{t}^{(i)} \leq y_{t}^{(i^*)}$ and thus the increment in $\REG$ is at most the increment in $\REG_i$.
\end{itemize}
Combining the two cases finishes the regret analysis as desired.
\end{proof}

By using the similar idea in \Cref{apx:pricing connection proof}, we can establish the connection between contextual principal-agent games with action-dependent contracts and contextual multi-item dynamic pricing and have the following result about action-dependent contracts.

\begin{theorem}
    For principal-agent games under the setting using action-dependent contracts, there exists an online algorithm with Stackelberg regret of 
    $$O(\dimension \log\log T + \dimension\log\log\dimension).$$
\end{theorem}

\begin{proof}
    Suppose we have an algorithm $\ALG^{\dagger}$ for the contextual multi-item dynamic pricing problem. We construct an algorithm $\ALG$ for the contextual principal-agent game as follows:
\begin{itemize}
    \item $\ALG$ receive contexts $\context_{\timeslot}^{(i)}\in \mathbb{R}^{\dimension}$ and the corresponding rewards $ \rewardAT{i}$ for each $i \in [n]$ and feed valuation contexts $\context_{\timeslot}^{(i),\dagger} \deq (\sqrt{2}\rewardAT{i},-\sqrt{2}\context_{\timeslot}^{(i)})\in \mathbb{R}^{\dimension+1}$ to $\ALG^{\dagger}$. 
    \item $\ALG$ receive a price vector $(p_{\timeslot}^{(1)},...,p_t^{(n)})$ from $\ALG^{\dagger}$ and propose an action-dependent contract $(\contract{\timeslot}^{(1)},...,\contract{\timeslot}^{(n)})$, where $\contract{\timeslot}^{(i)} = \max\left\{r_t^{(i)}-p_{\timeslot}^{(i)},0\right\}$.
    \item $\ALG$ observe the feedback. If the agent takes a non-trivial action $i_t$, tell $\ALG^{\dagger}$ that the buyer purchases the item $i_t$. Else, tell $\ALG^{\dagger}$ there is no purchase happens.
\end{itemize}

Algorithm $\ALG$ repeats the above procedure each round. If the contextual principal-agent game has a hidden vector $\hvecTrue\in \mathbb{R}^{\dimension}$, then the feedback received by algorithm $\ALG^{\dagger}$ is consistent with a $(\dimension+1)$-dimensional contextual multi-item dynamic pricing with the same hidden vector $\hvecTrue^{(i),\dagger}=(\frac{\sqrt{2}}{2}, \frac{\sqrt{2}}{2}\hvecTrue)\in \mathbb{B}_{\dimension+1}$ for each item $i \in [n]$. Note that $\lrangle{\context_{\timeslot}^{(i),\dagger}, \hvecTrue^{(i),\dagger}} = r_t^{(i)} - \lrangle{\context_{\timeslot}^{(i)},\hvecTrue}$ and $\ALG$ propose $\contract{\timeslot}^{(i)}=\max\{r_t^{(i)}-p_{\timeslot}^{(i)},0\}$. If the agent takes a non-trivial action $i_t$, it implies 
$x_t^{(i_t)} = r_t^{(i_t)}-p_t^{(i_t)} \geq c_t^{(i_t)}= \langle \mu_t^{(i_t)},\hvecTrue\rangle \geq 0$. 
On the other hand, we also have $i_t \in \argmax_{j \in [n]}\{ x_t^{(j)}-\langle \mu_t^{(j)},\hvecTrue\rangle\}$.
Thus, for every $j \in [n]$, we have $\langle \mu_t^{(i_t),\dagger},\hvecTrue^{(i_t),\dagger}\rangle-p_t^{(i_t)}=r_t^{(i_t)}-\langle \mu_t^{(i_t)},\hvecTrue\rangle -p_t^{(i_t)}=x_t^{(i_t)}-\langle \mu_t^{(i_t)},\hvecTrue\rangle \geq x_t^{(j)}-\langle \mu_t^{(j)},\hvecTrue\rangle \geq r_t^{(j)}-p_t^{(j)}-\langle \mu_t^{(j)},\hvecTrue\rangle = \langle \mu_t^{(j),\dagger},\hvecTrue^{(j),\dagger}\rangle-p_t^{(j)}$. Therefore, in contextual multi-item dynamic pricing, the buyer also purchases the item $i_t$. A similar analysis can be applied to the trivial zero action and the no-purchase action. By the correspondence relationship between the principal-agent game with action-dependent contracts and multi-item dynamic pricing we described, the feedback to $\ALG^{\dagger}$ is the same as the corresponding pricing problem and the utility of the principal from action $i$ equals to the utility of the seller from item $i$, which are both $p_{t}^{(i)}$. 
Furthermore, for $t$-th round, since the optimal utility of the principal in the principal agent game is $\max_{i \in [n]}\{ r_t^{(i)}-\langle \mu_t^{(i)},\hvecTrue\rangle\}$ by proposing an action-dependent contract with zero payoff for each action except for action $i^* \in \argmax \set{r_t^{(i)}-\lrangle{\mu_t^{(i)},\hvecTrue}}$, whose payoff is $x_t^{(i^*)}=\langle \mu_t^{(i^*)},\hvecTrue\rangle$. 
The optimal principal's utility is the same as the optimal utility of the seller, $\max_{i \in [n]}\{ \langle \mu_t^{(i),\dagger},\hvecTrue^{(i),\dagger}\rangle\}$, in multi-item dynamic pricing. The optimal utility is achieved by setting the price vector as $+\infty$ for each items except for the most valuable item $i^* \in \argmax_{i\in [n]}\set{\lrangle{\mu_t^{(i),\dagger},\hvecTrue^{(i^*),\dagger}}} = \argmax_{i\in [n]} \set{r_t^{(i)}-\langle \mu_t^{(i)},\hvecTrue\rangle}$, whose price is its valuation $p_t^{(i^*)}=\langle \mu_t^{(i^*),\dagger},\hvecTrue^{(i^*),\dagger}\rangle$.
Note that the principal's utility in $t$-th round, $r_t^{(i_t)}-x_t^{(i_t)}$, and the seller's utility, $p_t^{(i_t)}$, are also the same. 
We have the regret of $\ALG$ and $\ALG^{\dagger}$ are the same, which is 
\[O((\dimension+1)\log\log \timeHorizon+(\dimension+1)\log\log (\dimension+1)) = O(\dimension \log\log \timeHorizon + \dimension\log\log\dimension)\]
if we employ the algorithm in \Cref{thm:n loglog T upper bound} as $\ALG^{\dagger}$.
\end{proof} 
\section{Validity of the Non-Degenerating Contract Definition}
\label{apx:non-degenerating contract validity}
\begin{lemma}\label{lem:compactness}
$\set{\reward\in [0,1]\condition \contracts_{\instance{\actionsKnown_{\timeslot}\cup \set{\actionOptT}}{\reward},\actionOptT}\neq \emptyset}$ is non-empty and compact.
\end{lemma}
\begin{proof}
    For simplicity, denote $S:= \set{\reward\in [0,1]\condition \contracts_{\instance{\actionsKnown_{\timeslot}\cup \set{\actionOptT}}{\reward},\actionOptT}\neq \emptyset}$.
    Note that by the definition of $\actionOptT$, $\contracts_{\instance{\actionsKnown_{\timeslot}\cup \set{\actionOptT}}{\lrangle{\contextAT{\actionOptT},\hvecOptT}},\actionOptT}\neq \emptyset$. Therefore, $S$ is not empty.
    
    Given any converging sequence $\reward_n \rightarrow \reward_0$ with $\reward_n \in S$, we can find $\contractSimple_{n}\in \contracts_{\instance{\actionsKnown_{\timeslot}\cup \set{\actionOptT}}{\reward_n},\actionOptT}$ since $\contracts_{\instance{\actionsKnown_{\timeslot}\cup \set{\actionOptT}}{\reward_n},\actionOptT}\neq \emptyset$. Since $\contractSimple_n \in [0,1]$, we can find a converging subsequence $\contractSimple_{n_k}\rightarrow \contractSimple_0\in [0,1]$.
    Next, we show that $\contractSimple_0\in \contracts_{\instance{\actionsKnown_{\timeslot}\cup \set{\actionOptT}}{\reward_0},\actionOptT}$. By definition,

    \[\contractSimple_{n_k}\cdot\reward_{n_k}-\costA{\actionOptT}\geq \contractSimple_{n_k}\cdot\rewardA{\action^{\dagger}}-\costA{\action^{\dagger}}, \forall \action^{\dagger}\in \actionsKnown.\]
    Take limitation $k\rightarrow \infty$ both sides, we have 
    \[ \contractSimple_{0}\cdot\reward_{0}-\costA{\actionOptT}\geq\contractSimple_{0}\cdot\rewardA{\action^{\dagger}}-\costA{\action^{\dagger}}, \forall \action^{\dagger}\in \actionsKnown,\]
    which directly leads $\contractSimple_0\in \contracts_{\instance{\actionsKnown_{\timeslot}\cup \set{\actionOptT}}{\reward_0},\actionOptT}$. Thus $\contracts_{\instance{\actionsKnown_{\timeslot}\cup \set{\actionOptT}}{\reward_0},\actionOptT}\neq \emptyset$ and $\reward_0\in S$. Since our argument holds for any converging sequence $\set{\reward_n}_{n}$, $S$ is closed. Furthermore, $S\subseteq [0,1]$, so $S$ is compact. 
\end{proof}

\begin{lemma}\label{lem:non empty}
    We have $\contracts_{\instance{\actionsKnown_{\timeslot}\cup \set{\actionOptT}}{\nonDeReward{\actionOptT}},\actionOptT}\neq \emptyset$. Moreover, it is a closed interval in $\realNumbers$.
\end{lemma}
\begin{proof}
    By \Cref{lem:compactness}, $\nonDeReward{\actionOptT}\in \set{\reward\in [0,1]\condition \contracts_{\instance{\actionsKnown_{\timeslot}\cup \set{\actionOptT}}{\reward},\actionOptT}\neq \emptyset}$. Therefore, $\contracts_{\instance{\actionsKnown_{\timeslot}\cup \set{\actionOptT}}{\nonDeReward{\actionOptT}},\actionOptT}\neq \emptyset$. By definition, $\set{\reward\in [0,1]\condition \contracts_{\instance{\actionsKnown_{\timeslot}\cup \set{\actionOptT}}{\reward},\actionOptT}\neq \emptyset}$ is closed and convex. Since $$\set{\reward\in [0,1]\condition \contracts_{\instance{\actionsKnown_{\timeslot}\cup \set{\actionOptT}}{\reward},\actionOptT}\neq \emptyset}\subseteq [0,1],$$ it is a closed interval (or singleton). 
\end{proof}

\end{document}